\documentclass[aps,prb,notitlepage,onecolumn,citeautoscript]{revtex4-1}
\usepackage{amsmath,amssymb,mathrsfs,bm,feynmf,setspace}
\usepackage[normalem]{ulem} 
\usepackage{natbib}
\usepackage{graphicx}
\usepackage[tight]{subfigure}
\usepackage{color}
\usepackage{geometry}
\usepackage{hyperref}
\newcommand{\al}{\alpha}
\newcommand{\be}{\beta}
\newcommand{\g}{\gamma}
\newcommand{\de}{\delta}
\newcommand{\e}{\epsilon}
\newcommand{\ve}{\varepsilon}

\newcommand{\ka}{\kappa}
\newcommand{\la}{\lambda}
\newcommand{\La}{\Lambda}

\newcommand{\p}{\pi}

\newcommand{\s}{\sigma}

\newcommand{\w}{\omega}
\newcommand{\W}{\Omega}
\newcommand{\De}{\Delta}

\renewcommand{\S}{\Sigma}

\newcommand{\pd}{\partial}

\newcommand{\ang}[1]{\left\langle #1 \right\rangle}

\newcommand{\beq}{\begin{equation}}
\newcommand{\eeq}{\end{equation}}
\newcommand{\Beq}{\begin{eqnarray}}
\newcommand{\Eeq}{\end{eqnarray}}
\newcommand{\bml}{\begin{multline}}

\newcommand{\eeqm}{\end{multline}}

\newcommand{\bsp}{\begin{split}}
\newcommand{\esp}{\end{split}}

\renewcommand{\b}[1]{{\bm #1}}
\renewcommand{\t}{\tilde}

\newcommand{\inv}{^{-1}}

\newcommand{\mc}{\mathcal}

\renewcommand{\t}{\tilde}
\newcommand{\req}[1]{Eq.~(\ref{eq:#1})}

\newcommand{\rfig}[1]{Fig.~\ref{fig:#1}}
\newcommand{\nn}{\nonumber}

\DeclareMathOperator{\tr}{tr}

\DeclareMathOperator{\sgn}{sgn}

\newcommand{\largen}{large-$N$ }

\newcommand{\kapET}{$\ka$-(BEDT-TTF)$_2$Cu$_2$(CN)$_3$ }

\begin{document}

\title{Universal transport near a quantum critical Mott transition in two dimensions}
\author{William Witczak-Krempa$^1$, Pouyan Ghaemi$^2$, T. Senthil$^3$
and Yong Baek Kim$^{1,4}$}
\affiliation{
$^1$Department of Physics, University of Toronto, Toronto, Ontario M5S 1A7, Canada\\
$^2$Department of Physics, University of Illinois at Urbana-Champaign,
   Urbana, Illinois 61801-3080, USA\\
$^3$Department of Physics, Massachusetts Institute of Technology, Cambridge, Massachusetts 02139, USA \\
$^4$School of Physics, Korea Institute for Advanced Study, Seoul 130-722, Korea
}
\date{\today}
\begin{abstract}
We discuss the universal transport signatures near a zero-temperature \emph{continuous}
Mott transition between a Fermi liquid (FL) and a quantum spin liquid in two spatial dimensions.
The correlation-driven transition occurs at fixed filling and involves fractionalization
of the electron: upon entering the spin liquid,
a Fermi surface of neutral spinons coupled to an internal
gauge field emerges. We present a controlled calculation of the value of the
zero temperature universal resistivity jump predicted to occur at the transition.
More generally, the behavior of the universal scaling function
that collapses the temperature and pressure dependent resistivity is derived,
and is shown to bear a strong imprint of
the emergent gauge fluctuations. We further predict a universal jump of the thermal
conductivity across the Mott transition, which derives from the breaking of conformal invariance by
the damped gauge field, and leads to a violation of the Wiedemann-Franz law %\red{\sout{at the quantum critical point} 
in the quantum critical region. A connection to organic salts is made, where such a transition might occur.
Finally, we present some transport results for the pure rotor $O(N)$ CFT.
% We analyze the electric transport near a \emph{continuous} quantum phase transition between
% a Fermi liquid metal and a Mott insulator in two dimensions. We consider the case
% where the insulator is a gapless quantum spin liquid (SL) with a Fermi surface of fractional
% spinons. At zero temperature, the resistivity shows a universal jump at the critical point,
% whose precise value we determine. The evolution of the universal jump at finite temperature
% is also determined. A saliant feature is that the emergent gauge fluctuations associated
% with the fractionalized SL play a dominant role in the resistivity.
\end{abstract}
\maketitle

\tableofcontents
\section{Introduction}
Despite decades of study, the Mott metal-insulator transition remains a central problem in 
quantum many body physics\cite{imada98-rev}. %\red{(added citation to Imada's RMP)}.
In recent years attention has refocused on an old question: can the Mott transition at $T = 0$ be continuous? 
Usually the Mott insulating state also has magnetic long range order or in some cases broken lattice symmetry which doubles the unit cell. In such situations a continuous Mott transition between a symmetry unbroken metal and the Mott insulator requires not just the continuous onset of the broken symmetry but also the continuous destruction of the metallic Fermi surface. It is currently not clear theoretically if such a continuous quantum phase transition can ever occur.

Considerable theoretical progress\cite{florens04,sslee05,senthil08-1,senthil08-2,podolsky09,potter12,rahul} has been possible in situations in which the Mott insulator does not break any symmetries but rather is in a quantum spin liquid (SL) state\cite{anderson73}.
%\red{[edit:the next sentence was moved from the spot after which we mention Na4Ir3O8.]} 
The evolution from the Fermi liquid (FL) metal to the quantum spin liquid state is of interest because it 
provides an opportunity to understand the fundamental phenomenon of the Metal-Insulator
Transition (MIT) without the complications of the interplay with the onset of broken symmetry. 
Such a transition has acquired experimental relevance with the discovery of quantum spin liquid Mott insulators 
near the Mott transition in a few different materials, notably the 
quasi--two-dimensional triangular lattice organic 
salts\cite{kapEt_NMR_chi,sdmit_heat-cap,kato-prb08,kapEt_heat-cap,kapEt_thermal-transp,kato_kanoda11-rev} \kapET and EtMe$_3$Sb[Pd(dmit)$_2$]$_2$, and the three dimensional hyper-Kagom\'e material\cite{takagi} Na$_4$Ir$_3$O$_8$. Indeed upon application of hydrostatic pressure, these quantum spin liquids become
metallic\cite{kapEt_p-nmr-res}.(The phase diagram of \kapET in
addition has superconductivity at lower temperatures.)
The nature of the transition in the experiments is not currently understood but could be potentially
described by a SL-FL quantum critical MIT.

A wide variety of quantum spin liquid phases can exist theoretically. However a natural state\cite{motrunich05,sslee05} that emerges near the Mott transition is a gapless quantum spin liquid which has a Fermi surface of spin-1/2 neutral quasiparticles, the spinons, while the charge excitations are
fully gapped. Consequently there is a single particle gap in the electron spectral function even though there
are gapless spin-carrying excitations. This type of spin-charge separation where only the charge localizes can be
favored in systems (such as the organics)  where frustration and charge fluctuations suppress magnetic ordering.

\begin{figure}
\centering
\includegraphics[scale=.5]{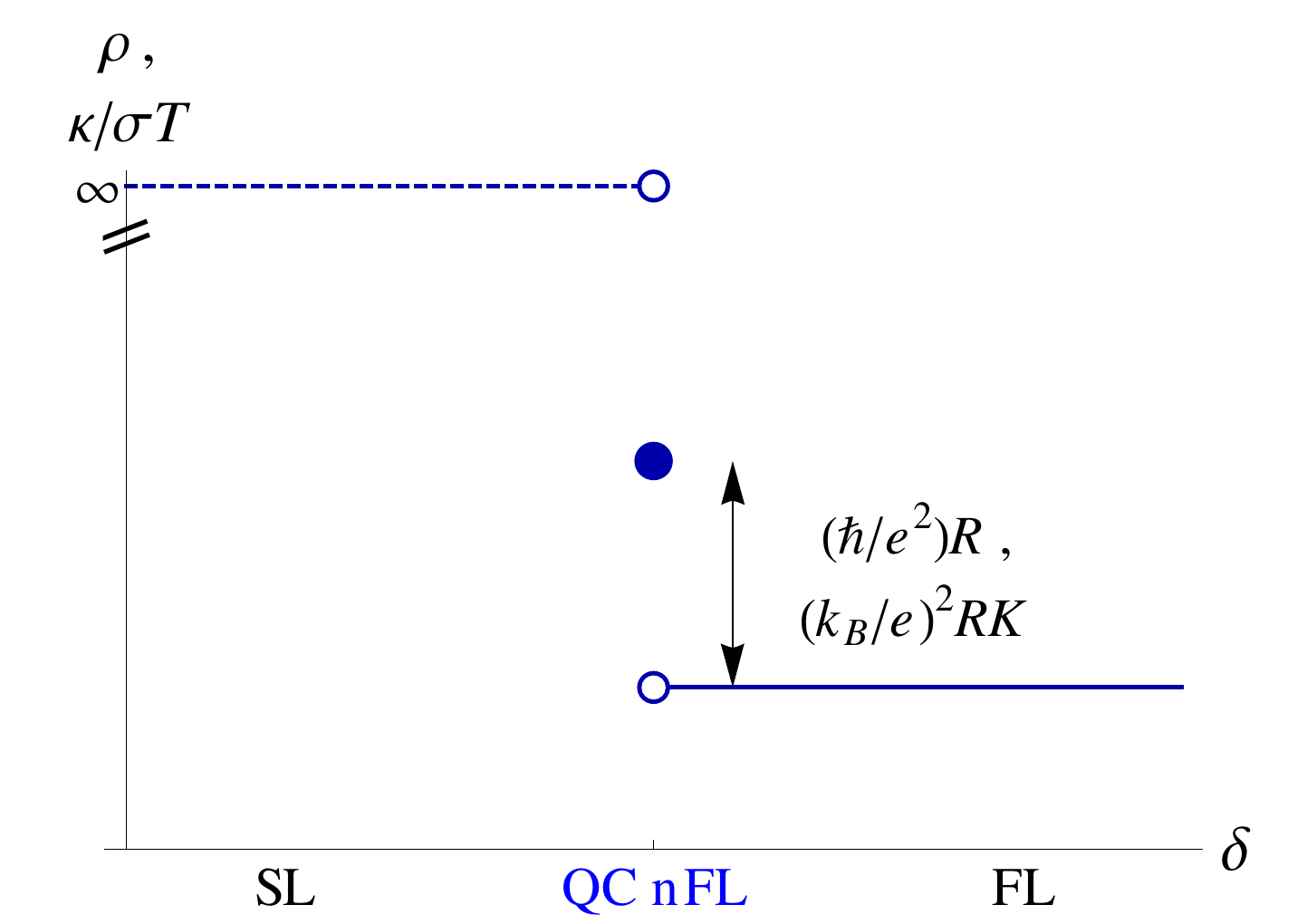}
\caption{\label{fig:jump}
Jump of the universal resistivity, $\rho$, and Lorentz number, $\ka/\s T$, at $T=0$ as a
function of $\de$, which is proportional to ratio of the bandwidth to the Hubbard repulsion, \req{de}.
The latter jump signals a violation of the Wiedemann-Franz law by the critical Fermi surface state.
$\ka$ is the thermal conductivity; $R, K$ are universal constants associated with the
Mott QCP. In particular, they strongly depend on the emergent gauge boson
associated with the electron fractionalization. The resistivity becomes infinite
in the SL, and as a consequence so does the Lorentz number.
}
\end{figure}

The MIT we consider was studied at the mean field level in Ref.~\onlinecite{florens04}.
A subsequent analysis\cite{senthil08-2} of the quantum fluctuations provided
evidence that the second order nature can survive the (inevitable) inclusion of many-body effects.
A rich set of properties associated with the
quantum critical point (QCP) was uncovered, many not present at the mean field level. For example, it
was found that quantum fluctuations lead to a divergence of the effective FL-mass
as one approaches the MIT. This gives rise to a divergence of the specific heat capacity coefficient,
$\g=C/T$. A key insight was the realization that the quantum critical state at the edge between
the FL and SL is actually a non-FL (nFL) metal where the Landau quasiparticle is destroyed but
the concept of a sharp Fermi surface persists. This was dubbed a
``critical Fermi surface''\cite{senthil08-1,senthil08-2}.
One prediction of this theory was that the zero-temperature resistivity
jumps by a finite amount as one goes from the FL to the quantum critical state, where the value
of the jump was predicted to be universal, $R\hbar/e^2$, $R$ being a dimensionless number
associated with the QCP. This resistivity jump is illustrated in \rfig{jump}.

The main purpose of this work is to provide a controlled calculation of the value of this jump
and, more generally, analyze the behavior of the electric resistivity in response to changes
in pressure (modifying the metallic bandwidth) and/or temperature in the vicinity of the QCP.
The resulting resistivity can be compared with experiments,
where the bandwidth can be changed by applying mechanical or chemical pressure.
The predictions we make, \emph{viz.} sharp resistivity variation on the order of $10 h/e^2\sim 100$ k$\W$,
quantum critical collapse of pressure and temperature dependence,
thermal conductivity jump and violation of Wiedemann-Franz
law, provide distinct signatures of the fractionalization at the MIT and of the critical Fermi surface
state intervening between the FL and SL.

A crucial ingredient of the theory is that, at zero temperature, 
the emergent gauge fluctuations associated with the fractionalization decouple from the quantum critical 
charge fluctuations. Despite this the gauge fluctuations are expected to play a crucial role for non-zero 
temperature transport properties. A similar phenomenon happens in the Kondo breakdown model studied in
Ref. \onlinecite{svslong}.
%although the reason for the suppression of the gauge fluctuations at $T=0$ is not the same as in the present
%work.
An important difference with the Kondo breakdown scenario is that the charge fluctuations near the Mott transition studied in this paper are described by an interacting field theory at low energies. Irrespective of this
difference, the same conclusion holds:
the gauge fluctuations become important for low frequency transport at non-zero temperature.
We show this explicitly by calculating the effects of these gauge fluctuations on the transport. In particular the precise value of the universal resistivity jump in the limit that the frequency of the
applied electric field goes to zero faster than temperature is strongly 
affected by the gauge fluctuations. In contrast, in the opposite order of limits the universal resistivity jump is
unaffected by the gauge field. We further predict a universal jump of the thermal
conductivity across the Mott transition, which derives from the breaking of low energy conformal invariance by
the gauge field, and leads to a violation of the Wiedemann-Franz law by the critical
Fermi surface.

The paper is organized as follows. In section \ref{sec:main-results}, we summarize our main findings;
section~\ref{sec:slave-rot} introduces the slave-rotor description for the Hubbard model.
In section \ref{sec:transport}, we formulate the transport via a quantum Boltzmann equation
for the critical charge fluctuations, the rotors, and present its solution at criticality.
Section \ref{sec:transport-qc-region} extends the resistivity calculation to the
entire QC region. In \ref{sec:largeT-w}, we discuss the behaviour of the resistivity at large
temperatures and frequencies. Signatures relating to thermal transport are discussed in
section~\ref{sec:thermal}. The Appendices give further details regarding the critical rotors,
with a focus on their transport properties in the absence of the emergent gauge field, i.e.
in the pure $O(N)$ non-linear sigma-model (NL$\s$M) for $N\geq 2$.

\begin{figure}
\centering
\includegraphics[scale=.6]{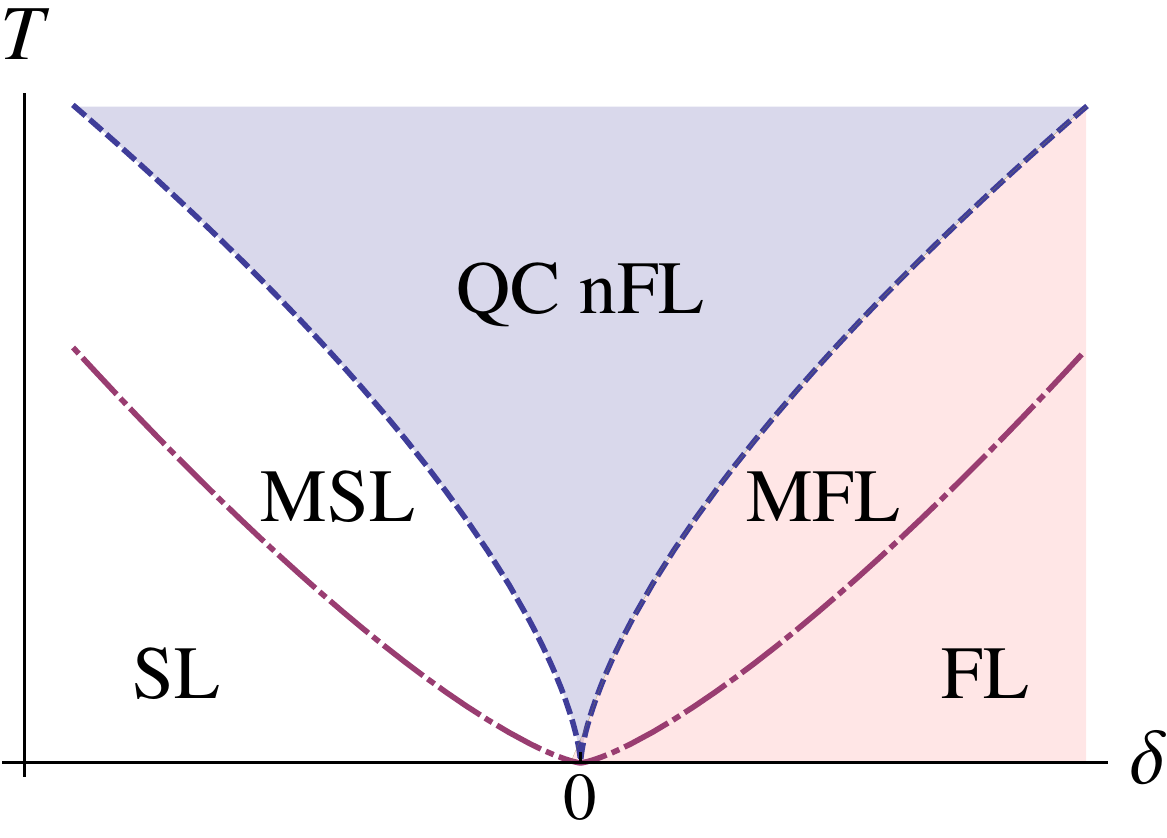}
\caption{\label{fig:pd} The phase diagram of the quantum critical Mott
transition. $\de$ tunes the ratio of onsite repulsion to the bandwidth
away from its critical value, and can be put in correspondence with $P-P_c$,
the deviation from the QC pressure, $P_c$.
The dark shaded (blue) region is the quantum critical region, where
the Landau quasiparticle is destroyed but a ``critical Fermi surface'' nonetheless exists.
It separates the spin liquid (SL) and the Fermi liquid (FL). The intermediate-$T$ states (with prefix `M'), the
marginal SL and FL, differ from the low temperature ones by the fact that the spinons and gauge bosons still behave
as in the QC region.
}
\end{figure}

\section{Main results}
\label{sec:main-results}
\begin{figure}
\centering
\subfigure[]{\label{fig:pd_R-vs-T}\includegraphics[scale=.52]{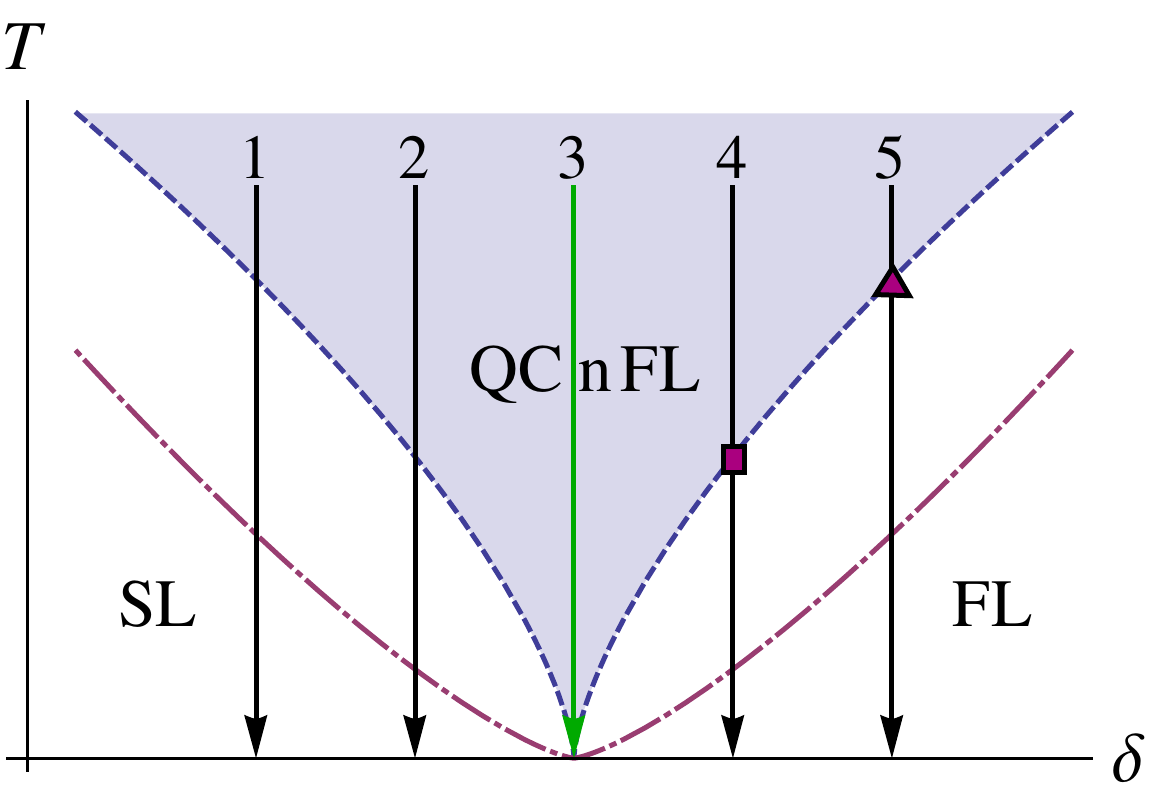}}
\subfigure[]{\label{fig:pd_R-vs-P} \includegraphics[scale=.53]{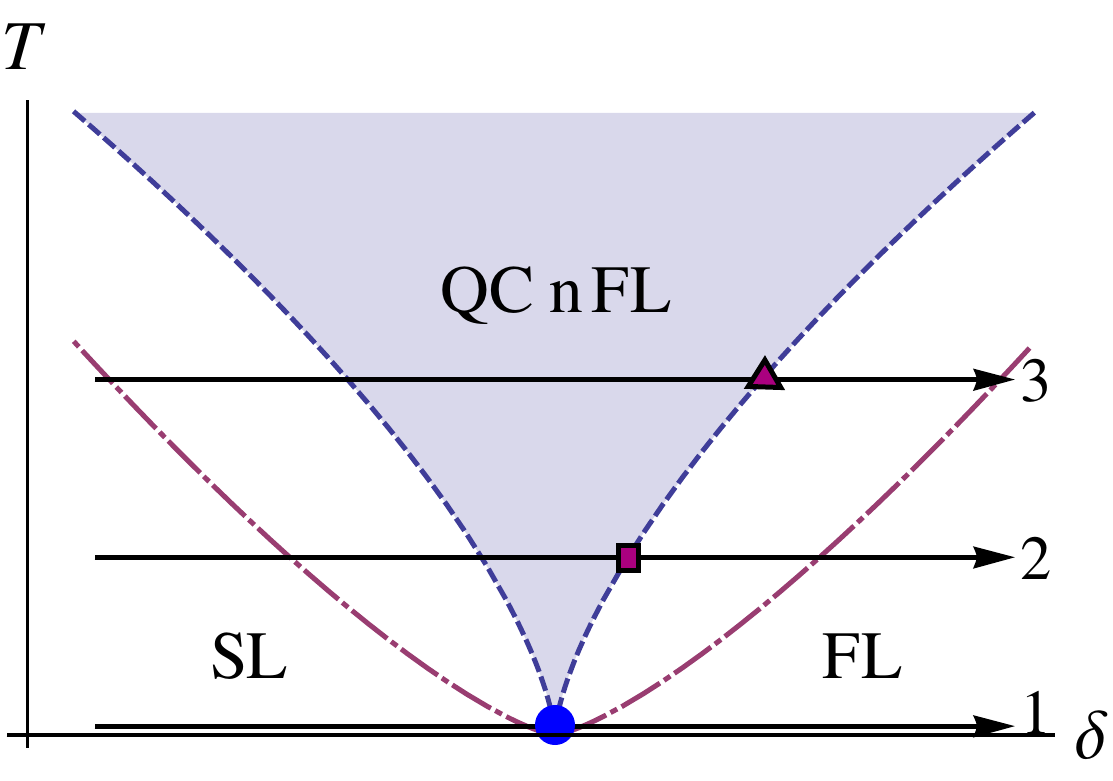}}\\
\subfigure[]{\label{fig:R-vs-T}\includegraphics[scale=.44]{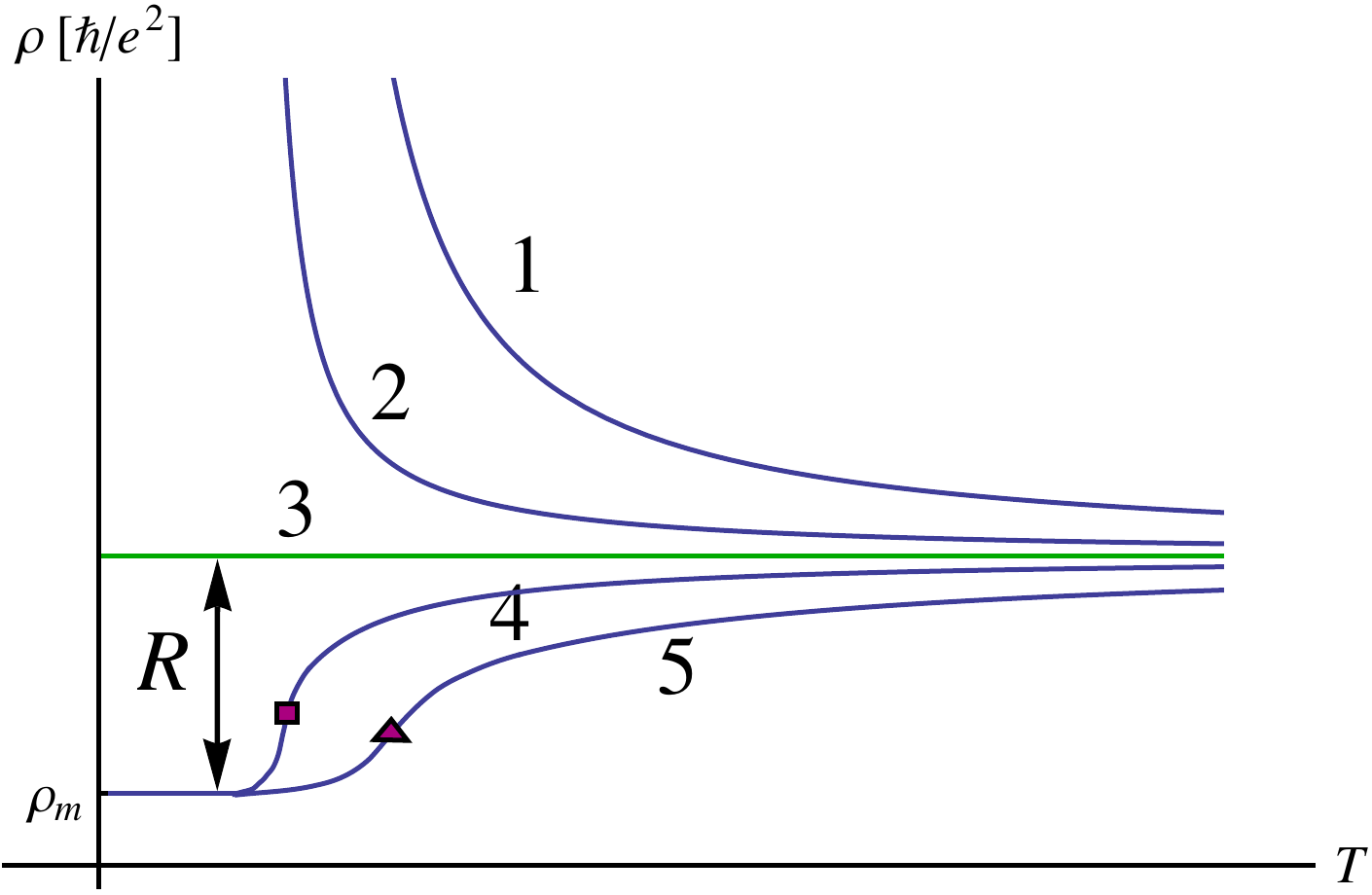}}
\subfigure[]{\label{fig:R-vs-P} \includegraphics[scale=.46]{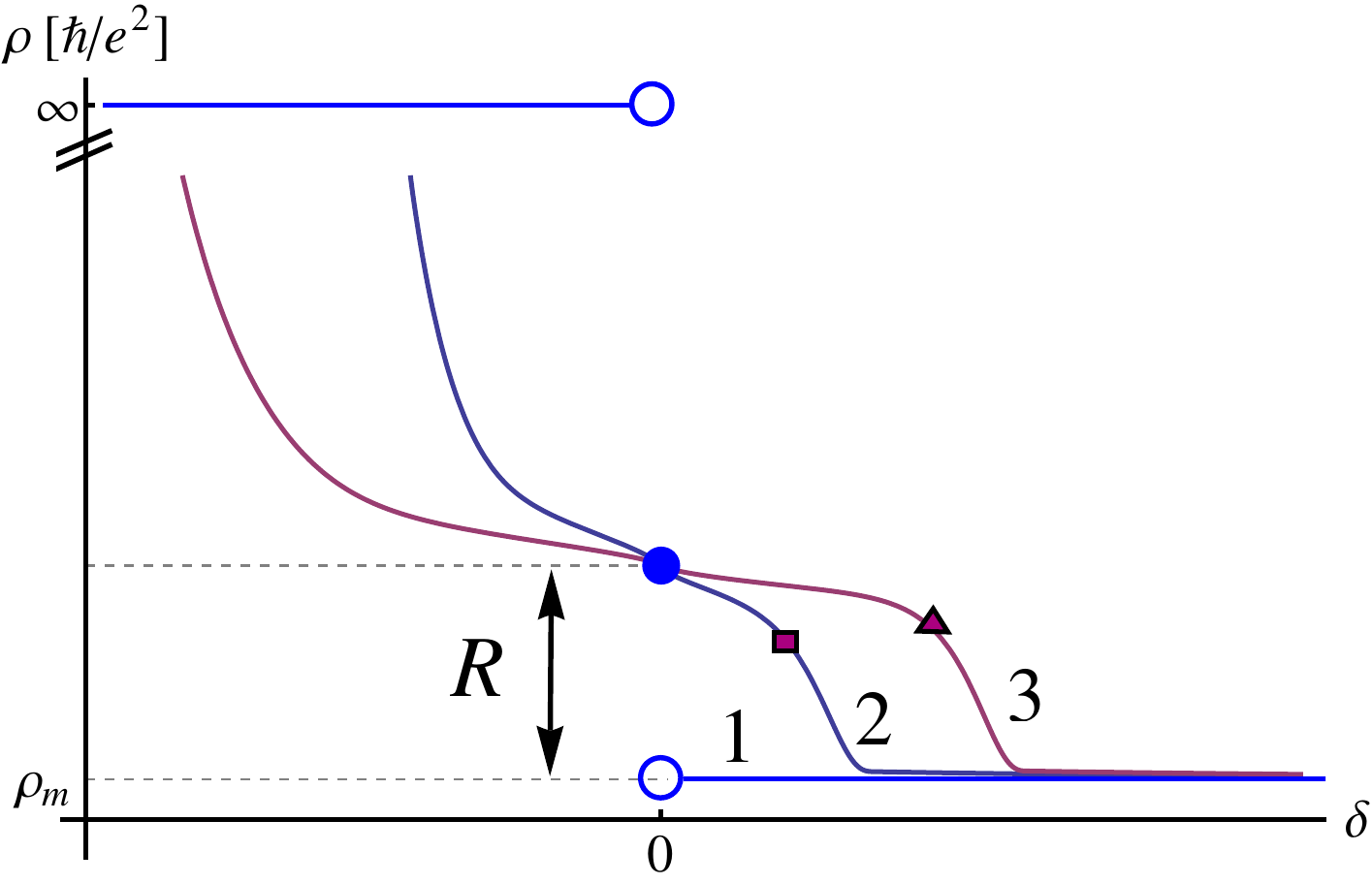}}
\caption{\label{fig:R}
Sketch of low temperature behaviour of the resistivity near the quantum critical (QC) Mott transition.
Panel c) shows the resistivity vs $T$ for different values of the onsite repulsion over the
bandwidth (tuned by $\de$), with the corresponding cuts shown in the phase diagram in a).
Panel d) shows the resistivity vs $\de$ at different temperatures,
with the corresponding cuts shown in the phase diagram in b). In c) and d), the
markers correspond to the location of the resistivity jump upon entering
the QC state from the FL. The value of the jump is universal: $R\hbar/e^2$. Our calculations
yield $R=49.8$, which translates to a jump of $\sim 8 h/e^2$.
}
\end{figure}

In this section we summarize our main results. 
We first need to briefly describe the finite temperature phase diagram obtained in
Ref.~\onlinecite{senthil08-2}, which is reproduced in \rfig{pd}.
The parameter $\de$ tunes the ratio of the  
electronic bandwidth to the onsite repulsion away from its critical value, and can be put in correspondence with $P-P_c$,
the deviation from the quantum critical pressure, $P_c$:
\begin{align}\label{eq:de} 
  \de \propto t/U - (t/U)_c \sim P-P_c\, .
\end{align}
At $T>0$, the metal-insulator transition becomes a crossover
due to the presence of the emergent gauge boson. We distinguish three main phases in \rfig{pd}:
the spin liquid (in white in the figure), the Fermi liquid (pink/light shading),
and the quantum critical state bridging the two (blue/dark shading).
The latter state is a non-FL where the Landau quasiparticle has been destroyed, yet
a sharp Fermi surface persists: an instance of a ``critical Fermi surface''.
In exiting the QC region, one enters two intermediate phases: a marginal spinon liquid (MSL) or
a marginal Fermi liquid (MFL). These are similar to their low temperature counterparts, the SL
and FL, except that the spinons and gauge bosons still behave as in the QC region.
As these correspond to fluctuations in the spin degrees of freedom, the two crossovers may be interpreted as corresponding to spin and charge degrees of freedom exiting criticality at parametrically different temperatures. At sufficiently
low temperature they crossover to the usual SL and FL states.
The behavior of the electric resistivity as one tunes across the phase diagram
is illustrated in \rfig{R}. Panels a) and c) correspond to the $T$-dependent
behaviour at fixed $\de$ (i.e. pressure), and vice-versa for b) and d).
The important crossovers for low temperature transport are the boundaries of the QC region: there, the
charge degree of freedom either localizes (SL) or condenses (FL). At the former crossover the
resistivity becomes thermally activated, $\sim e^{\De_+/T}$, because of the finite Mott charge gap $\De_+$.
This can be seen in curves 1-2 in \rfig{R-vs-T}. At the crossover to the FL,
it abruptly drops to its residual metallic value $\rho_m$ (curves 4-5 in \rfig{R-vs-T}).
The regime of interest for transport
corresponds to the QC non-FL, where the resistivity
relative to its residual value in the metal, $\rho_m$, is purely universal: 
$\rho-\rho_m\approx (\hbar/e^2)R$, where
$R$ is a universal dimensionless constant. Our controlled calculation of $R$ in a large-$N$ approximation gives the estimate $R = 49.8$. $R$ sets the size of the
jump shown in \rfig{jump}, which is reproduced in \rfig{R-vs-P}, curve 1.
At finite temperature, this jump becomes a steep increase, as shown in
curves 2-3 of \rfig{R-vs-P}. We emphasize that the low
temperature resistivity above the QCP, $\de=0$, is $T$-independent and
takes the value $\rho=\rho_m +(\hbar/e^2)R$.

The diverse behavior shown in \rfig{R} can be obtained from a single-variable function.
Indeed,  the temperature and pressure dependent resistivity (relative to its constant residual value in the FL)
can be collapsed by a universal scaling function $G$ associated with the Mott QCP:
\begin{align}
  \rho-\rho_m= \frac{\hbar}{e^2} G\left(\frac{\de^{z\nu}}{T}\right)\, ,
\end{align}
where the dynamical and correlation length exponents correspond to those of the
3D XY universality class: $z=1$ and $\nu\approx 0.672$. Indeed,
the critical charge degrees of freedom can be effectively described by a Bose-Hubbard model
at half-filling near its insulator-superfluid transition, which belongs to that universality
class. We show that although the spin fluctuations encoded in the emergent gauge field associated
with the electron fractionalization
do not alter these exponents, they have strong effects on the scaling function, and thus
on the value of the universal jump, $(\hbar/e^2)R$.

We predict that thermal transport also shows signatures of the critical Fermi
surface. In particular, the thermal conductivity divided by temperature, $\ka/T$,
has a universal jump at criticality, by an amount $(k_B^2/\hbar)K$, where $K$ is a dimensionless number
just like $R$. As we explain in section~\ref{sec:thermal}, the emergent gauge fluctuations play an
important role
by breaking the conformal invariance present in their absence, thus reducing $\ka/T$ from a formally
infinite value to a finite, universal one. Finally, combining the electric resistivity and thermal conductivity jumps,
we predict that the QC non-FL violates the Wiedemann-Franz law by a universal amount: the Lorentz number
differs from its usual value in the FL by $(k_B/e)^2RK$, as shown in \rfig{jump}.

\section{Mott transition in the Hubbard model: a slave-rotor formulation}
\label{sec:slave-rot}
To set the stage we briefly review the description of the insulating quantum spin liquid with a spinon Fermi surface\cite{sslee05}, and the continuous bandwidth-tuned Mott transition\cite{senthil08-2} to it from a Fermi liquid.
We consider a single-band Hubbard model at half-filling on a 2D non-bipartite lattice
(for e.g. triangular):
\begin{align}
  \label{eq:Hubbard}
  H=-t\sum_{\langle rr'\rangle}(c_{\s r}^\dag c_{\s r'}+{\rm h.c.})+U\sum_r(n_r-1)^2,
\end{align}
where $c_{\s r}$ annihilates an electron with spin $\s$ at site $r$, and
$n_r=c_{\s r}^\dag c_{\s r}$. In the small $U/t$ limit, the ground state is
a Fermi liquid metal, while in the opposite limit a Mott insulator results.
The interplay of frustration and strong charge fluctuations can lead to
a quantum spin liquid ground state instead of a conventional antiferromagnetic
Mott insulator. We shall focus on the transition to such a state.

The slave-rotor construction\cite{florens04} is tailor-made to describe the spin-charge
separation that occurs as the charge localizes when the electronic repulsion becomes sufficiently large,
yet weak enough for the spins to remain disordered, even at $T=0$.
At the level of the microscopic Hubbard model, \req{Hubbard}, the slave-rotor construction
is a change of variables to degrees of freedom better suited to describe the SL, in which the electron is
fractionalized into spin and charge carrying ``partons'':
\begin{align}
  c_{r\s}=\psi_{\s r}b_r\, .
\end{align}
The fermionic spinons, $\psi_{\s r}$, carry the spin, while the bosonic rotors, $b_r = e^{-i\theta_r}$, the charge
of the original electron.
The projection from the enlarged Hilbert space to the physical one is obtained from the operator identity
relating the rotor charge or ``angular momentum'', $l_b$, to the fermion number, $n_f$:
$l_{b}=1-n_{f}$, which is enforced at each site, where $n_{f}=n$ is the actual electronic occupation number (because
$|b|=1$). By virtue of Pauli exclusion, the charge relative to half-filling
at each site can only be $-1$ (double occupancy), $+1$ (hole) and $0$ (single occupancy). Hence, the positive (holon)
and negative (doublon) charge excitations encoded in the rotors relate to the holes
and doubly-occupied sites of the half-filled
Hubbard model, see \rfig{holon-doublon}. Moreover, since the system
is at half-filling, there is a low-energy particle-hole symmetry between these positive and
negative charge excitations.

In the long-wavelength limit, a U(1) gauge structure emerges\cite{sslee05}.
The temporal component of the gauge field results from the above constraint
necessary to recover the physical Hilbert space, while the spatial components derive
from the fluctuations of spinon bilinears about their saddle-point configuration.
After coarse-graining, the low-energy effective action for the
Hubbard model in terms of the fractionalized degrees of freedom can be written as
\begin{align}
  S &= S_{b,a} + S_{f,a}+S_a\, , \\
  S_{b,a} &= \frac{1}{2g}\int_x \left(|(\pd_\nu-ia_\nu)b|^2 +i\la(|b|^2-1)\right)\, , \\
%  S_a &= \frac{1}{2}\int_q |a(q)|^2\left(\Pi_f^{\rm j}+\Pi_b^{\rm j}\right) \\
  S_{f,a} &= \int_x \bar \psi_\s\left(\pd_\tau -\mu -ia_0 +\frac{(\b \nabla-i\b a)^2}{2m_f}\right)\psi_\s\, , \\
  S_a &= \frac{1}{e_0^2}\int_x (\e^{\nu\g\be}\pd_\g a_\be)^2\, .
\end{align}
We work in units where the rotor velocity, $c$, is set to one, unless otherwise specified.
The complex boson field $b$ is constrained to lie
on the unit circle via the Lagrange multiplier field $\la$. The indices $\nu,\g,\be$ run over
imaginary time and the two spatial dimensions; $\mu$ is the electronic chemical
potential. We have used the shorthand $\int_x=\int_0^{1/T}d\tau\int d^2x$.
The parameter that tunes the Mott transition of the rotors (and hence of the whole electronic liquid) is
$g\propto U/t$, where $U$ is the Hubbard repulsion from the original electronic Hamiltonian,
while $t$ is proportional to the electronic bandwidth. To make the action dimensionless,
the parameter $g$ carries the dimension of length, as given by a real space cutoff scale.
We can relate it to the parameter $\de$ introduced in \req{de}
via $\de = g\inv -g_c\inv\propto t/U - (t/U)_c$.
For small coupling, $g<g_c$, the rotors spontaneously condense, corresponding to the metallic phase of the original
Hamiltonian; while in the opposite limit, $g>g_c$, the rotor field is disordered leaving the system
in a SL ground state. 
The gauge fluctuations have a Maxwellian action resulting from the elimination
of high energy fluctuations; $e_0$ is the corresponding bare gauge charge.

The emergence of a relativistic action for the rotors, which have a dynamical exponent
$z=1$, is a consequence of the emergent low-energy particle-hole symmetry of the Hubbard model at
half-filling noted above. This low-energy symmetry will be important when we examine the effect of the
gauge field on the critical rotors. It will lead to a strong suppression
of the dynamical gauge fluctuations in the charge (rotor) sector.

The field theory above is strongly interacting. Indeed, in two dimensions the rotor NL$\s$M
considered separately flows to a strong coupling fixed point where even the $b$-field quasiparticles
are ill-defined. The spinons and gauge fluctuations do not alter this.
One perturbative approach to the problem
extends the field theory to include a large number of flavors of the matter fields:
when that number is very large, we have weakly interacting quasiparticles, at least in the boson sector.
We will use this extension, which we now describe in more detail, to bring the calculation under control.

\subsection{Low energy theory and large-$N$ extension}
\label{sec:largeN}
We consider the slave-rotor field theory extended to have
a large number of rotor and spinon flavors, allowing for
a systematic study of transport\cite{sachdev-book}.
The number of copies of the complex rotor is taken to be $N/2$, 
yielding an even number, $N$, of real scalar fields. The case of physical interest has $N=2$.
In this large-$N$ extension, the effective actions of the rotors and spinons read:
\begin{align}
   S_{b,a} &= \frac{N}{2g}\int_x \left(|(\pd_\nu-ia_\nu)b_\al|^2 +i\la(|b_\al|^2-1)\right)\, , \\
  S_{f,a} &= \int_x \bar \psi_\s\left(\pd_\tau -\mu -ia_0 +\frac{(\b \nabla-i\b a)^2}{2m_f}\right)\psi_\s\, ,
\end{align}
where each $b_\al$ is a complex scalar, and $\al$ runs from 1 to $N/2$, while
there are $N$ copies of spinons, $\s=1,\dots, N$. Repeated indices are summed over,
for example $|b_\al|^2=\bar b_\al b_\al=\sum_\al |b_\al|^2$.
$\la$ is the Lagrange multiplier field enforcing the constraint that the $O(N)$
real field be unimodular. The coupling $g$ has been rescaled $g\rightarrow g/N$.
Note that the gauge field reduces the rotor global symmetry from $O(N)$ to $U(N/2)$.
In the $N\rightarrow \infty$ limit, the fluctuations of the $\lambda$ and gauge bosons are unimportant.
Formally integrating out the rotors yields the following form for the
partition function\cite{sachdev-book} $\mc Z=\int\mc D\la e^{-S_{\rm eff}[\la]}$, with
\begin{align}
  S_{\rm eff}[\la]=\frac{N}{2}\left[\tr\ln(-\pd^2+i\la)-\frac{i}{g}\int_x \la\right]\, .
\end{align}
The overall factor of $N$ plays the role of $1/\hbar$ so that in the \largen
limit the quantum fluctuations are suppressed and we only need to consider
the classical equation of motion\cite{sachdev-book}
\begin{align}\label{eq:mf-eq}
  \int_q \frac{1}{q^2+m^2}=\frac{1}{g}\, ,
\end{align}
where $\int_q=T\sum_{\w_n}\int d^dq/(2\pi)^d$ and the mass squared, $m^2=i\la_0\in R$,
corresponds to the saddle point value
of the uniform component of $\la$, $\la_0$.
This mean-field value of $\la$ plays the role of the mass for the rotors
in their insulating phase at large $g$. It can be alternatively seen as the
inverse correlation length, $\xi\sim 1/m$. At sufficiently small $g$, \req{mf-eq} has no solution,
and a different approach must be used to describe the condensation.
In the following, we shall focus mainly on the quantum critical regime
as well as on the insulating phase.
The solution of the saddle-point equation, \req{mf-eq}, in the \largen limit and
at finite temperature directly above the quantum critical point,
$g=g_c$, yields $m=\Theta T$, where
$\Theta=2\ln(\tfrac{1+\sqrt{5}}{2})\approx 0.96$, twice the logarithm of the golden ratio\cite{damle97}.
The mass vanishes linearly as $T\rightarrow 0$, i.e. the correlation length of the charged rotors
diverges upon approaching the QCP: $\xi\sim 1/T$.
The full dependence of $m$ on $g$ and $T$ at $N=\infty$ is given
in section~\ref{sec:transport-qc-region}.

Corrections at order $1/N$ to the $N=\infty$ saddle-point correspond to interactions mediated by the
$\la$ and $a$ bosons, which develop dynamics when $N$ is finite. The rotor action,
including the effective mass corresponding to the saddle-point value of
the $\la$ boson, now reads
\begin{align}
  S_{b,a} &= \frac{1}{2g}\int_x \left[|(\pd_\nu-ia_\nu)b_\al|^2
    +m^2 |b_\al|^2 +i\la |b_\al|^2
    \right]
\end{align}
We are using the Coulomb or transverse gauge
so that it is understood that we only include configurations where $\b\nabla\cdot \b a=0$.
The transverse and temporal component of $a_\mu$ are decoupled in this gauge.
Further, we can omit the latter in the low-energy limit
because it is screened by the spinon Fermi surface. The transverse part of the
gauge field remains gapless because the currents remain unscreened, as opposed to the charge.
In the remainder, we shall use $a$ to represent the transverse component.
The $1/N$ corrections to the saddle point will generate $\mc O(1/N)$ propagators
for both the gauge and $\la$ bosons, which acquire the following effective action:
\begin{align}
 \frac{1}{2} \int_q \left[|a(q)|^2\frac{N}{2}\left(\Pi_f^{\rm j}+\Pi_b^{\rm j}\right)
+ |\la(q)|^2\frac{N}{2}\Pi_b\right]
\end{align}
where the finite temperature, imaginary-time polarization functions read
\begin{align}
  \Pi_b(i\nu_l,q) &=T\sum_{n}\int \frac{d^2\b p}{(2\pi)^2}
  \frac{1}{(\w_n+\nu_l)^2+\e_{p+q}^2}\frac{1}{\w_n^2+\e_p^2}\, , \\
  \Pi_b^{\rm j}(i\nu_l,q) &=-T\sum_{n}\int \frac{d^2\b p}{(2\pi)^2}
  \frac{(2\hat q\times\b p)^2}{(\w_n+\nu_l)^2+\e_{p+q}^2}\frac{1}{\w_n^2+\e_p^2}\, ,
\end{align}
where the superscript ``j'' identifies the current-current correlator;
we have defined the rotor dispersion relation: $\e_p=\sqrt{p^2+m^2}$.
Details about the computation of $\Pi_b,\Pi_b^{\rm j}$ can be found in
Appendices \ref{ap:pure-rot} and \ref{sec:Pi_b_j}, respectively. $\Pi_b^{\rm j}$ is
discussed in the next subsection, \ref{sec:role_gf}.
The spinon Fermi surface contributes
\begin{align}
  \Pi_f^{\rm j} &= \mu\left(c_1\frac{|\w_n|}{v_Fk} +c_2 \frac{k^2}{k_F^2}\right)\, , \quad |\w_n|<v_F k
\end{align}
where the $c_i$ are real numbers, while $k_F,v_F$ are the Fermi momentum and velocity, respectively.
As we work in units where the velocity of the rotors is set to 1,
we need to keep $v_F$ explicitly. Because of the term $|\w_n|/v_Fk$ in the fermionic polarizability,
the gauge fluctuations are Landau damped.
\subsubsection{Role of gauge fluctuations}
\label{sec:role_gf}
We now examine the role of the gauge fluctuations. The Landau damped dynamics due to the Fermi surface
dominate those induced by the rotor polarization function, $\Pi_b^{\rm j}$, and we can evaluate the
latter in the static limit\cite{senthil08-2}. Note that $\Pi_b^{\rm j}$, just as $\Pi_b$, depends on the temperature and
$g$ via the mass of the rotors, $m$.
%that vanishe at the critical point, $\Delta_\pm$. $\Delta_+$ is the spin or phase stiffness
%of the rotors in the FL ($g<g_c$), while $\Delta_-$ is the charge gap in the Mott
%insulator ($g>g_c$).
%This dependence arises through the mass, $m/T=X_\pm(\Delta_\pm/T)$, where $X_\pm$ are universal functions.
%Recall that $m=\Theta T$ directly above the critical point. The dependence
%on the \emph{ratio} of $T$ and $\Delta_\pm$ is purely universal in $\Pi_b$ and $\Pi_b^{\rm j}$.
For $g=g_c$, as shown in Appendix \ref{sec:Pi_b_j}, we get
\begin{align}\label{eq:Pib-j}
  \Pi_b^{\rm j}(0,q) =  \begin{cases}
   \g_2 \frac{q^2}{T} & \text{if } q \ll T \\
   \s_b^0 q       & \text{if } q \gg T
  \end{cases}
\end{align}
where $\g_2\approx 0.031$ is a dimensionless constant, while $\s_b^0\approx 0.063$
is the rotor conductivity in units of $e^2/\hbar$ in the large frequency or $T\rightarrow 0$ limit $\w/T\gg 1$.
As discussed in sections \ref{sec:qbe} and \ref{sec:large-w}, it differs from the DC conductivity we are seeking and
can be obtained from a simple $T=0$ analysis.
The $q^2$ behaviour at $q\ll T$ results from including the mass
of the rotors, $m=\Theta T$, when computing the current polarization function. It has
the behaviour expected from massive modes since for $q\ll T$, the fluctuations exceed the
correlation length, $1/m\sim 1/T$, and must be gapped. The important term for the low-temperature behaviour
is the linear $q$ contribution.
This non-analytic dependence arises because of the gaplessness of the critical rotors
and gives the gauge field a $z_a=2$ dynamical exponent, making it less singular than
deep in the SL where the rotors are gapped and the gauge bosons have $z_a=3$.
Indeed, the $z_a=2$ damped gauge fluctuations give the
spinons a self-energy $\sim \tfrac{1}{N}i\w\ln(\mu/|\w|)$, which is weaker than
in the usual SL, where we have $\tfrac{1}{N}i|\w|^{2/3}$. Thus, at criticality,
as well as in the MFL and MSL phases, the spinons form a marginal Fermi liquid,
leading to the usual logarithmic corrections. We emphasize that we do not need to worry about
possible subtleties with the breakdown of the naive \largen expansion for a Fermi
surface coupled to a gapless boson\cite{lee-largeN}. Indeed, we are mainly concerned
with the quantum critical region, where, as stated above, the fermions only acquire
logarithmic corrections due to gauge fluctuations. Such a marginal Fermi liquid
of spinons can be controlled by a simple perturbative renormalization group (RG)
approach\cite{Nayak94,Mross10}. In contrast, deep in the
SL, one might need to take the limit of small $z_a-2$ simultaneously with $1/N\rightarrow 0$ to
make the expansion controlled\cite{Mross10}. Further the main transport properties will derive from the rotor sector for which the \largen works reliably.
The spinons affect the rotors only via the damping of the gauge bosons, which we believe
is a robust feature, independent
of the expansion scheme.

Regarding the rotor or charge sector, it was shown\cite{senthil08-2} that the gauge fluctuations
do not alter the nature of the rotor excitations, i.e. the rotor self-energy
is subleading compared to the bare dynamics.
Insofar as the thermodynamic critical properties of the charge sector are concerned,
they belong to the 3D XY universality class, unaffected by the gauge bosons or
spinons. The importance of the damping at quenching the gauge fluctuations
can be heuristically understood by examining the dominant rotor fluctuations, which have $\w\sim q$ ($z=1$).
Substituting this dispersion relation into the Landau damping term we get $\mu |\w|/q \sim \mu$.
Thus, the dominant rotor fluctuations see the gauge bosons as screened.
%This can be interpreted as
%a ``fermionic Higg's mechanism'', where a Fermi surface renders a gauge
%boson effectively massive, at least from the point of view of the
%rotors.
Such an effect was also identified in Ref.~\onlinecite{kaul08}, in the context of
a quantum critical transition between a N\'eel-ordered Fermi-pocket metal to a
non-FL algebraic charge liquid, called a ``doublon metal''. This suppression mechanism
of the gauge field due to a Fermi surface was referred to as a ``fermionic Higgs effect''.

In the next section, we shall show that although the gauge fluctuations are not effective at
influencing the thermodynamic critical properties in the charge sector,
they can have strong effects on non-zero temperature transport, down to arbitrarily low
temperatures.
% The original fermionic Hubbard model near its Mott
% transition has its critical charge fluctuations described by a Bose-Hubbard model, which is
% well-known to have a continuous insulator-superfluid transition in the 3D XY universality class.
%As we shall soon see
%this will affect the small momentum scattering rate due to the gauge fluctuations, hence this finite
%$T$ effect needs to be accounted for. As one tunes away from $g=g_c$, the dependence
%becomes a bit richer as show in Appendix \ref{sec:Pi_b_j}. Indeed, for $g\ll g_c$, the mass
%differs from the temperature,
%$m/T\ll 1$, and this introduces a scale in addition to $T$. This will lead to three regimes
%instead of the two found at $g=g_c$, where $m\approx T$.
\begin{figure}
\centering
\subfigure[]{\label{fig:holon-doublon} \includegraphics[scale=.53]{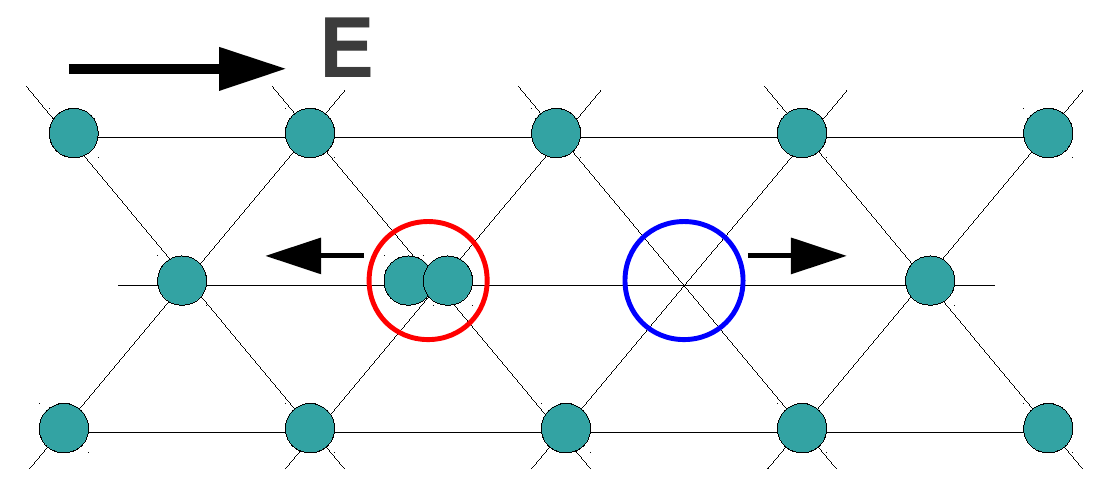}}
\subfigure[]{\label{fig:doublon-current}\includegraphics[scale=.33]{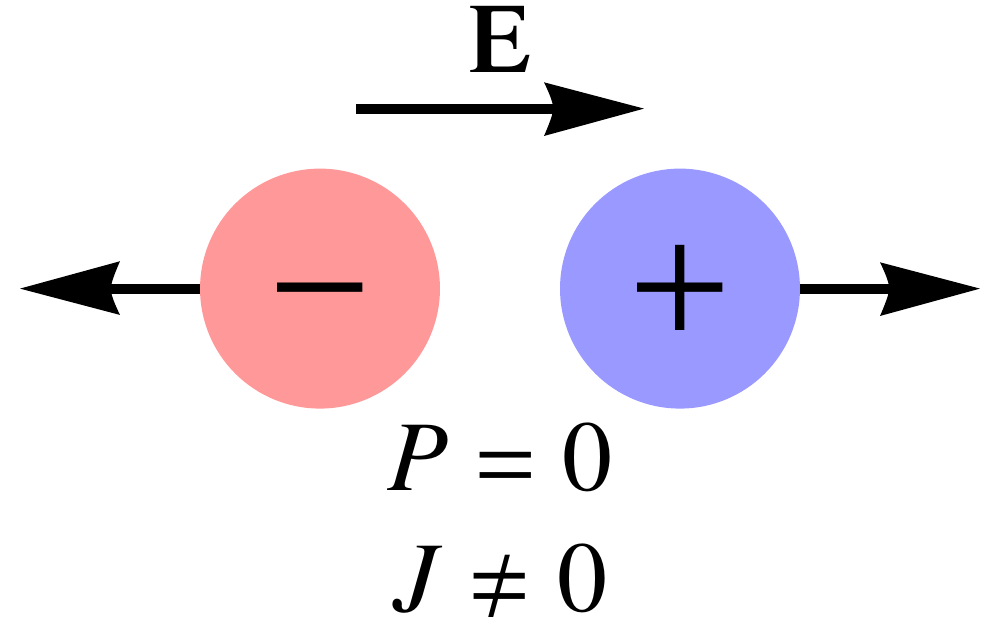}}
\caption{\label{fig:holon-doublon}
Charge excitations near the Mott transition. a) Triangular lattice at half-filling;
the small shaded disks represent electrons. The double-occupied (empty) sites
are identified by a red/left (blue/right) circle. These are encoded in the
charge rotor excitations, the holons and
doublons, respectively. Under an applied electric field, they will move in opposite
directions. b) By virtue of the emergent particle-hole symmetry between doublons and holons,
it is possible to have a state with zero momentum, $P$, but finite current, $J$. This
allows interactions to dissipate current while conserving momentum. }
\end{figure}
\section{Critical transport near the Mott transition}
\label{sec:transport}
In a slave-particle theory such as the one under consideration, many
observables can be determined from the separate responses of the partons.
These relations generally go under the name
of Ioffe-Larkin composition rules. For example, the one for the resistivity reads\cite{ioffe89}
\begin{align}
  \label{eq:IL}
  \rho=\rho_b+\rho_f \, ,
\end{align}
where $\rho_{b,f}$ is the resistivity of the spinons and rotors, respectively.
The resistivities ``add in series'' because of the constraint relating the
spinons and rotors to recover the original Hamiltonian/Hilbert space:
The electric field induces a motion of the electrically charged rotors, forcing
spinons to flow as well. Alternatively, we can say that the external
electric field induces an internal one. It follows that the parton with the highest
resistivity governs the entire electric response. Near the Mott transition,
the rotors have the most singular response as they undergo a quantum phase
transition, while the spinons form a Fermi surface throughout. We thus anticipate that
the strong variation of $\rho_b$ across the transition will give the entire resistivity
its key dependence.

Let us first discuss the $T=0$ and clean limit, in which case the spinon Fermi surface has vanishing resistivity
throughout. The rotors also have vanishing resistivity in their condensed phase, such
that the FL has $\rho=\rho_b+\rho_f=0+0=0$ as expected. On the Mott side,
the rotors are gapped hence the whole system has infinite resistivity: $\rho_b=\infty=\rho$.
The interesting feature happens directly at criticality, where although we still have
$\rho_f=0$, the rotors have a finite universal resistivity induced purely by interactions\cite{cha91,damle97},
$\rho_b=R\hbar/e^2$.
It is possible for systems with particle-hole symmetry or equivalently emergent
relativistic invariance to have a finite resistivity in the absence of disorder
or Umklapp scattering. For such systems, the momentum
and electric current operators need not be proportional, allowing interactions to dissipate
the latter while preserving the former. More physically, these systems have independent and
symmetry-related positive and negative charge excitations that flow in opposite directions under an applied
electric field yielding a state with a finite current but with zero momentum. This is
schematically illustrated in \rfig{holon-doublon}. The finite, interaction-driven rotor resistivity at criticality
leads to a discrete jump at $T=0$, as illustrated in \rfig{jump}.

This scenario naturally extends to finite but low temperatures: the fermions still
contribute only a constant $\rho_f$, which is zero for a clean system, or finite
in the presence of weak disorder. Instead of discontinuously jumping,
the rotor resistivity increases rapidly upon entering
the quantum critical region, where it slowly increases until the growth becomes exponential
at the crossover to the spin liquid, as is shown in \rfig{R-vs-P}. We shall thus focus on the resistivity of
the rotors, $\rho_b$, for which we perform a $1/N$ expansion. 
In the simplest limit, $N=\infty$, the rotors are free because they
decouple from the $\la$ and gauge bosons. The DC
resistivity thus vanishes in the absence of scattering, $\rho_b=0$.
At order $1/N$, the rotors begin colliding with
the constraint field $\la$ and the emergent gauge boson, leading to a finite
resistivity. For sufficiently large $N$, the system
has well-defined quasiparticles, whose transport properties can be unambiguously studied by a quantum
kinetic (or Boltzmann) equation, to which we now turn.

\subsection{Quantum Boltzmann equation for critical charge fluctuations}
\label{sec:qbe}
We formulate the quantum Boltzmann
equation (QBE) for the distribution functions of the rotor excitations in the presence
of an oscillating electric field, $\b E(t)$, with driving frequency $\w$.
This frequency plays an important role as it introduces an energy scale that
divides the critical frequency-dependent resistivity into two regimes:
$\w<T$ and $\w>T$. Indeed, the
frequency $\w$ must be
compared with the dominant scale for the rotors in the quantum critical regime,
which is the temperature.
As was established in seminal work by Damle and Sachdev\cite{damle97}, the limits $\w\rightarrow 0$
and $T\rightarrow 0$ do not generally commute for the response functions of critical systems.
For example, the $T=0$ DC resistivity is obtained by first taking $\w/T\rightarrow 0$,
then $T\rightarrow 0$, so that one must necessarily perform a finite temperature analysis
to obtain the correct DC response. A $T=0$ calculation, which is equivalent to taking $T\rightarrow 0$
first, yields the response in the $\w/T\rightarrow \infty$ limit, which generically differs
from the DC behaviour. This non-commutativity of the $\w\rightarrow 0$
and $T\rightarrow 0$ limits can be explained on physical grounds: The small frequency
resistivity ($\w<T$) is dominated by the incoherent scattering of thermally excited critical fluctuations;
it corresponds to the hydrodynamic limit. In contrast, the large frequency resistivity ($\w>T$)
arises from the coherent motion of field generated excitations; it is mainly collisionless.
The dichotomy is even more striking in our case due to the presence of the gauge bosons:
We shall show that although the gauge fluctuations do not affect the transport in the
large $\w/T$ limit, they actually dominate the DC resistivity!
% In the \largen limit, the rotor quasiparticles are well-defined.
% This results in a clear separation of the frequency dependent conductivity
% into 2 regimes, at small and large frequencies. The small frequency ($\w/T<1/N$)
% component, $\s_I$, results from the incoherent scattering of thermally excited carriers.
% While the large frequency ($\w>2m$) conductivity, $\s_{II}$, derives mainly from excitations that
% are created by the external field. While the former is collision-dominated and incoherent,
% the latter is phase coherent and collision-less. In the following, we concentrate
% on the small frequency component, $\s_I$, which comprises the DC conductivity.

We assign the electric charge to a single rotor flavor, $b_1$, which couples to
the oscillating electric field $\b E(t)$.
The standard mode expansion for the electrically-charged rotor operator reads:
\begin{align}
  b_1(x) = \int_{\b k} \al_+(t,\b k) e^{i\b k\cdot\b x} +  \al_-^\dag(t,\b k) e^{-i\b k\cdot\b x}\, ,
\end{align}
where we have defined $\al_\pm/\al_\pm^\dag$ as the annihilation/creation operators for holons ($+$)
and doublons ($-$),
i.e. the positive and negative electric charge excitations. The expectation value of the
current can be decomposed into two pieces:
$\b J(t)= \b J_I(t)+\b J_{II}(t)$, where
\begin{align}
  \b J_{I}(t) &=\int_{\b k}\sum_{s=\pm}s\frac{\b k}{\e_k}\langle \al_s^\dag(t,\b k) \al_s(t,\b k)\rangle \\
  &= \int_{\b k}\sum_{s=\pm}s\frac{\b k}{\e_k} f_s(t,\b k)\\
  \b J_{II}(t) &=\int_{\b k}\frac{\b k}{2\e_k}\langle
\al_+^\dag(t,-\b k) \al_+^\dag(t,\b k)- \al_-^\dag(t,-\b k) \al_-^\dag(t,\b k)
-2\al_+^\dag(t,-\b k) \al_-^\dag(t,\b k)\rangle +{\rm h.c.} \label{eq:J_II}
\end{align}
We have defined the distribution functions of positive and negative charge excitations:
$f_s=\langle \al_s^\dag(t,\b k) \al_s(t,\b k)\rangle$, $s=\pm$; $\e_k=\sqrt{m^2+k^2}$ is the rotor dispersion.
From \req{J_II}, it should be apparent that as
$\b J_{II}$ involves pair production, it will only contribute when the driving frequency is above
the pair-production threshold, $\w>2m$, where
$m\sim T$ in the QC region. We shall
concern ourselves with the determination of $\b J_{I}$, i.e. $f_s(t,\b k)$, which governs the transport
in the small frequency limit. The asymptotic high frequency resistivity in the limit $\w\gg T$
can be obtained from a $T=0$ calculation and we leave its analysis to section~\ref{sec:large-w}.

The QBE for the distribution function of
holon (doublon) rotor excitations, $f_s$ with $s=\pm$, respectively, reads
\begin{align}
\label{eq:qbe}
(\pd_t &+s \b E\cdot\pd_{\b k})f_s(\b k,t) = \frac{1}{N}(I_\la[f_\pm]+I_a[f_\pm]) \\
&\quad= \frac{2}{N}\int_0^\infty \frac{d\W}{\pi}
\int \frac{d^2q}{(2\p)^2}
\left[\Im\left(\frac{1}{\Pi_b(\W,q)}\right)+ (2\b k\times \hat{\b q})^2\Im D(\W,q)\right] \\\nn
&\qquad\times \Bigg\{\frac{(2\p)\de(\e_k-\e_{k+q}-\W)}{4\e_k\e_{k+q}}\Big[f_s(\b k,t)
(1+f_s(\b k+\b q,t))(1+n(\W)) \\ \nn
&\qquad\hskip10em -f_s(\b k+\b q,t)(1+f_s(\b k,t))n(\W)\Big]\\ \nn
&\qquad+\frac{(2\p)\de(\e_k-\e_{k+q}+\W)}{4\e_k\e_{k+q}}\Big[f_s(\b k,t)
(1+f_s(\b k+\b q,t))n(\W) \\ \nn
&\qquad\hskip10em -f_s(\b k+\b q,t)(1+f_s(\b k,t))(1+n(\W))\Big]\\ \nn
&\qquad+\frac{(2\p)\de(\e_k+\e_{-k+q}-\W)}{4\e_k\e_{-k+q}}\Big[f_s(\b k,t)
f_{-s}(-\b k+\b q,t)(1+n(\W)) \\ \nn
&\quad\hskip10em- (1+f_s(\b k,t))(1+f_{-s}(-\b k+\b q,t))n(\W)\Big] \Bigg\}\, .
\end{align}
We have absorbed the magnitude of the rotor charge into $\b E$. The right hand
side of the QBE, the collision term, can be obtained for instance
by invoking Fermi's Golden Rule\cite{sachdev98,sachdev-book}. The first two $\de$-functions enforce energy
conservation for absorption and emission of $\la$ and gauge bosons by the rotors,
while the last one corresponds to pair creation/annihilation of holon-doublon pairs.
%These last processes correspond to the thermal creation/annihilation of boson pairs, which
%affect $f_s$ and are not to be confused with the neglect of the pair distribution
%functions contributing to the current for $\w>T$, $\b J_{II}$.

The propagators of the $\la$ and gauge bosons, $2/N\Pi_b$ and $2D/N=2/N(\Pi_j^{\rm j}+\Pi_b^{\rm j})$, respectively,
enter into the QBE via their spectral functions which dictate the
density of states the rotor excitations can scatter into.
They are evaluated in equilibrium.
This is justified in the \largen limit since the external field couples to a single rotor flavor
such that the associated non-equilibrium corrections to the polarization functions sublead in $1/N$.
In other words, the drag of the $\la$ and gauge fields by the electric field, an analogue of ``phonon drag'',
is negligible in the \largen regime. The rotor flavors that do not directly couple to the electric field, $b_{\al>1}$,
play the role of an effective bath at equilibrium, from which the constraint field acquires its dynamics.

The scattering terms on the RHS of \req{qbe} all scale like $1/N$ because
the gauge and $\la$ bosons have propagators of that order. As $N\rightarrow \infty$,
the scattering terms vanish and the rotors become free, displaying a sharp Drude peak in the real part of the
small frequency conductivity: $\s_b\propto \de(\w)$, $\w<T$. As we shall see, the finite $1/N$ effects will cure this singularity
yielding a finite DC conductivity.
%Finally, note that the vertex between the rotors and the gauge bosons has a momentum dependence.
\begin{figure}
\centering
\includegraphics[scale=.35]{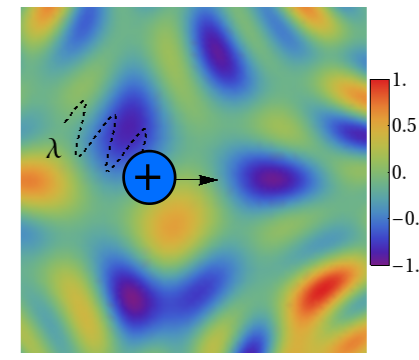}
\caption{\label{fig:rand-b}
Illustration of the main scattering mechanisms determining the
resistivity in the QC region. The blue disk corresponds to a
holon excitation with charge $+e$. In addition to the usual scattering
between critical charge fluctuations
(mediated by the $\la$ field), the static emergent gauge fluctuations
generate a random
``magnetic field'', $\b\nabla\times\b a$, that scatters the holons and doublons.
A schematic configuration of this emergent magnetic field (which is always perpendicular
to the plane) is shown, where
the scale gives its strength and direction, the latter dictated by the sign.
}
\end{figure}

\begin{figure}
\centering
\includegraphics[scale=.5]{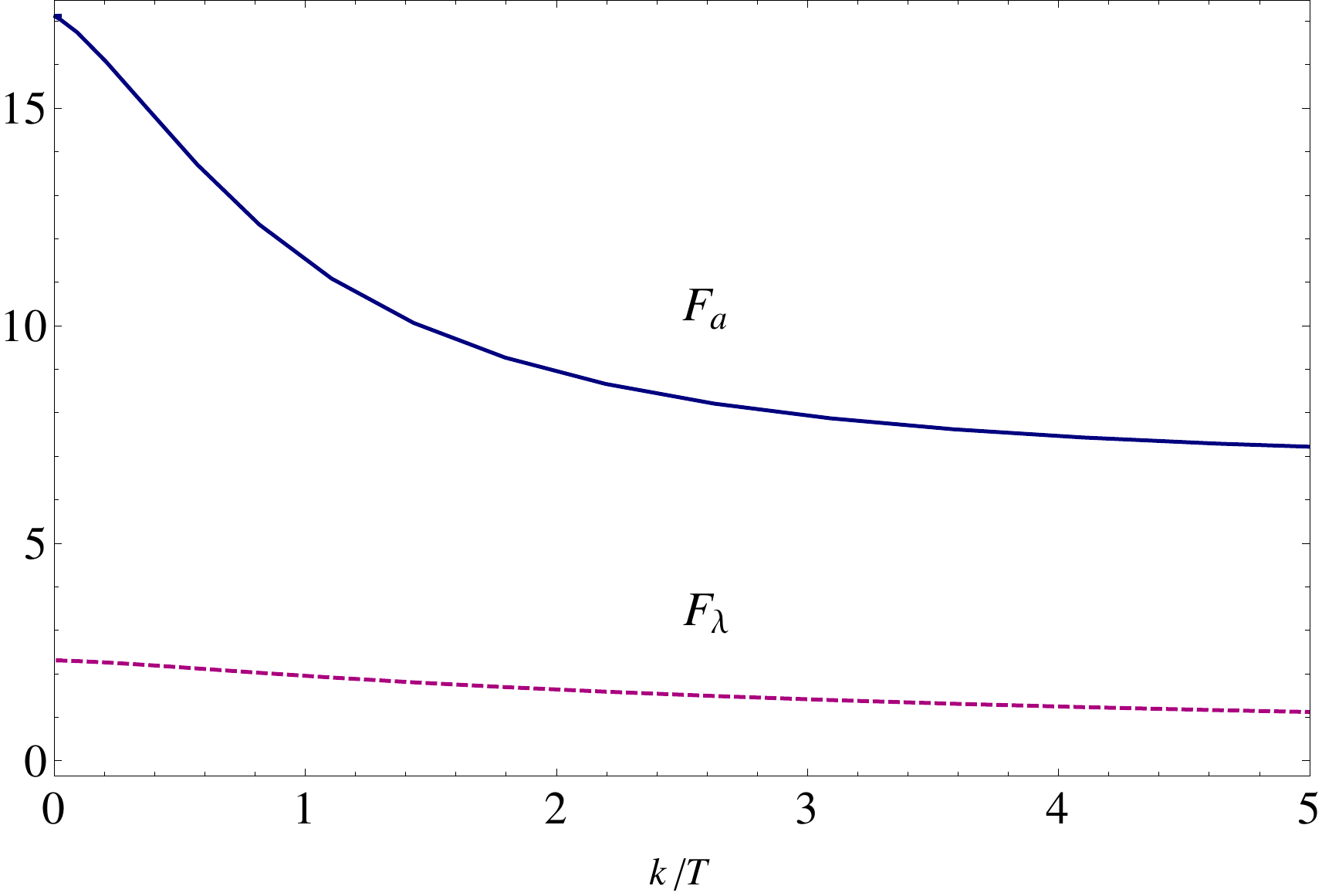}
\caption{\label{fig:scatt_rate} The scattering rate due to gauge fluctuations (solid line),
\req{scatt_rate}, and due to the $\la$-bosons, (dashed). The latter, $F_\la$,
appears in \req{simplified-qbe} and is defined in \req{scatt-rate-la}.
We see that the rate due to the gauge fluctuations
is a factor of $\sim 8$ larger than that due to the $\la$ field and is thus the dominant
scattering mechanism.
}
\end{figure}

We now proceed to the solution of the rotor QBE, \req{qbe}, by first expanding
the distribution function to linear order in $\b E$,
\begin{align}
f_s(\b k,\w)=n(\e_k) 2\pi\de(\w)+s\b E\cdot\b k \varphi(k,\w)\, ,
\end{align}
where we have
Fourier transformed from time to frequency. The deviation function, $\varphi(k,\w)$, only depends
on the magnitude of $\b k$, since $\b E\cdot\b k$ fully encodes the rotational symmetry breaking in the
presence of the external electric field.
The unknown function $\varphi$ parametrizes departures from the
equilibrium Bose-Einstein distribution, $n(\e_k)$, with $\e_k^2=m^2+k^2$.
The result of the linearization of the collision term due to the $\la$
bosons can be found in Appendix~\ref{ap:pure-rot}; we cannot make significant
simplifications there. Indeed, we need
to perform a careful numerical evaluation of the retarded polarization function $\Pi_b(\W,q)$
for all frequencies and momenta.
On the other hand, we can significantly simplify the scattering term due to gauge bosons. As mentioned above,
dynamical gauge fluctuations are suppressed in the rotor sector at $T = 0$.
%because of the fermionic Higgs
%mechanism.
Although a static ($\W=0$) gauge mode, $a(0,\b q)$, escapes the Landau damping, at $T=0$ it constitutes
a set of measure zero in the continuum of excitations and is thus unimportant. At finite $T$,
the static $\W_n=0$ Matsubara frequency is well separated from the others and thus provides a
viable scattering channel. As long as the hierarchy $\w\ll T\ll \mu$  is maintained,
where $\mu$ is the electronic chemical potential, this effect remains. We can interpret the situation as follows:
the rotors are scattered by a static random magnetic field, $\b\nabla\times \b a(\b x)$, generated
by the emergent gauge fluctuations, which increases the resistivity compared with
the usual insulator-superfluid transition of rotors. The width of this random distribution of static magnetic fields
is proportional to the temperature $T$. This is illustrated in \rfig{rand-b}.
To determine the corresponding
scattering rate, let us first rewrite the gauge-boson--rotor scattering term:
\begin{align}
I_a[f_s] &= \frac{2}{N}\int_0^\infty \frac{d\W}{\pi}
\int \frac{d^2q}{(2\p)^2}
 (2\b k\times \hat{\b q})^2\Im D(\W,q) \\\nn
&\times \Bigg\{\frac{(2\p)\de(\e_k-\e_{k+q}-\W)}{4\e_k\e_{k+q}}\Big[f_s(\b k,t)
(1+f_s(\b k+\b q,t))(1+n(\W)) \\ \nn
&\hskip10em -f_s(\b k+\b q,t)(1+f_s(\b k,t))n(\W)\Big]\\ \nn
&+\frac{(2\p)\de(\e_k-\e_{k+q}+\W)}{4\e_k\e_{k+q}}\Big[f_s(\b k,t)
(1+f_s(\b k+\b q,t))n(\W) \\ \nn
&\hskip10em -f_s(\b k+\b q,t)(1+f_s(\b k,t))(1+n(\W))\Big]\Bigg\}\, .
\end{align}
We have neglected the particle-antiparticle production term, the one with $\de(\e_k+\e_{-k+q}-\W)$,
because it requires an energy $\W>2m$, which renders it subleading due to the suppression of
dynamical gauge fluctuations. Linearizing yields the simple relaxational form 
\begin{align}
  I_a = -\frac{s\b E\cdot\b k \varphi(k,\w)}{\tau_a(k)} =-\frac{\de f_s}{\tau_a(k)}
\end{align}
with the momentum-dependent scattering rate
\begin{align}\label{eq:scatt_rate}
  \frac{1}{\tau_a(k)} = \frac{T}{N}\times\frac{8k^2}{\pi \e_k}\int_0^1 dy \frac{y^2\sqrt{1-y^2}}{\Pi_b^{\rm j}(0,2ky)}
\end{align}
This result for the scattering rate due to the emergent gauge bosons is valid for temperatures
$T\ll \mu$. 
This \emph{elastic} scattering rate obtained in the static regime is universal
in the sense that it does not depend on the Fermi surface information, such as $k_F$ and $v_F$.
It follows that we can express it in terms of a single-parameter scaling function:
\begin{align}\label{eq:Fa}
  \frac{1}{\tau_a}=\frac{T}{N} F_a\left(\frac{k}{T}\right)%, \frac{\Delta_\pm}{T}\right)
\end{align}
%(The dependence on $\Delta_\pm$ arises entirely from the mass.)
The exact momentum dependence can be determined numerically and is shown in \rfig{scatt_rate}.
In evaluating \req{scatt_rate}, had one simply used the $T=\de=0$ result $\Pi_b^{\rm j}(0,q)\sim q$, the rate
would have vanished as $k\rightarrow 0$.

After linearizing \req{qbe}, we get the following
equation for the deviation, $\varphi$, from the equilibrium distribution:
\begin{align}\label{eq:simplified-qbe}
  -i\w \varphi(p,\w)+ g(p)/T^2 = \frac{T}{N}\left\{-\left[F_\la(p)+F_a(p)\right]
\varphi(p,\w) +\int dp' \, K_\la(p,p')\varphi(p',\w)\right\}
% \frac{g(\e_k/T)}{T^2} = \frac{T}{N}\left[-F(\bar k)\varphi(k,\w)
%     +\int d\bar q K_\la(\bar k,\bar q)\varphi(q,\w)\right]
\end{align}
where we have Fourier transformed from time to frequency and have introduced
dimensionless momentum variables $p,p'$ that have been normalized by temperature:
$p=k/T$. On the LHS,
the term that makes the equation non-homogeneous reads $g(p)=\pd_{\e_p} n(\e_p)/\e_p= -e^{\e_p}/\e_p(e^{\e_p}-1)^2$,
where it is understood that when we use the dimensionless momentum, the rotor mass is scaled by $T$,
$\e_p=\sqrt{p^2+(m/T)^2}$, and the Bose function does not contain the usual
factor of temperature: $n(\e_p)=1/(e^{\e_p}-1)$.
We have also introduced the dimensionless kernel $K_\la(p,p')$ describing
the non-elastic processes in which the rotors exchange energy with the $\la$ bosons.
It is independent of the gauge field, the driving frequency and $N$.
In analogy with \req{Fa}, we have further defined the dimensionless function $F_\la$ that corresponds to the scattering
rate due to interactions with the $\la$ field. More details about $F_\la$ and
$K_\la$ can be found in Appendix~\ref{ap:pure-rot}.
% \begin{align}
%   g(\e) &= \frac{\pd_\e n(\e)}{\e} \\
%   F(k) &= F_\la(k) + F_a(k) \label{eq:scatt_rate_F}
% \end{align}
\subsection{Solution of QBE and rotor conductivity}
\label{sec:qbe-sol}
By performing the rescalings
\begin{align}
  \tilde\w &=\frac{\w N}{T}\, , \\
  \Phi\left(p,\t\w \right) &= \frac{T^3}{N}\varphi(p,\w) \, ,
\end{align}
where again $p=k/T$ is the dimensionless momentum,
we obtain a universal, parameter-free equation:
\begin{align}
  -i\tilde\w \Phi(p,\tilde\w)+g(p) = -F(p)\Phi(p,\tilde\w)
    +\int dp' K_\la(p,p')\Phi(p',\tilde \w)\, .
\end{align}
The gauge fluctuations do not spoil the existence of such a universal equation, which
arises for the pure rotor theory\cite{sachdev-book}, because
they contribute a universal scattering rate modifying $F_\la\rightarrow F_\la + F_a=:F$.

\begin{figure}[h]
\centering
\includegraphics[scale=.6]{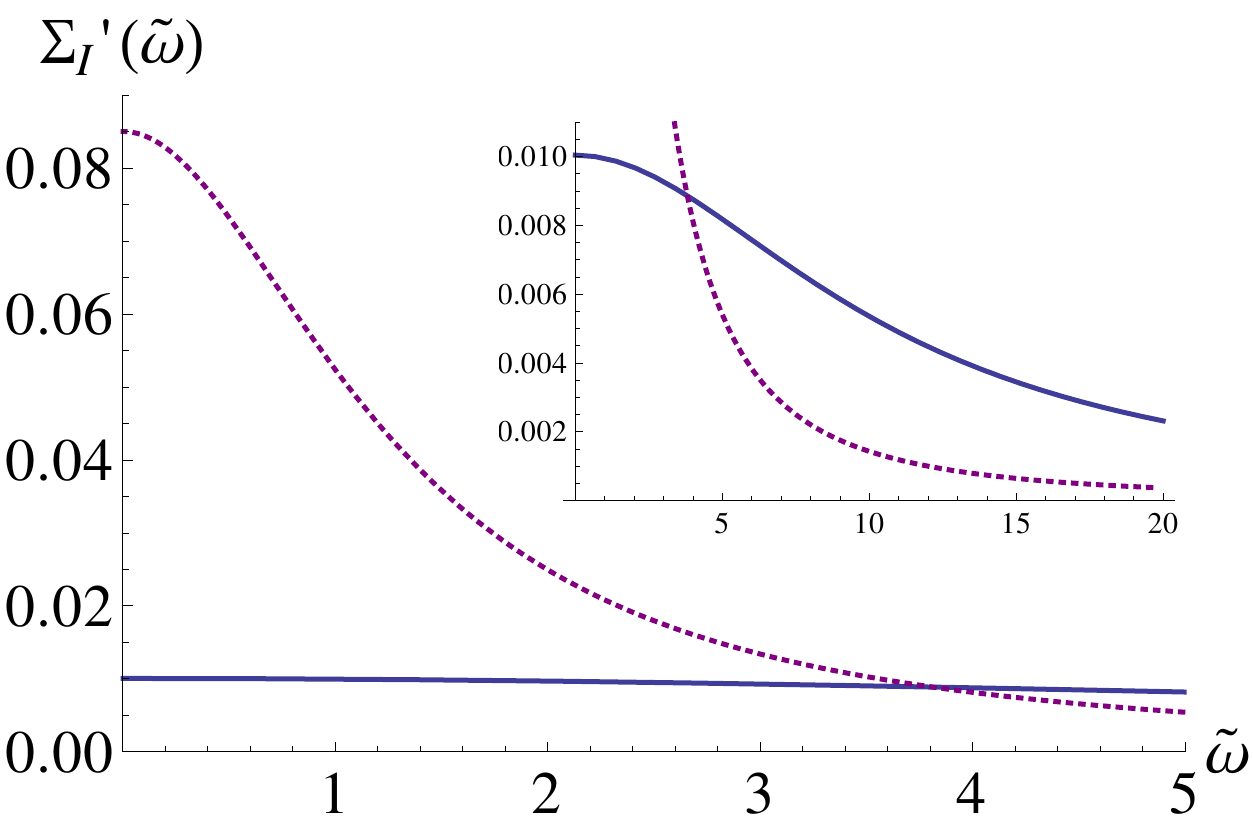}
\caption{\label{fig:Sigma_re_crit} Universal scaling function for the real
  part of the conductivity above the QCP ($\de=0$). $\t\w=N\w/T$ is
the rescaled frequency. The dotted line shows
the scaling function in the absence of the gauge field. The inset shows $\S_I'$
for a larger range of $\tilde\w$.
}
\end{figure}

The above integral equation needs to be solved numerically. Once we obtain
$\Phi$, we can compute the low-frequency conductivity, $\s_{bI}$, from the expression for the
current in terms of the distribution function of the rotors
\begin{align}
  \ang{\b J_I(\w)} &= \int\frac{d^2k}{(2\pi)^2} \sum_{s=\pm}s \frac{\b k}{\e_k}f_s(\b k,\w) \label{eq:e-current}\\
  &= \sum_s s\int\frac{d^2k}{(2\pi)^2}\frac{\b k}{\e_k} s\b E\cdot\b k\varphi(k,\w)
\end{align}
Assuming the $\b E$ field is in the $x$-direction, we get:
\begin{align}
   \sigma_{bI}(\w) &= \ang{J_{Ix}(\w)}/E_x(\w) \\
   &= \frac{1}{2\pi}\int_0^\La dk \frac{k^3\varphi(k,\w)}{\e_k} \\
   &= N\times\frac{1}{2\pi}\int_0^{\La/T} dp \frac{p^3\Phi(p,\t\w)}{\e_p}  \label{eq:sig_Phi-int}
\end{align}
where $\La$ is the momentum cutoff used in the numerical solution. The last equality
makes use of the scaling function for $\varphi$, so that the small-frequency conductivity, $\w/T\ll 1$,
can be written as
\begin{align}
  \s_{bI}(\w)=\frac{e^2}{\hbar} N\times\S_{I}\left(\frac{N\w}{T}\right)\,,%,\frac{\Delta_\pm}{T}\right)
\end{align}
where the fundamental constants $e$ and $\hbar$ were re-introduced;
$\S_{I}$ is a complex valued universal function defined by \req{sig_Phi-int},
the real part of which is
shown in \rfig{Sigma_re_crit}.
As shown there, the conductivity is substantially reduced \emph{at small frequencies} $\t \w$ compared
with the pure $O(N)$ model due to the presence of the emergent gauge field.
The universal number that determines the DC conductivity is $\S_I(0)=0.010$.
It can be compared with $\S_I^{\rm O(N)}(0)=0.085$ in the absence of the
gauge field, i.e. for the pure $O(N)$ model.
Extrapolating to the case of physical interest, $N=2$,
the conductivity reads:
\begin{align}
  \s_b(0)&= \frac{e^2}{\hbar}\times 0.020, \quad {\rm  O(2) + damped\; gauge\; field} \\
  &= \frac{e^2}{\hbar}\times 0.170, \quad {\rm pure\; O(2) }
\end{align}
We note \emph{en passant} that this last number for the pure $O(2)$ model, $0.170$, is very close to
what was obtained in the small-$\e$ expansion\cite{damle97}: $0.1650$. Both these numbers lie near the
self-dual value $1/2\pi\approx 0.159$ which is associated with a conductivity equal to
the quantum of conductance, $e^2/h$.

As the quantity that has a universal jump at the MIT is the resistivity and not the conductivity
(unless we restrict ourselves to clean systems),
we here give the expression for the rotor DC resistivity, $\rho_b=1/\s_b$,
\begin{align}
  \rho_b = \frac{\hbar}{e^2} R, \quad R = 49.8
\end{align}
or $\rho_b= \tfrac{h}{e^2}\times 7.93 = 205\; {\rm k\W}$. This constitutes one of our main results:
the value of the universal resistivity jump estimated in the large-$N$ approximation, as shown in \rfig{jump}.

\begin{figure}
\centering
\subfigure[ $N\gg 1$]{\label{fig:sig-vs-w-largeN}\includegraphics[scale=.49]{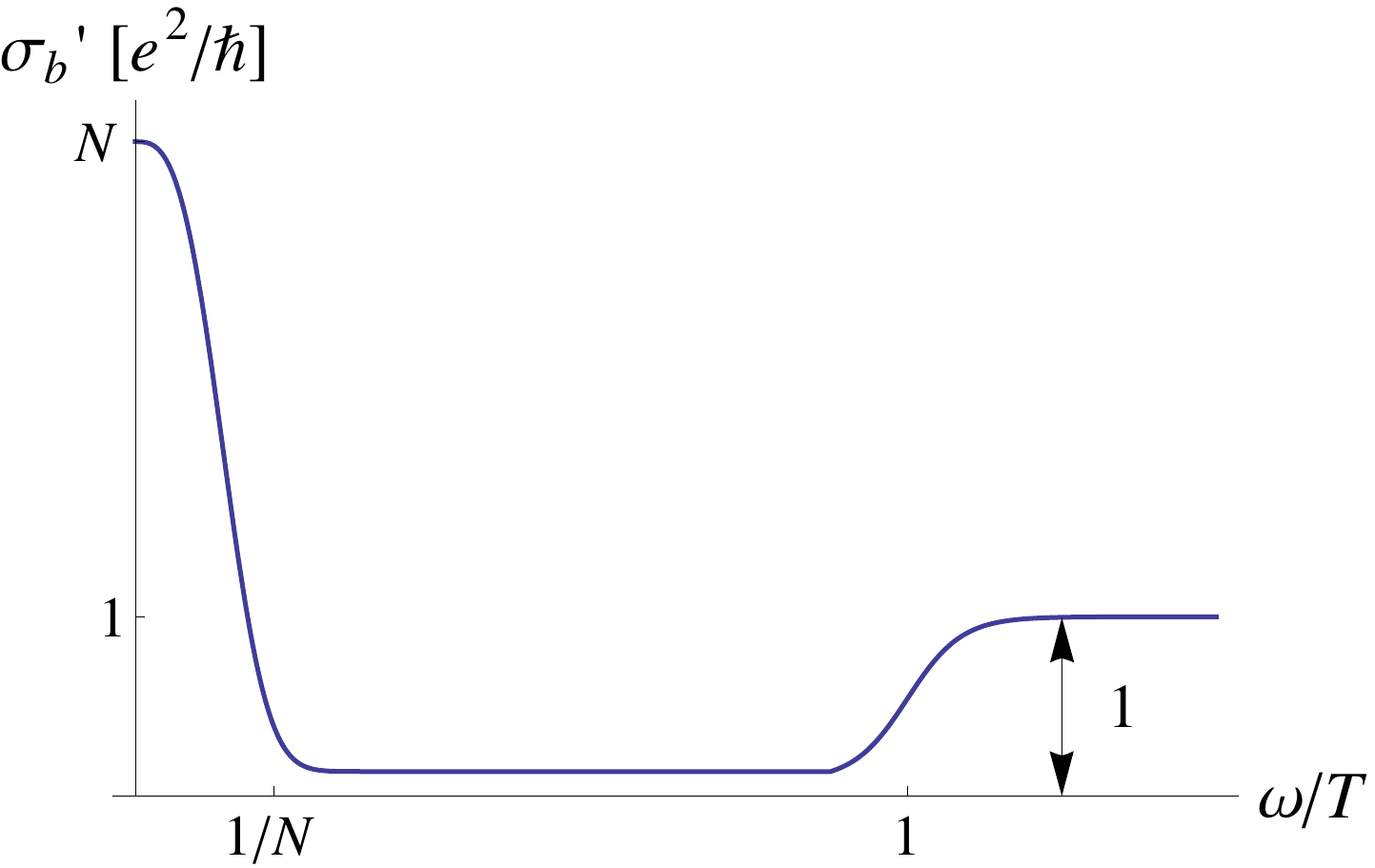}}
\subfigure[ $N= 2$]{\label{fig:sig-vs-w-phys} \includegraphics[scale=.44]{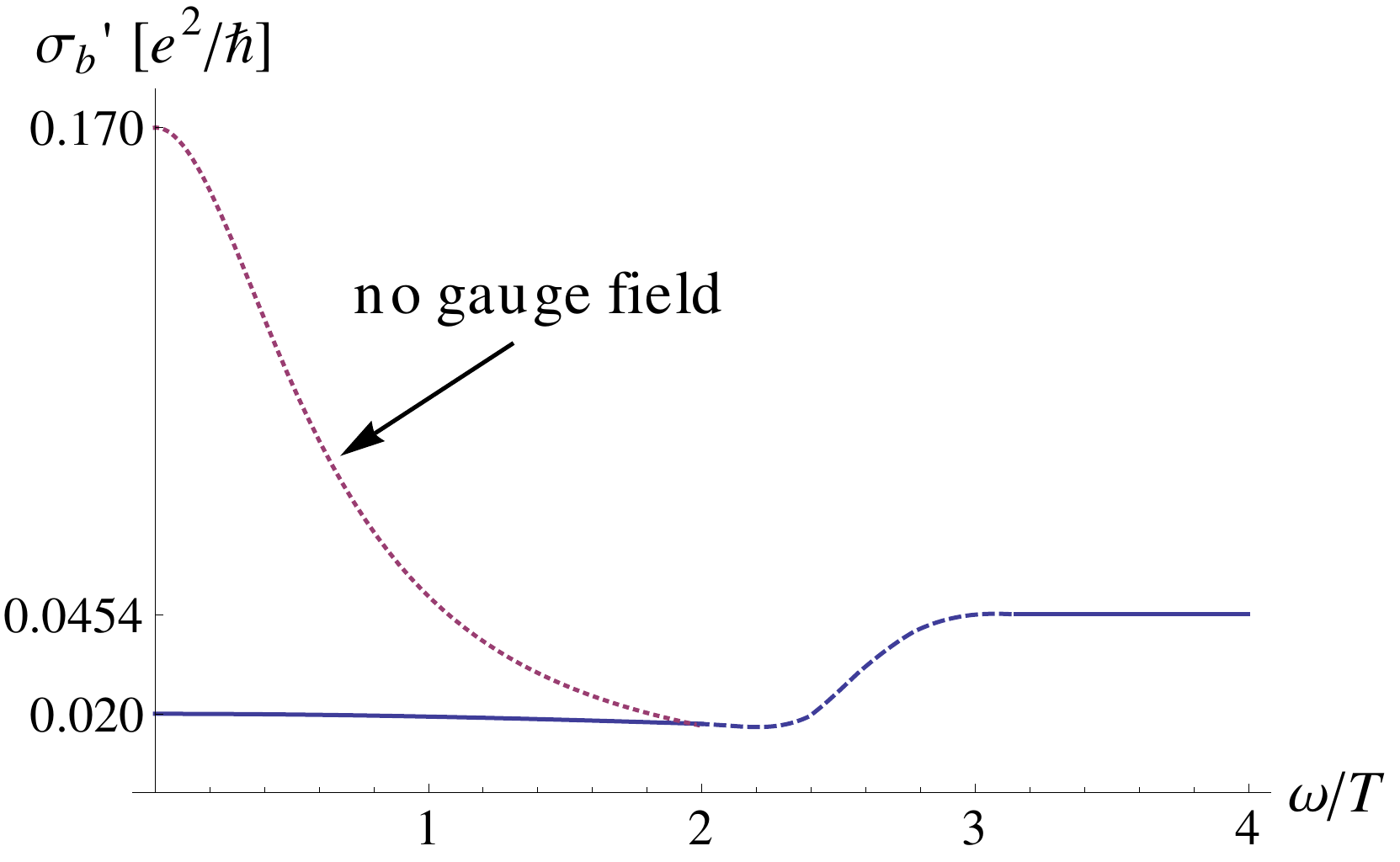}}
\caption{\label{fig:sig-crit}  Frequency and temperature dependent rotor conductivity (real part).
a) Sketch of $\s_b'$ in the \largen limit. There is a Drude-like peak of width $1/N$ and height $N$. While it is due 
to incoherent scattering of thermally excited quasiparticles, the large-frequency $\mc O(1)$
conductivity is from coherent and elastic scattering of field-excited carriers. b) Large-$N$
result extrapolated to $N=2$, the case of physical interest for the Mott transition. The
solid blue (dotted purple) line corresponds to the conductivity with (without) the
emergent gauge field. The dashed line at intermediate frequencies is a sketch
of the expected crossover to the high frequency regime, where the gauge field becomes unimportant.
}
\end{figure}

Although the exact numbers can only be trusted in the
\largen limit, we expect that some semi-quantitative features are captured
in our extension to $N=2$. First, it is clear that the damped gauge field will necessarily
make the conductivity smaller because it adds an additional scattering channel.
The decrease in the rotor conductivity will lead to an increase
in the resistivity, which translates into a bigger jump as one approaches
the QCP from the metallic side, compared to a treatment that neglects gauge fluctuations.
In our calculation, the increase is by a factor of $\sim 8$, which is substantial.
Although, the actual enhancement might not be as large, our result suggests that the gauge fluctuations are
the dominant source of current dissipation. Moreover, the strong scattering by the gauge bosons leads to the
frequency dependence found in \rfig{sig-vs-w-phys}, where the Drude-like peak
occurring in the pure $O(N)$ model disappears. Indeed, the small frequency conductivity
is smaller than in the $\w/T\gg 1$ limit. This ``inverse Drude peak'' might
naively suggest that vortex excitations, whose conductivity is the
quasiparticle \emph{resistivity}, $\s_b^{\rm vortex}=\rho_b=1/\s_b$, would be better suited,
at least to describe electric transport. However, the vortices are known
to give unreliable perturbative results for the superfluid-insulator
transition in the pure $O(2)$ model. As the presence of the damped
gauge boson does not alter the thermodynamic universality class, we suspect that
this remains true in our model and that
the dual vortex formulation does not offer any numerical advantage.

We briefly comment on the precise frequency dependence of the conductivity at
low frequency, as shown in \rfig{Sigma_re_crit}.
It is possible to determine almost exactly the analytic form of this frequency
dependence for both the pure and gauged $O(N)$ models\cite{will-subir}, which can be surprising
given the complicated form of the QBE. Using the analytic expression, one can evaluate 
a low-frequency sum rule for the real part of the conductivity: $\int_0^\infty d\w\, \s_b'(\w)={\rm constant}$. 
It is found that this integral equals the weight of the delta-function Drude peak obtained in the pure $O(N)$ NL$\s$M at $N=\infty$, 
even in the presence of the gauge field. In particular this means that the inclusion of the
interactions at $1/N$ ($\la$ or gauge boson mediated) only spreads the spectral weight of the delta function over a finite range
of frequencies, this range being broader in the presence of the gauge field, as can be seen in the
inset of \rfig{Sigma_re_crit}.

\section{Conductivity in the entire QC region}
\label{sec:transport-qc-region}
\begin{figure}
\centering
\includegraphics[scale=.5]{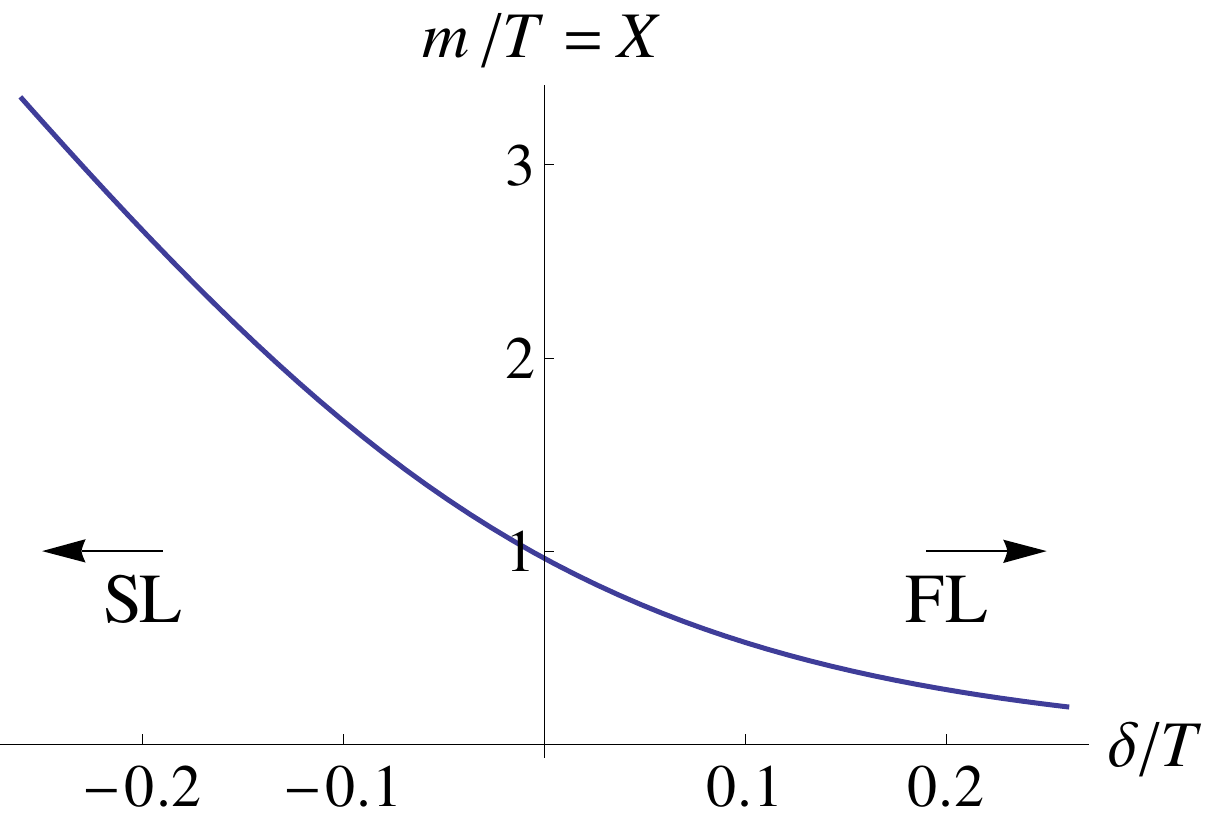}
\caption{\label{fig:mass} Rotor mass scaling function.}
\end{figure}

We now examine how the conductivity changes as the system is tuned away from
$g=g_c$ within the QC region, with $g\propto U/t$, the ratio of the Hubbard repulsion to
the electronic bandwidth. The QC regions is defined
by $T>\De_\pm$, where $\De_\pm$ are the two energy scales that vanish at the QCP. $\De_+$,
defined for $g>g_c$, is the Mott gap of the bosons, while $\De_-$, $g<g_c$, is the
phase stiffness of the rotors in their condensed phase.
These two scales vanish approaching the QCP according to the power law
\begin{align}
  \De_\pm\sim |\de|^{z\nu}\,,
\end{align}
where we are again using the signed energy scale associated with tuning the non-thermal parameter:
\begin{align}
  \de = g\inv -g_c\inv\propto t/U -(t/U)_c\, .
\end{align}
The dynamical exponent, $z$, is unity for all $N$, while the correlation length exponent, $\nu$,
depends on $N$: in the $N\rightarrow \infty$ limit, $\nu=1/(d-1)$ so that in
$d=2$, $\nu=1$. In the QC region, the effective mass of the rotors, the saddle-point value
of the $\la$ field, will change as $\de$ is varied. This will naturally
affect the conductivity: as one approaches the SL, the effective mass of the charge
excitations increases and this leads to
a larger electric resistivity. The mass will depend on the ratio of $\De_\pm/T$:
$\frac{m}{T} = X_\pm\left(\frac{\De_\pm}{T} \right)$.
%which enters in the polarization functions $\Pi_b$ and $\Pi_b^{\rm j}$.
At $N=\infty$, we simply have $\De_\pm\propto |\de|$, with the proportionality constants:
\begin{align}
  \De_+ &= 4\pi \de, \quad  g>g_c\\
  \De_- &= -\de,\quad g<g_c
\end{align}
There, an analytic solution can be obtained for the mass scaling function\cite{chubukov94}
\begin{align}\label{eq:m-scaling}
  \frac{m}{T} &= X\left(\frac{\de}{T}\right)\, , \\
  X(\bar\de) &= 2 \sinh\inv\left(\frac{e^{2\pi\bar\de}}{2} \right)\, ,
\end{align}
which is plotted in Fig.~\ref{fig:mass}.

\begin{figure}
\centering
\subfigure[]{\label{fig:pd_R-vs-T_num}\includegraphics[scale=.52]{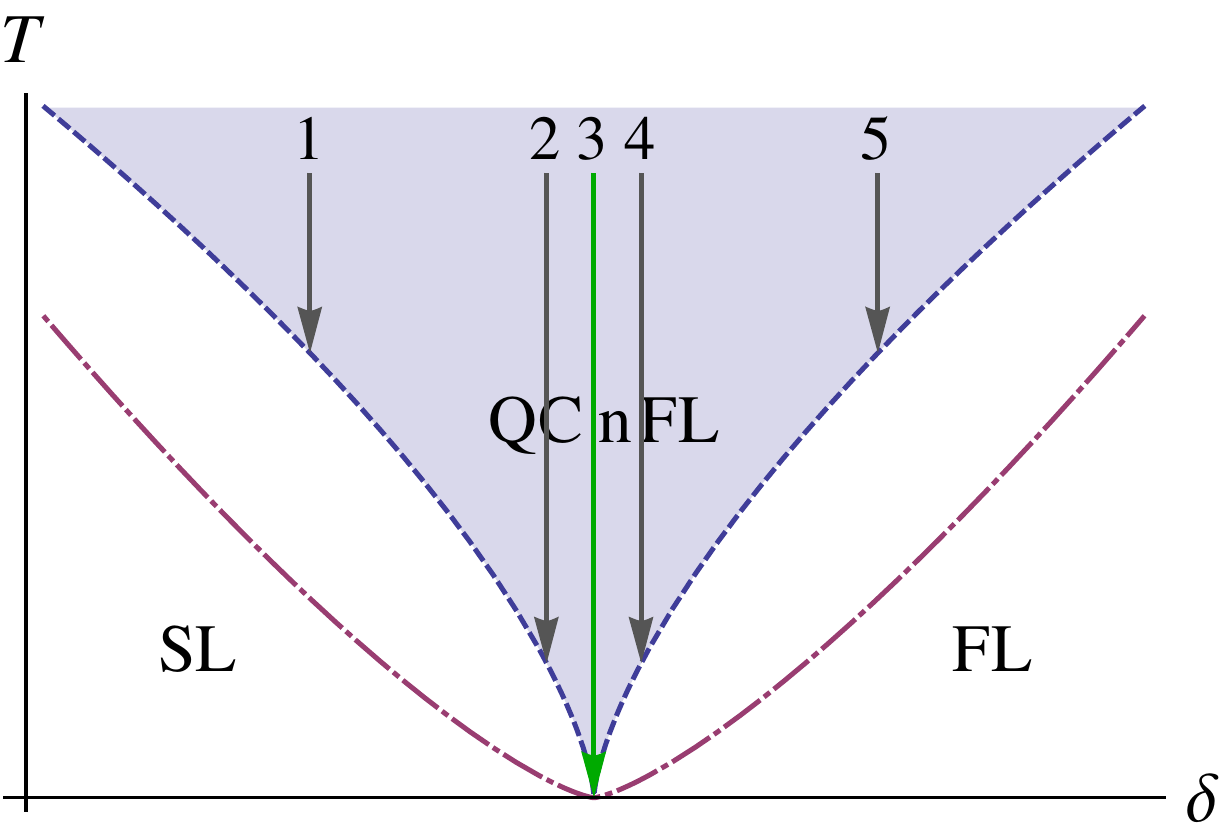}}
\subfigure[]{\label{fig:pd_R-vs-P_num} \includegraphics[scale=.53]{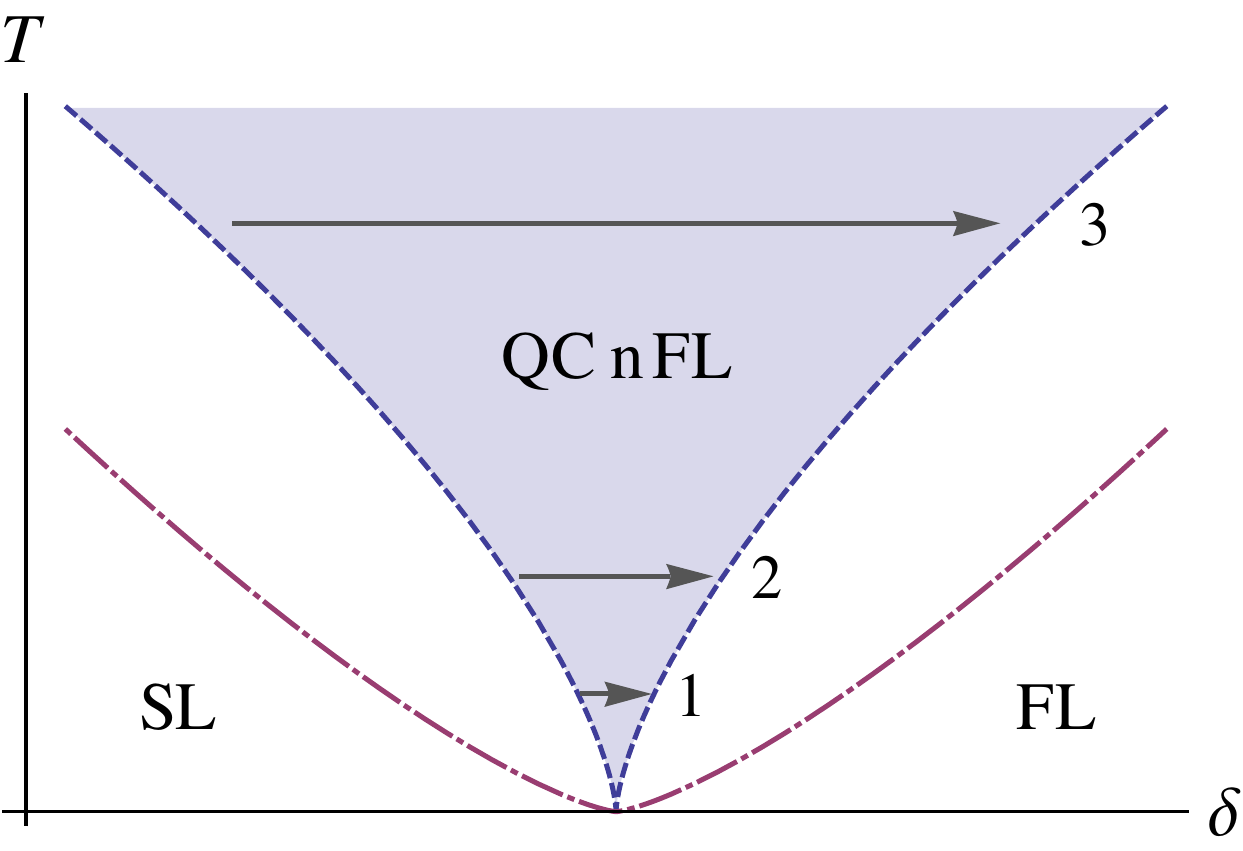}}\\
\subfigure[]{\label{fig:R-vs-T_num}\includegraphics[scale=.46]{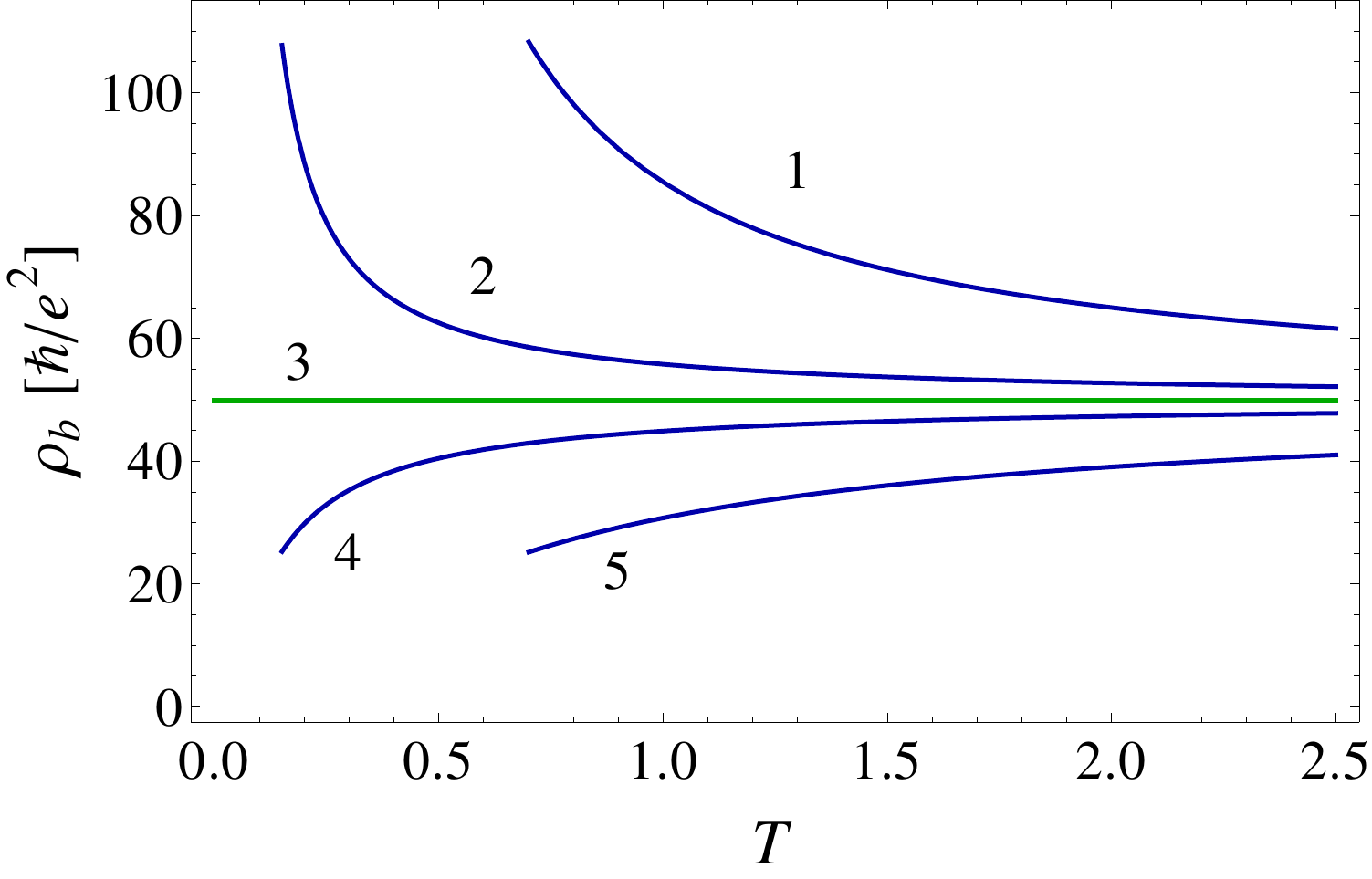}}
\subfigure[]{\label{fig:R-vs-P_num} \includegraphics[scale=.46]{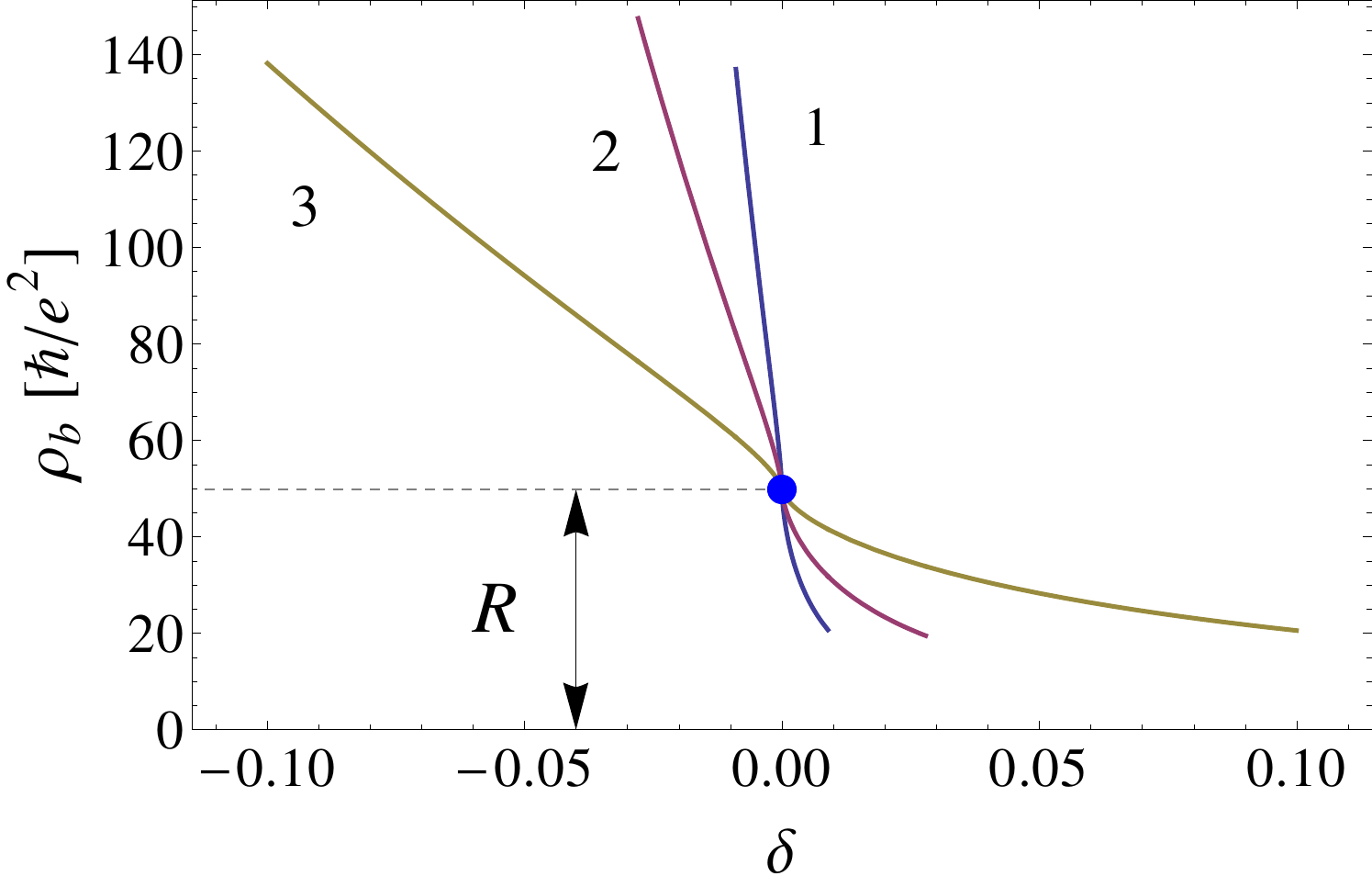}}
\caption{\label{fig:R_num}
Behaviour of the low temperature DC resistivity near the quantum critical (QC) Mott transition
as obtained from the solution of the quantum Boltzmann equation.
Panel c) shows the resistivity vs $T$ for different ratios of the onsite repulsion over the
bandwidth (tuned by $\de$). The corresponding cuts are shown in the phase diagram in a)
and correspond to $\de=-0.01, -0.001, 0, 0.001, 0.01$ going from curve 1 to 5.
Panel d) shows the resistivity vs $\de$ at different temperatures,
with the corresponding cuts shown in the phase diagram in b). Curves 1,2,3 correspond to
$T=0.5, 1.0, 2.5$, respectively.
The universal rotor resistivity at criticality is $\rho_b=R\hbar/e^2$, with $R=49.8$.
($\de$ and $T$ are given in a common and arbitrary unit of energy.)
}
\end{figure}

\subsection{Results}
We have solved for the frequency dependent conductivity at different values of $\de/T$. The latter
affects the polarization functions $\Pi_b$ and $\Pi_b^{\rm j}$ via the rotor effective mass.
The polarization functions determine the propagators of the $\la$ and gauge boson,
and thus the associated density of states the rotor excitations can scatter into. The universal
scaling function for the low-frequency conductivity, $\S_I$, was numerically obtained in the QC region:
\begin{align}
  \s_b=\frac{e^2}{\hbar}N \Sigma_I\left(\frac{N\w}{T},\frac{\de}{T}\right)\,.
\end{align}
In \rfig{R_num}, we show the behaviour of the corresponding DC resistivity extrapolated to
$N=2$: $\rho_b(0)=1/\s_b(0)=(\hbar/e^2)/2\S_I(0,\de^{z\nu}/T)$. These numerical results should be
compared with \rfig{R}, where a sketch of the resistivity near the QCP was given, not restricted
to the QC fan. It should also be compared with \rfig{R_no-gf}, in Appendix \ref{ap:pure-rot},
showing the rotor contribution without the gauge field, as relevant
for the conventional superfluid-insulator transition.
Let us first examine \rfig{R-vs-T_num}, which shows the resistivity along constant-$\de$
cuts (\rfig{pd_R-vs-T_num}). These curves can be naturally compared with sheet resistivity versus pressure
experimental data, for instance, where $\de\sim P-P_c$, plays the role of the deviation from the quantum critical pressure.
As mentioned above, the rotor resistivity corresponds to the total electronic resistivity relative to the residual
value in the FL: $\rho_b=\rho-\rho_m$.
Curve 3 shows the resistivity at the critical pressure ($\de=0$): it is constant at
low temperatures and takes a universal value $\frac{\hbar}{e^2}R$, with $R=49.8$. As one goes down in temperature
at pressures differing from $P_c$, the resistivity decreases approaching the FL (curves 4-5), or
increases near the SL Mott insulator (curves 1-2). The more pronounced resistivity jump $\rho_b\rightarrow 0$
occurs upon exiting the critical Fermi surface state and going to the (marginal) FL,
as is shown in \rfig{R-vs-T}.
Figure~\ref{fig:R-vs-P_num} presents the results from a complementary perspective: by fixing $T$ and tuning pressure
following cuts shown in \rfig{pd_R-vs-P_num}.
This illustrates how the $T=0$ resistivity jump becomes smooth at finite temperature. By virtue of the scaling
nature of the resistivity data, all curves cross at $\de=0$, where the resistivity is universal, $\frac{\hbar}{e^2}R$.
The resistivity data was extracted from a single universal scaling function
\begin{align}
  \label{eq:scal-rho}
  \rho_b(\w=0)=\frac{\hbar}{e^2} G\left(\frac{\de^{z\nu}}{T}\right)\,.
\end{align}
We emphasize that, in experiments, it is
the resistivity relative to its residual metallic value should accordingly be examined for scaling in the vicinity of a
quantum critical Mott transition.
\emph{Notes}: In obtaining the results shown in \rfig{R_num}, we have replaced the
$N=\infty$ scaling, $\de/T$, by the one appropriate for $N=2$: $\de^{z\nu}/T$, with
$\nu=0.67$ and $z=1\; \forall N$. (When $\de<0$, it is understood that $\de^{z\nu}=\sgn(\de)|\de|^{z\nu}$.)
A caveat with the extrapolation is that we have used the mass scaling function obtained at $N=\infty$.
Although the specific form of $m/T=X(\de^{z\nu}/T)$ will be different for $N=2$, we expect it to be qualitatively
similar, at least near criticality.

%We see that it is possible to use a single analytic scaling function, $X$, instead of two, $X_\pm$,
%when we use $\de$ instead of $\De_\pm$.

\section{Conductivity at large frequencies and temperatures}
\label{sec:largeT-w}
\subsection{Large frequencies}
\label{sec:large-w}
In this section we discuss the behaviour of the rotor conductivity in the
large frequency limit $\w\gg m\sim T$. As discussed above, in the
\largen limit, the finite universal conductivity in that region mainly
results from the elastic, coherent transport
of charged excitations created by the external field, as opposed to the
incoherent transport of thermally excited quasiparticles relevant at small frequencies.
In our above treatment of the quantum Boltzmann equation we cannot obtain
that part of the conductivity as we have neglected precisely those processes
that are dominant for $\w\gg T$.
Rather, the large-frequency conductivity $\s_{bII}=(e^2/\hbar)\S_{II}(\w/T)$ can be obtained from
a $T=0$ calculation, i.e. in the limit $\w/T\rightarrow \infty$.
At $N=\infty$, the $\la$ and gauge fields do not contribute and we recover the pure rotor
contribution known from previous works\cite{cha91}:
\begin{align}
  \label{eq:sig_large-w_N_infty}
  \S_{II}'(\w/T \gg 1) =\frac{\pi S_d}{2^d d}\left|\frac{\w}{c}\right|^{d-2}
  \xrightarrow{d=2}\frac{1}{16} = 0.0625
\end{align}
where $S_d=2/[\Gamma(d/2)(4\pi)^{d/2}]$. Contrary to the DC conductivity, which was infinite at
$N=\infty$, the fact that the high frequency conductivity is already finite in the free limit
testifies about the different mechanisms at play, namely, its collisionless nature in contrast
to the hydrodynamic transport at small frequencies.
Including the $1/N$ correction that arises due
to interactions of the rotors with the $\la$ field yields\cite{cha91}
\begin{align}
  \S_{II}'(\w/T\gg 1) = \frac{1}{16}\left(1-\frac{8}{3}\eta\right) \xrightarrow{N=2} 0.03998
\end{align}
where $\eta=N\inv (8/3\pi^2)$ is the leading correction to the
anomalous dimension of the rotor field. We do not include the gauge fluctuations
because they are made ``massive'' by the Landau damping. At $T=0$, even the static
component is ineffective, lying within the continuum of dynamical excitations.
For a more precise estimate of $\S_{II}'(\infty)$, we quote the Monte Carlo
results of Cha et al.\cite{cha91}: $\S_{II}'(\infty)=0.0454$. This is the number we use
for the large frequency conductivity in \rfig{sig-vs-w-phys}.
\subsection{Temperature dependence of electrical transport at criticality}
In our analysis of the quantum critical resistivity,
we have so far limited our discussion to the universal quantum critical resistivity of the bosons.
As the temperature is decreased to zero right at the quantum critical point how is this universal value approached? To address this we first note that the full resistivity at low $T$ is given by the Ioffe-Larkin formula as the sum of boson and fermion resistivities. At low-$T$ the dependence of the fermion resistivity is dominated by spinons scattering off the gauge fluctuations.
 As mentioned in Ref.~\onlinecite{senthil08-2},  this leads to
\begin{align}
  \label{eq:rho_f-largeT}
  \rho_f\sim \left(\frac{T}{\mu}\right)^2 \ln \left(\frac{\mu}{T}\right)\,.
\end{align}
To obtain this dependence the gauge propagator at criticality,
$D\inv\sim i\mu\frac{\w}{v_Fq} + q$, was used.
Regarding the subleading temperature dependence of the rotor conductivity,
we have indications (see next subsection) that the treatment of the QBE requires more care,
as in the case of a Fermi surface of spinons coupled to an emergent gauge
field\cite{lee92,yb-qbe,nave}. We leave such analysis for future investigation.
Notwithstanding, the leading non-constant $T$-dependence of the
full resistivity will show a departure
from the usual $T^2$ behavior because the rotor contribution is not expected to cancel the non-FL
term provided by the spinons.
This departure from a $T^2$-dependence constitutes a further signature of the non-FL nature
of the critical Fermi surface state. We note that the logarithmic correction due to the spinons will persist in the
MFL phase, because the gauge field only becomes Higgsed by the rotor condensate
at lower temperatures, where the usual FL is recovered.

\subsubsection{Necessity of careful treatment of subleading $T$-dependence of rotor conductivity}
To estimate the temperature dependence of the rotor conductivity at $\de=0$
beyond its universal constant value, one can try to compute the transport
scattering rate, just as was done for the spinon contribution, \req{rho_f-largeT}.
We need to evaluate the imaginary part of the rotor self-energy due to the gauge
fluctuations, with the usual additional factor of $1-\cos\theta$ in the integrand.
The leading order dependence is universal and linear in $T$, in agreement with
our QBE calculation. The subleading term is negative and goes like $-T^2/\mu$, where
$\mu$ is the fermionic chemical potential. We believe that such a negative contribution
points to the inability of such a simple approach to capture the correct $T$-dependence.
Indeed, this scattering rate would imply a conductivity that
increases with temperature via $\s_b\sim T\tau_{\rm tr}^b\sim 1/(1-T/\mu)$.

Instead of using this semiclassical approach, we can turn to the full quantum Boltzmann equation.
However, a straight-forward approach fails because the QBE contains terms that diverge due to
scattering of low energy gapless gauge bosons. This can be related to the
divergence of the rotor self-energy at finite $T$, which
displays a logarithmic singularity in the infrared. It cannot be
naively cutoff because the low energy properties of the gauge field and rotors were
properly treated. This situation is exactly analogous to the case of fermionic spinons or
non-relativistic bosons coupled to an emergent gauge field studied in the
context of U(1) spin liquids\cite{lee92}. It was found that the divergence of the
self-energy (or vanishing of the Green's function) is a natural consequence
of the gauge fluctuations on a non--gauge-invariant operator, which in this case is
the rotor Green's function.

This type of singularity in the QBE was previously encountered in the context
of the spinon--gauge-field problem by Kim \emph{et al.}\cite{yb-qbe},
revisited latter by Nave and Lee\cite{nave}. Kim \emph{et al.} found that a separation of the gauge
fluctuations into static (with frequency less than the temperature) and dynamic ones together with the
use of a gauge invariant momentum remove the singularities and allow the extraction of physical quantities.
It was also noted that the static gauge fluctuations act as a random magnetic field that contributes to the
conductivity by increasing the scattering of spinons, a situation very analogous to the effects of the gauge
field on the critical rotors analyzed in this work. However,
in the present system, the static gauge fluctuations dominate the low temperature transport and
we did not need to take into account the dynamical or quantum gauge fluctuations, at least for the low
temperature transport. The latter are essential to determine the higher temperature behaviour
and one needs to perform a careful treatment analogous to Ref. \onlinecite{yb-qbe}, a task
we leave for the future.

\section{Thermal conductivity}
\label{sec:thermal}
The thermal conductivity also bears a signature of the critical Fermi surface, in which
the emergent gauge field plays an even more important role than for electric transport.
%The thermal conductivity, $\kappa$, is defined by $\b J_h=\kappa \b\nabla T$, where $\b J_h$ is
%the heat current in response to a thermal gradient $\b\nabla T$.
According to the Ioffe-Larkin
composition rule, the thermal \emph{conductivities} of the spinons and rotors add\cite{lee92}:
$\kappa=\kappa_b+\kappa_f$.
In other words, the thermal resistivities add in parallel, which is in contrast to the rule for
electrical resistivity
(addition in series) because of the absence of charge flow in response to the thermal gradient. In this sense,
the thermal current is oblivious to the slave-rotor constraint relating the rotor charge to the spinon number,
and the \emph{relative} flow of the partons can proceed unconstrained.
% This removes the
% constraint on the relative flow of the partons because the constraint necessary to remove unphysical
% states doesn't need to be explicitly enforced in the heat current.

Let us first consider the rotor thermal conductivity, $\kappa_b$. In the absence of the gauge field,
the rotors decouple from the spinons and their action reduces to the critical theory of the pure $O(N)$ model.
This is a conformal field theory (CFT). On symmetry grounds, a CFT has infinite thermal conductivity because the
energy current is conserved\cite{senthil-cft,vojta00}. Indeed, by virtue of conformal
invariance, there exists a conserved and symmetric
energy-momentum tensor: $\pd_\mu T_{\mu\nu}=0$ and $T_{\mu\nu}=T_{\nu\mu}$, where $\mu,\nu$ are space-time indices.
These two conditions imply that $\tfrac{d}{d\tau}\int d\b x\; T_{i\tau}=0$, with spatial indices
$i=x,y$. Put in words, the energy current, $T_{i\tau}$,
is conserved and will not be dissipated by interactions, contrary to the charge current.
In our critical theory, this situation is avoided because of the damped gauge fluctuations, which naturally break conformal invariance and
lead to a finite and universal critical thermal conductivity:
\begin{align}\label{eq:kap_b}
  \kappa_b=\frac{k_B^2}{\hbar} K T\, ,
\end{align}
where $K$ is a dimensionless number
associated with the Mott QCP, just like $R$. The gauge fluctuations are more detrimental
in the determination of the thermal conductivity than for the electric conductivity:
whereas the later was already a finite universal
constant without the gauge bosons, the former is formally infinite without the gauge fluctuations.
In reality, this would not be the case due to the presence of irrelevant (in the RG sense) Umklapp scattering
by the lattice, which would lead to a large, non-universal but finite conductivity. In the case under consideration,
we do not need to refer to such processes because the gauge
scattering is stronger and leads to a universal answer, \req{kap_b}.

Although the full calculation of $\kappa_b$ is beyond the scope of this work, we mention some
of the important aspects. First, in the electric resistivity calculation performed above, a key
simplification in the \largen framework is that we can neglect the effect of the electric field
on the $\la$. As was explained in section~\ref{sec:qbe}, only one rotor
flavor is directly coupled to the
electric field so that its effects on the rotor polarization functions, which are obtained by
summing over all rotor flavors, are subleading in $N$. This is no longer true when a thermal gradient
is present, as all rotor flavors inexorably transport energy/entropy in the same way. Hence, the
non-equilibrium corrections to the polarization functions cannot be neglected. To obtain the correct
QBE describing the heat transport, one should use the Keldysh formalism, a task beyond the scope of the
present work.

We now turn to the spinon conductivity $\ka_f$.
In the presence of weak disorder, the low temperature spinon thermal conductivity will scale like
$\kappa_f={\rm const}\times T$, a form valid on both
sides of the transition. Approaching from the FL, the constant $\ka/T$ is simply $L\s_m$,
where $L=\frac{\pi^2}{3}\left(\tfrac{k_B}{e}\right)^2$ is the usual Lorentz number while
$\s_m$ is the residual metallic conductivity, by virtue of the Wiedemann-Franz law obeyed
in the FL. As the critical point is reached, $\ka/T$ jumps by a universal amount $\frac{k_B^2}{\hbar} K$.
Note that contrary to the electric conductivity, the thermal one is finite on the SL side, and
is dominated by the spinon--gauge-field sector.

\subsection{Violation of Wiedemann-Franz}
Combining the electric resistivity and $\kappa/T$ jumps,
we predict the Lorenz number, $L=\kappa/T\s=\rho\kappa/T$, will also
jump at the transition, indicating a violation of the Wiedemann-Franz law. Approaching from
the FL side, the Lorenz number will obey the Wiedemann-Franz law
$L=\frac{\pi^2}{3}\left(\tfrac{k_B}{e}\right)^2$ until it jumps by a universal
amount  $\left(\tfrac{k_B}{e}\right)^2KR$ directly at the transition.

\section{Discussion}
\label{sec:discussion}
\subsection{Experiments}
A preliminary analysis of unpublished pressure and
temperature dependent resistivity data on both \kapET\cite{kanoda-up} and EtMe$_3$Sb[Pd(dmit)$_2$]$_2$\cite{kato-up} 
provide encouraging
hints regarding the presence of a quantum critical Mott transition as described in this work.
Indeed, at sufficiently high temperatures, the \emph{sheet resistivity} saturates to a constant
$\sim h/e^2$, as we predict. The qualitative pressure and temperature dependence is also roughly in agreement
with our results. A closer analysis of the data will be needed to make a stronger statement.
Not to mention the potential complications with the inhomogeneous effects of pressure on the relatively soft organic salts.
%\subsection{3D Mott transition}
\subsection{Disorder}
We discuss the effects of disorder and provide an estimate for the temperature at which we expect it to
become important in the context of the organics.
We shall assume the disorder is weak as is appropriate for these materials.
Then its main effect will be to modify the gauge field's inverse Green's function to (we work in imaginary time)
\begin{align}
  k_F \frac{|\Omega_n|}{\sqrt{q^2 + 1/l^2}} + \sigma_b^0 \sqrt{\W_n^2 + c^2 q^2}
\end{align}
Here, $l$ is the spinon elastic mean free path, and we have reinstated $c$, the rotor velocity,
which is on the order of $v_F$. $\s_b^0$ is a shorthand for
the $T=0$ rotor conductivity, $\s_b(\w/T\gg 1)$.
The zero Matsubara frequency gauge mode does not respond to the change in the Landau damping, which vanishes
at zero frequency.
So the question is about the effect on the non-zero Matsubara frequency components.
These will notice the finite mean free path when $q<1/l$ or relating the gauge
boson energy ($\sim T$) to the momentum via the $z=2$ scaling $\W\sim q^2$, we get
$T \lesssim \frac{\sigma_b^0 c}{k_F l^2}$.
Using $c\approx v_F$, this can be cast as
\begin{align}
  T \lesssim \frac{\s_b^0 \mu}{(k_F l)^2}\, ,
\end{align}
where $\mu$ is the Fermi energy or chemical potential.
The organics are good metals, so we take $k_F l \approx 10$, although the actual mean
free path is probably larger. Also, $\mu\approx 10^3$ K and $\s_b\approx 0.05$, in units
of $e^2/\hbar$. 
%This corresponds to the value of the rotor conductivity in the $T/\w\rightarrow 0$ limit.
We thus conclude that within our framework, using parameters relevant to the organics of interest,
disorder will start playing an important role at temperatures
\begin{align}
  T \lesssim 0.5\, \rm{K}\, .
\end{align}
Above that temperature, we can treat the higher Matsubara modes without including disorder,
and the results of this paper should provide a valid description.

\section{Conclusion}
We have analyzed the transport signatures of a quantum critical Mott transition between
a Fermi liquid metal and a paramagnetic Mott insulator in two spatial dimensions, at fixed filling.
In this scenario, the Mott phase is characterized by a Fermi surface of spin-only quasiparticles,
resulting in a gapless U(1) spin liquid. The physics of such a transition is conveniently captured
by a slave-rotor field theory of the electronic Hubbard model, in which the charge fluctuations are
described by a quantum XY model of rotors,
coupled to an emergent gauge field. The superfluid phase of the rotors corresponds to the metal,
while a Mott insulator results in the disordered phase, where only spin fluctuations remain gapless.
Directly at the transition, a strongly correlated non-Fermi liquid metal emerges where the electronic Fermi
surface is on the brink of disappearance; it is an instance of a critical Fermi surface.
The zero temperature electric resistivity of such a critical state was
predicted to be greater than that of the Fermi liquid by a universal amount, $R\hbar/e^2$, where $R$ is a
dimensionless number associated with the Mott quantum critical point, see \rfig{jump}. We have obtained an estimate for
this number, $R\approx 49.8$, via the solution of a quantum Boltzmann equation for the charge fluctuations, analyzed in a
\largen limit. We found that the emergent gauge fluctuations strongly contribute to the universal resistivity jump,
albeit being ineffective at changing the universality
class of the charge fluctuations from 3D XY. This is so because the \emph{static} gauge fluctuations escape the Landau damping,
and are responsible for the strong, and universal, scattering of the critical charge fluctuations. Though we have focused on transport properties which are best accessed in experiments, we may anticipate that the static gauge field fluctuations might also affect other dynamical quantities (for instance the low frequency form of the electron self energy at non-zero temperature) in the quantum critical regime. 

We have further examined how this resistivity jump evolves at finite temperature and as one changes the ratio
of the Hubbard repulsion to the electronic bandwidth (experimentally tunable by applying pressure). We have obtained
a universal scaling function that can be used to collapse the temperature and pressure dependent
resistivity in the quantum critical Fermi surface state: $\rho-\rho_m=(\hbar/e^2)G(\de^{z\nu}/T)$,
where $\rho_m$ is the residual resistivity in the Fermi liquid, $\de$ can be mapped to the
deviation from the critical pressure, and $z$ and $\nu$ are the usual
dynamical and correlation length exponents, respectively, of the 3D XY universality class. In particular, $G(0)=R$.

Turning to thermal transport, we make the prediction that the low temperature
thermal conductivity (divided by temperature), $\kappa/T$,
has a universal jump at the transition, \rfig{jump}. The gauge fluctuations are responsible for this jump by
breaking the conformal invariance of the charge fluctuations.
Together with the jump of the electric resistivity, the thermal conductivity jump leads to
a violation of the Wiedemann-Franz law by the critical Fermi surface state.

Regarding experiments, the organic salts \kapET and EtMe$_3$Sb[Pd(dmit)$_2$]$_2$ are candidate materials for
the transition covered in this work. Indeed, at ambient pressure they display numerous
signatures characteristic of a gapless quantum spin liquid, and become metallic under the
application of hydrostatic pressure.
As mentioned in section \ref{sec:discussion}, recent unpublished data for the temperature and pressure
dependent resistivity near the Mott transition seem to indicate a jump on the order of $h/e^2$ near the transition,
as well as qualitative agreement with the scaling we propose. A confirmation would point towards the first experimental
example of a quantum critical Mott transition (of fermions), as well as a non-trivial signature of spin-charge
separation in two dimensions.

On the theoretical side, it would be interesting to investigate the charge transport within a controlled dimensional
expansion, with small parameter $3-d$ where $d$ is the spatial dimension, and see how the results compare with
the \largen expansion. Numerical simulations directly in $d=2$ and at $N=2$ would also be desired.
Regarding the jump of the thermal conductivity, a full treatment within the \largen expansion is still missing,
due to the complication of drag of the constraint field of the $O(N)$ non-linear sigma model. A first step would be to
establish the divergence of the thermal conductivity in the pure XY model, as required by conformal invariance, then to
include the emergent gauge fluctuations. Taking a broader perspective, we envision that
the results regarding the effects of the damped gauge fluctuations on the transport of relativistic
or particle-hole symmetric quasiparticles discussed in this work can be applied to other
systems.
%, the ``doublon-metal''\cite{kaul08} being one example.

\section*{Acknowledgements}
We are very grateful for many insightful discussions with S. Sachdev, and for his assistance in
helping verifying numerical results for the pure $O(N)$ model. We are indebted to K. Kanoda and
R. Kato for sharing their unpublished data, and for enlightening
discussions. We also wish to thank K. Damle for early discussions and for providing
a copy of his PhD thesis, as well as
A. G. Green for sharing his knowledge of the thermal response of relativistic bosons.
Finally, we acknowledge insightful discussions with M. Barkeshli, S.-S. Lee,
K. Michaeli, L. Motrunich, D. Mross, A. Paramekanti, D. Podolsky, A. Potter, J. Rau, G. Refael, and X.-G. Wen.
We acknowledge the hospitality of MIT (WWK) and Perimeter Institute (TS, WWK),
where significant portions of this work were done. This research
was funded by NSERC, the Canada Research Chair program,
and the Canadian Institute for Advanced Research (WWK, YBK),
FQRNT and the Walter Sumner Foundation (WWK), an ICMT fellowship
at UIUC (PG), and NSF Grant DMR-1005434 (TS).
\appendix
\section{Transport in the pure $O(N)$ rotor theory}
\label{ap:pure-rot}
In this appendix we present details regarding transport properties
of the pure $O(N)$ NL$\s$M,
where the $N=2$ theory describes the superfluid-insulator transition of
bosons in two dimensions, as present in the Bose-Hubbard model at integer filling for instance.
Some of these results correct previous ones, or cannot be found in the literature.
In addition, they are important as a comparison ground with the rotor theory used to
describe charge criticality in the Mott transition under consideration. The latter
is a gauged version of the $O(N)$ model, with a Landau damped gauge boson.

The bare action for the pure $O(N)$ theory is:
\begin{align}
  S_{b} = \frac{1}{2g}\int_x \left(|\pd_\nu b_\al|^2 +i\la(|b_\al|^2-1)\right)\, ,
\end{align}
where are $\al$ runs from 1 to $N/2$, so that the $N/2$ complex fields form
a real $O(N)$ vector field. The above action has full $O(N)$ global invariance.
(The notational choice of using $N/2$ complex fields instead of $N$ real ones
is more convenient from the point of view of the slave-rotor formulation.)
Again, $\la$ is a Lagrange multiplier field enforcing the constraint that
the rotor is unimodular: $\sum_\al |b_\al|=1$. In the $N=\infty$ limit, the
field theory becomes free, allowing a $1/N$ perturbative expansion.
At $N=\infty$, the expectation value of the $\la$ field plays
the role of an effective mass for the rotors in their insulating phase, as described in section~\ref{sec:largeN}.
The saddle point equation for $\la$ is still given by \req{mf-eq},
because the fluctuating gauge field does not affect the $N=\infty$ rotors. In the
symmetry broken phase, the saddle point equation for the effective mass has no solution,
indicating long range order (at $T=0$ only for $N>2$).

\begin{figure}
\centering
\includegraphics[scale=.5]{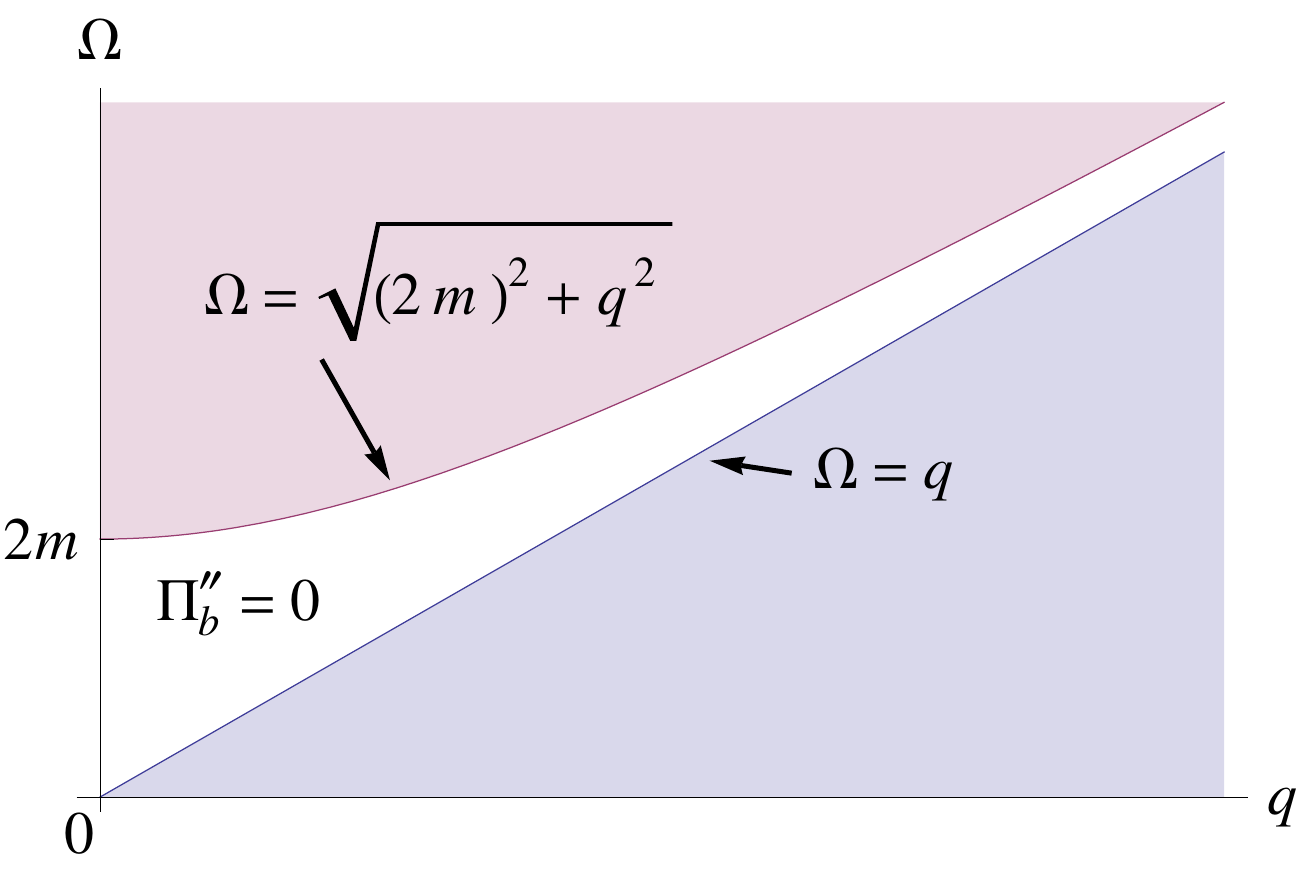}
\caption{\label{fig:supp-ImPi}  The shaded regions correspond to $\Pi_b''\neq 0$.
The upper one (purple) is due to the on-shell pair-production of holons
and doublons, each with mass $m$. }
\end{figure}

We now turn to the charge transport near the quantum critical point. Out of the many charges present at
large $N$,
we couple the electric field to a single one, say the complex component $b_1$.
We focus on the regime where the external field
has a small driving frequency compared with temperature; this corresponds to the
collision dominated hydrodynamic regime. At $N=\infty$, the DC conductivity is infinite
because interactions are entirely suppressed in that limit. At order $1/N$, the $\la$ field
acquires a non-zero propagation amplitude and can scatter the positive and negative
charge excitations, the holons and doublons, respectively.
The quantum Boltzmann
equation for the holons and doublons can be read off \req{qbe}, where one
simply has to set the gauge propagator, $\propto D(\W,q)$, to zero. As explained in the main text,
we expand the holon and doublon distribution functions to linear order in the electric field:
$f_\pm(\b k,\w)=2\pi\de(\w)n(\e_k)\pm\b E\cdot \b k \varphi(k,\w)$. Linearizing the QBE, we obtain
\begin{align}\label{eq:simplified-qbe-la}
  -i\w \varphi(p,\w)+ g(p)/T^2 = \frac{T}{N}\left\{-F_\la(p)
\varphi(p,\w) +\int dp' \, K_\la(p,p')\varphi(p',\w)\right\}
\end{align}
which is the same as \req{simplified-qbe}, except that we have set the scattering rate due to gauge fluctuations, $F_a$,
to zero. As before, $g(p)=\pd_{\e_p} n(\e_p)/\e_p$. The functions $F_\la$ and $K_\la$ can be written in the simple form:
\begin{align}
  F_\la(p)&=\int_0^\infty \frac{p'dp'}{2\pi}\frac{1}{\e_p\e_{p'}}\left[\mc A(p,p')|\g(p,p')|
    +\mc A_{\rm dh}(p,p')\g_{\rm dh}(p,p')\right] \label{eq:scatt-rate-la} \\
  K_\la(p,p')&=\frac{p'}{2\pi}\frac{1}{\e_p\e_{p'}} \left[\mc A_{\rm c}(p,p')|\g(p',p)|
    -\mc A_{\rm c,dh}(p,p')\g_{\rm dh}(p',p)\right]\frac{p'}{p}
\end{align}
where we have defined the two $\g$ functions:
\begin{align}
  \g(p,p') &= n(\e_{p'}-\e_p)-n(\e_{p'}) = \frac{1-e^{-\e_{p'}}}{(1-e^{-\e_p})(1-e^{\e_p-\e_{p'}})} \\
  \g_{\rm dh}(p,p') &= n(\e_{p'})-n(\e_p+\e_{p'})
\end{align}
and the four $\mc A$ functions:
\begin{align}
  \mc A(p,p') &= \int_0^{2\pi}\frac{d\theta}{2\pi}A(|\e_p-\e_{p'}|,|\b p+\b p'|) \\
  \mc A_{\rm c}(p,p') &= \int_0^{2\pi}\frac{d\theta}{2\pi}A(|\e_p-\e_{p'}|,|\b p+\b p'|)(-\cos\theta) \\
  \mc A_{\rm dh}(p,p') &= \int_0^{2\pi}\frac{d\theta}{2\pi}A(\e_p+\e_{p'},|\b p+\b p'|) \\
  \mc A_{\rm c,dh}(p,p') &= \int_0^{2\pi}\frac{d\theta}{2\pi}A(\e_p+\e_{p'},|\b p+\b p'|)(-\cos\theta)
\end{align}
where $\theta$ is the angle between $\b p$ and $\b p'$. We have also defined the scaled spectral function
of the $\la$ field: $A(\W,q)=-\Im[1/\Pi_b(\W,q)]$, where the factor of $2/N$ in the $\la$ propagator,
$2/N\Pi_b$, was omitted from the definition of $A$. The sign of $A$ is such that $A(\W>0,q)>0$.
The $\mc A$ functions can be readily seen to be positive and symmetric. The ``c'' indicates a cosine in the
angle integral, while ``dh'' stands for doublon-holon. To understand this last piece of notation, we
need to examine $\Pi_b$, specifically its imaginary part,% $\Pi_b''(\W,q)$,
\begin{multline}
  \Pi_b''(\W,q)=\int \frac{d^2k}{16\p\e_k\e_{k+q}}\Big\{|n(\e_{k+q})-n(\e_{k})|
\de(|\e_{k+q}-\e_k|-\W) \\
+(1+n(\e_{k})+n(\e_{k+q}))\de(\e_k+\e_{k+q}-\W)\Big\}\, ,
\end{multline}
which is non-zero in two separate regions,
a low and high energy one: $\W<q$
and $\W>\sqrt{(2m)^2+q^2}$, respectively, as is illustrated in \rfig{supp-ImPi}. The latter region arises
from the production of holon-doublon pairs, which requires an energy beyond a certain threshold. The
$\la$-boson spectral function has support in the same regions, being proportional to $\Pi_b''$:
$A=-\Im[1/\Pi_b]=\Pi_b''/[(\Pi_b')^2+(\Pi_b'')^2]$. We plot the numerically-evaluated spectral function in
\rfig{A_la}. We note that $A$ vanishes at the boundaries of the region
where $\Pi_b''=0$. Going back to the $\mc A$ functions, it can be seen that
those labelled ``dh'' receive contributions precisely from the doublon-holon pair production region.

\begin{figure}
\centering
\subfigure[]{\label{fig:A_la}\includegraphics[scale=.37]{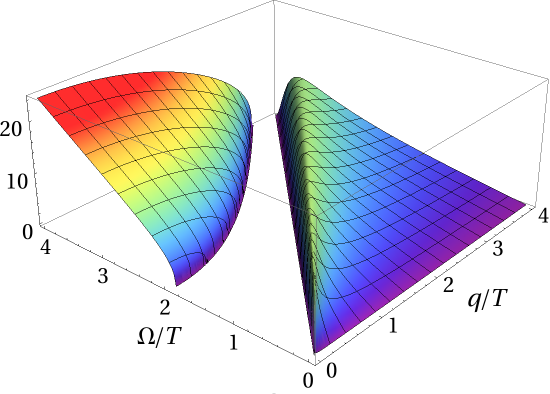}}
\subfigure[]{\label{fig:A_la-densityPlot} \includegraphics[scale=.3]{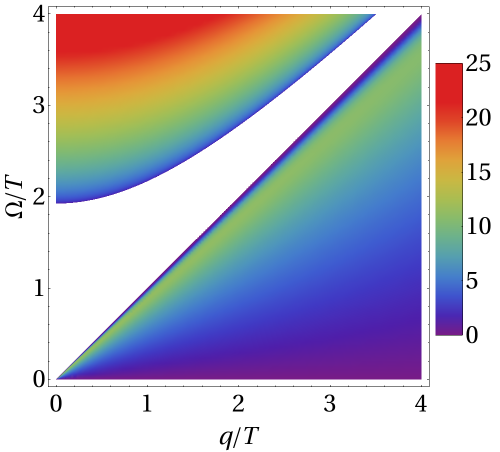}}
\caption{\label{fig:A_la}
a) Spectral function for the $\la$-bosons: $A(\W,q)=-\Im[1/\Pi_b(\W,q)]$. b) The corresponding
density plot.
}
\end{figure}

By performing the rescaling $\tilde\w=\w N/T$ and
\begin{align}
  \Phi\left(p,\t\w\right) = \frac{T^3}{N}\varphi(p,\w)
\end{align}
we obtain a universal, parameter-free equation:
\begin{align}
  -i\tilde\w \Phi(p,\tilde\w)+g(p) = -F_\la(p)\Phi(p,\tilde\w)
    +\int d p' K_\la(p,p')\Phi(p',\tilde \w)
\end{align}

\begin{figure}
\centering
\includegraphics[scale=.5]{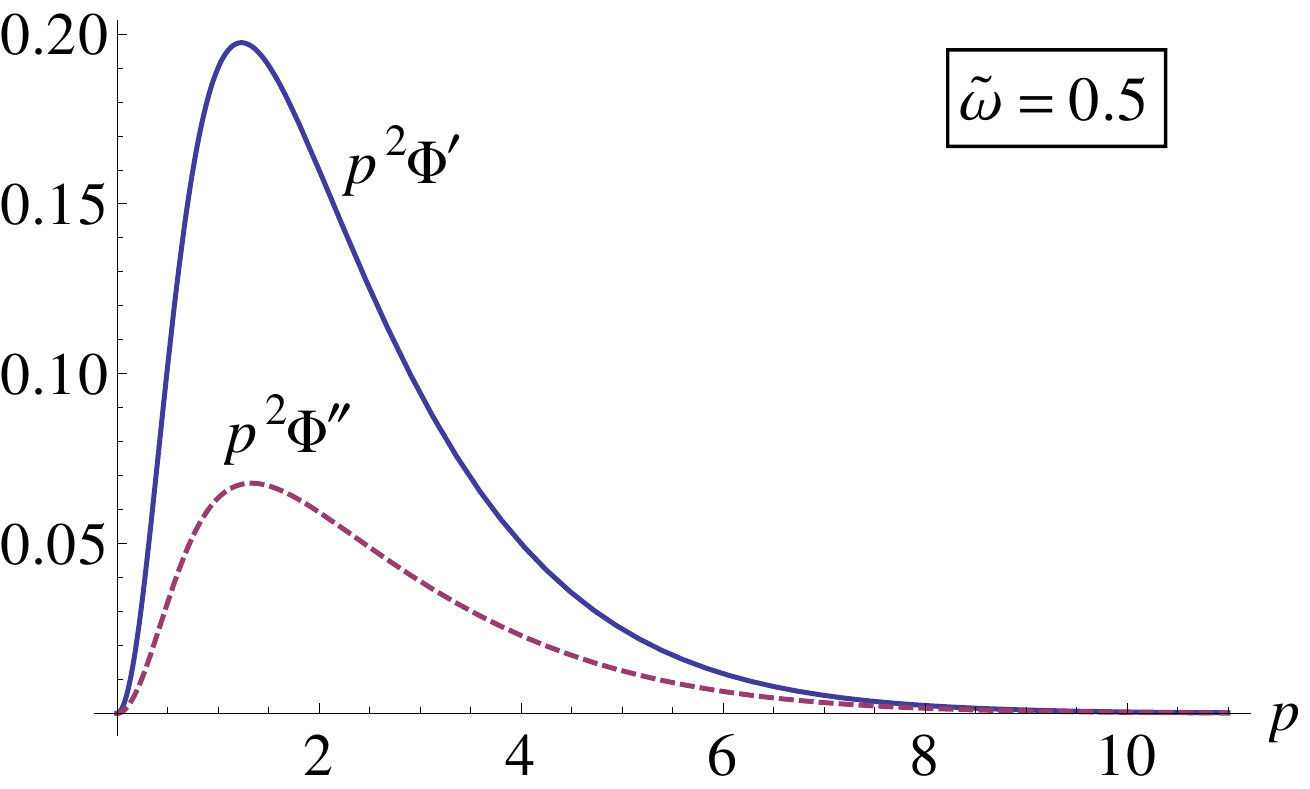}
\caption{\label{fig:Phi-sol}
Solution for electric-field induced deviation to the rotor distribution
function, $\Phi(p,\t\w)$, for $\t\w=N\w/T= 0.5$. The solid (dashed) line
shows the real (imaginary) part, multiplied by $p^2$.
}
\end{figure}

We solve the integral equation numerically by discretizing the momentum variables
and expanding the unknown function $\Phi$ in terms of Chebyshev polynomials. This
procedure converts the integral equation into a matrix equation, which is solved by
simply inverting the matrix corresponding to the kernel $K_\la$. We plot the solution in
\rfig{Phi-sol}. $\Phi(p,\t\w)$ goes to a constant as $p\rightarrow 0$ and decays
exponentially for $p\gg 1$, i.e. for momenta much greater than the temperature.
Also, as the driving frequency $\w\propto\t\w$ goes to zero, the imaginary part, $\Phi''$,
vanishes as expected since the DC conductivity is purely real.

\begin{figure}
\centering
\subfigure[]{\label{fig:Sig-vs-w_no-gf} \includegraphics[scale=.6]{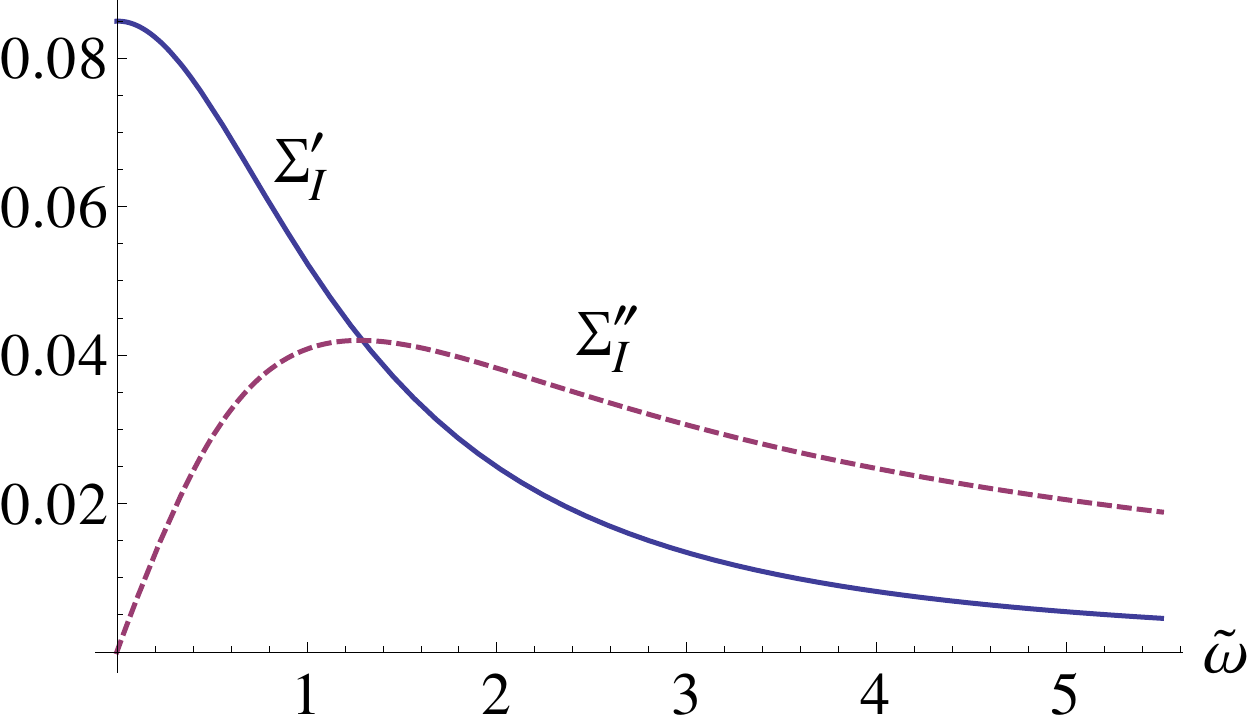}}
\subfigure[]{\label{fig:ReSigCurves}\includegraphics[scale=.5]{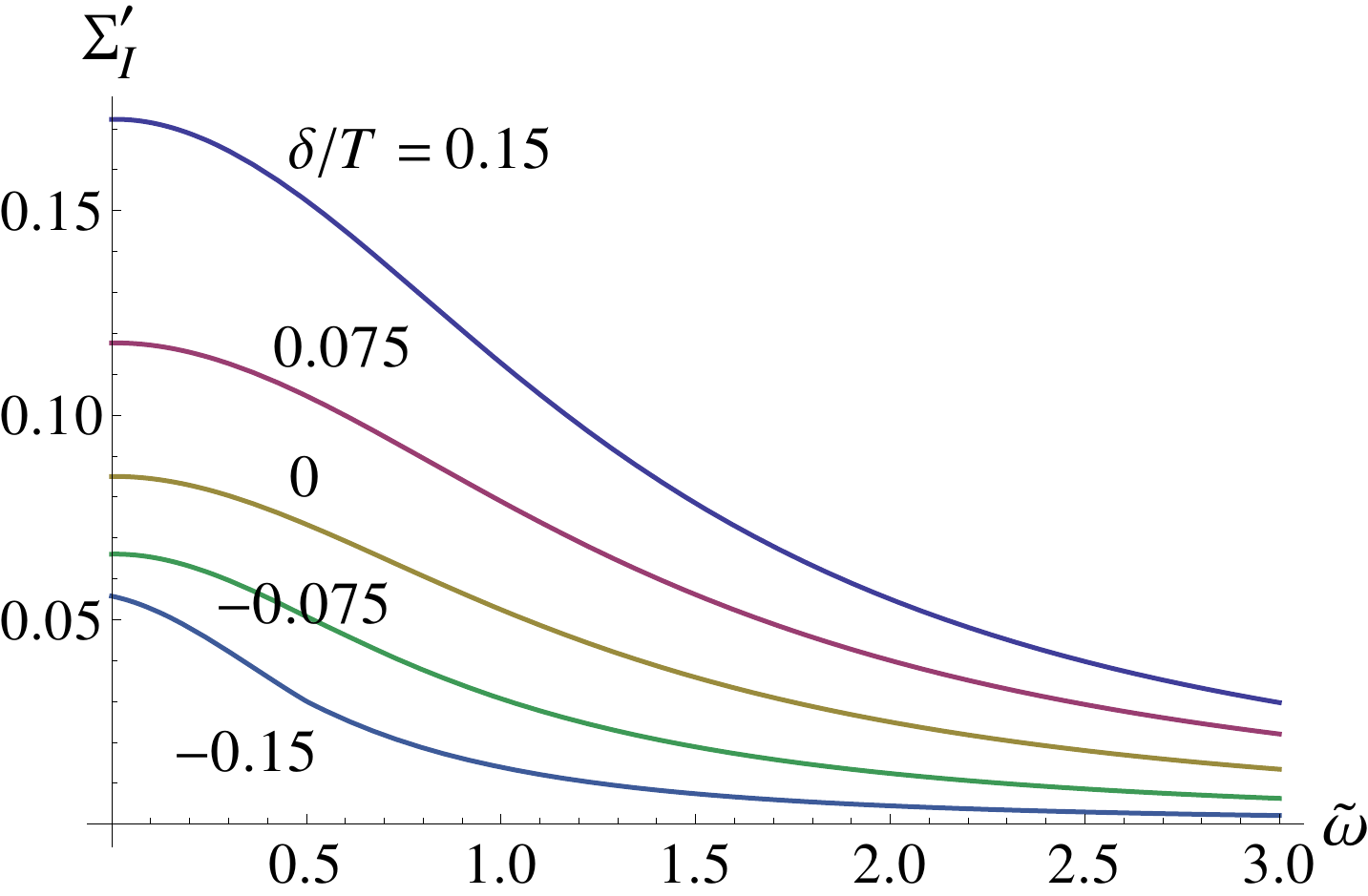}}
\caption{\label{fig:Sig_no-gf}
Conductivity universal scaling function for the pure $O(N)$
theory. a) Frequency-dependent scaling at finite $T$
above the QCP, $\de=0$: $\s_b=N\S_I(\t\w)$, where the argument
is $\t\w=N\w/T$. The solid (dashed) line corresponds to the real (imaginary) part.
b) The scaling function (real part) as a function of the departure from criticality, $\de/T$.
}
\end{figure}

The resulting frequency-dependent conductivity can be obtained by integrating $\Phi(p,\t\w)$ over
momentum, as shown in \req{sig_Phi-int}. As a result we obtain a scaling function
for the conductivity: $\s_b=(e^2/\hbar)N\S_I(\t\w,\de^{z\nu}/T)$. The numerical solution for the scaling
function is shown in \rfig{Sig_no-gf}, where a) shows the frequency dependence of both the real
and imaginary parts of the scaled conductivity for $\de=0$, while b) shows the real part away
from criticality at five different values of $\de/T$. \rfig{ReSigCurves} shows that the conductivity 
increases going away from the disordered to the symmetry-broken phase, corresponding to sweeping $\de$ 
from negative to positive values.

For $N=2$, the conductivity is given by
$\s_b=(e^2/\hbar)2\S_I(2\w/T,\de^{z\nu}/T)$, so that the curves in \rfig{Sig_no-gf} give
the conductivity (in units of $2e^2/\hbar$) versus $2\times \w/T$. In general, the scaled frequency
$\t\w$ on the horizontal axis should be multiplied by $1/N$ to get the dependence on $\w/T$. In the \largen
limit, we recover a Drude-like peak, as the charged quasiparticles interact more weakly.
As mentionned at the end of Section~\ref{sec:qbe-sol}, the essentially exact frequency dependence  
of the small frequency conductivity, cf. \rfig{Sig-vs-w_no-gf}, was recently determined\cite{will-subir}.
This result can be used to prove a sum rule constraining the integral of the real part of the conductivity to the
weight of the delta-function Drude peak at $N=\infty$, i.e. in the free limit of the theory.
This provides an excellent check on the numerics.

%At $N=2$, this can be phrased as going from the Mott insulator (MI) to the superfluid (SF).

\begin{figure}
\centering
\subfigure[]{\label{fig:pd_R-vs-T_no-gf}\includegraphics[scale=.52]{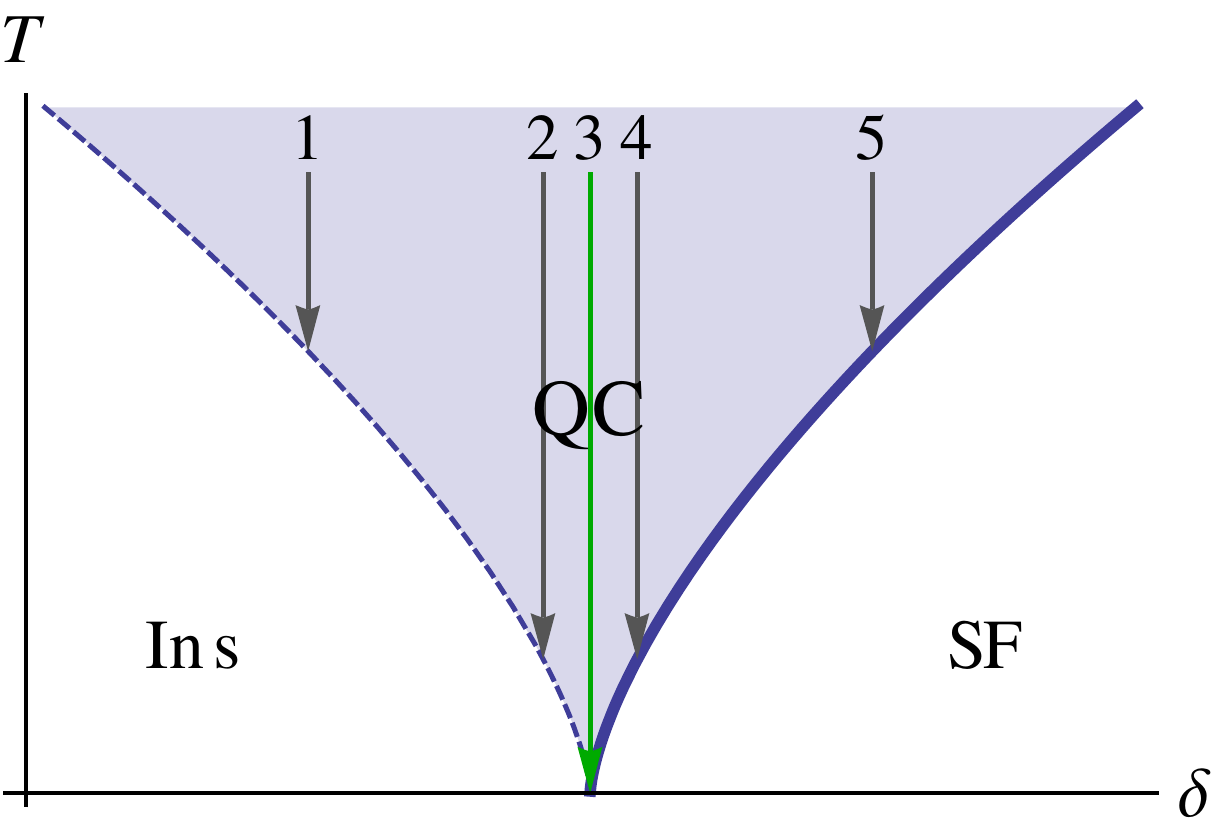}}
\subfigure[]{\label{fig:pd_R-vs-P_no-gf} \includegraphics[scale=.53]{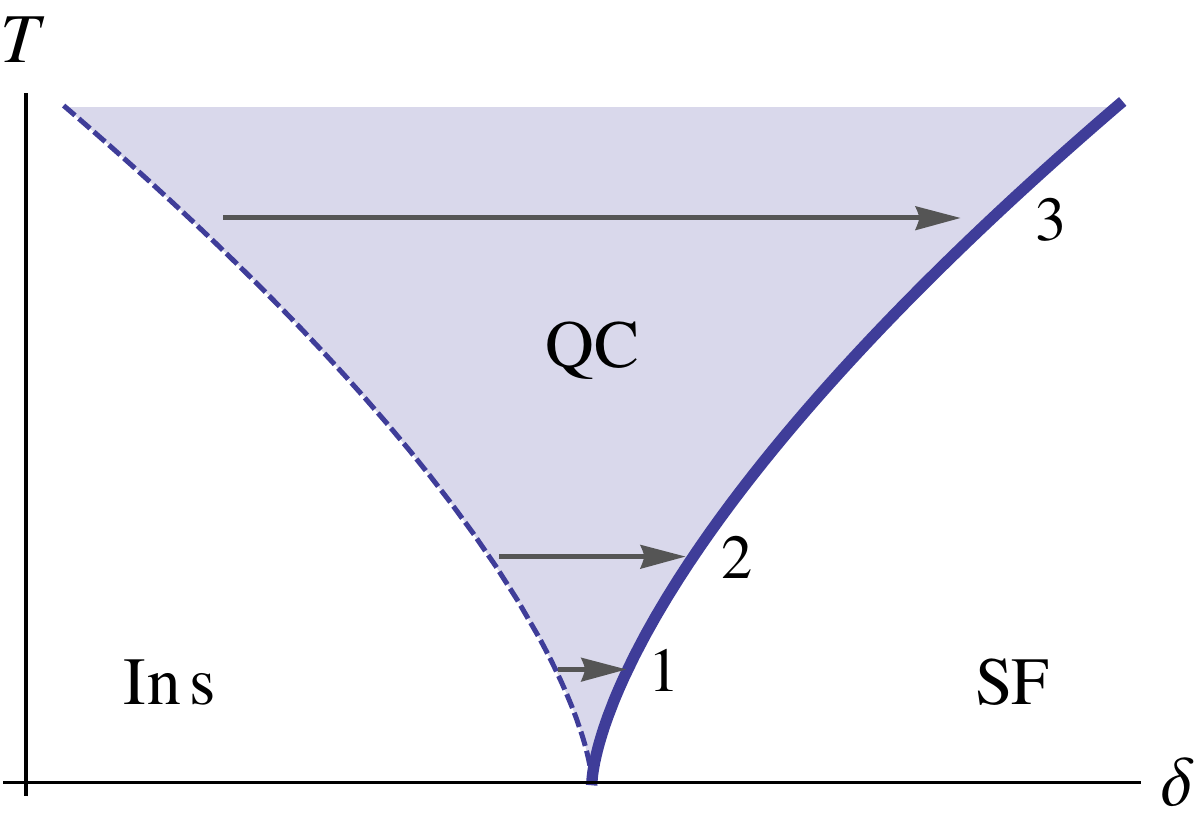}}\\
\subfigure[]{\label{fig:R-vs-T_no-gf}\includegraphics[scale=.44]{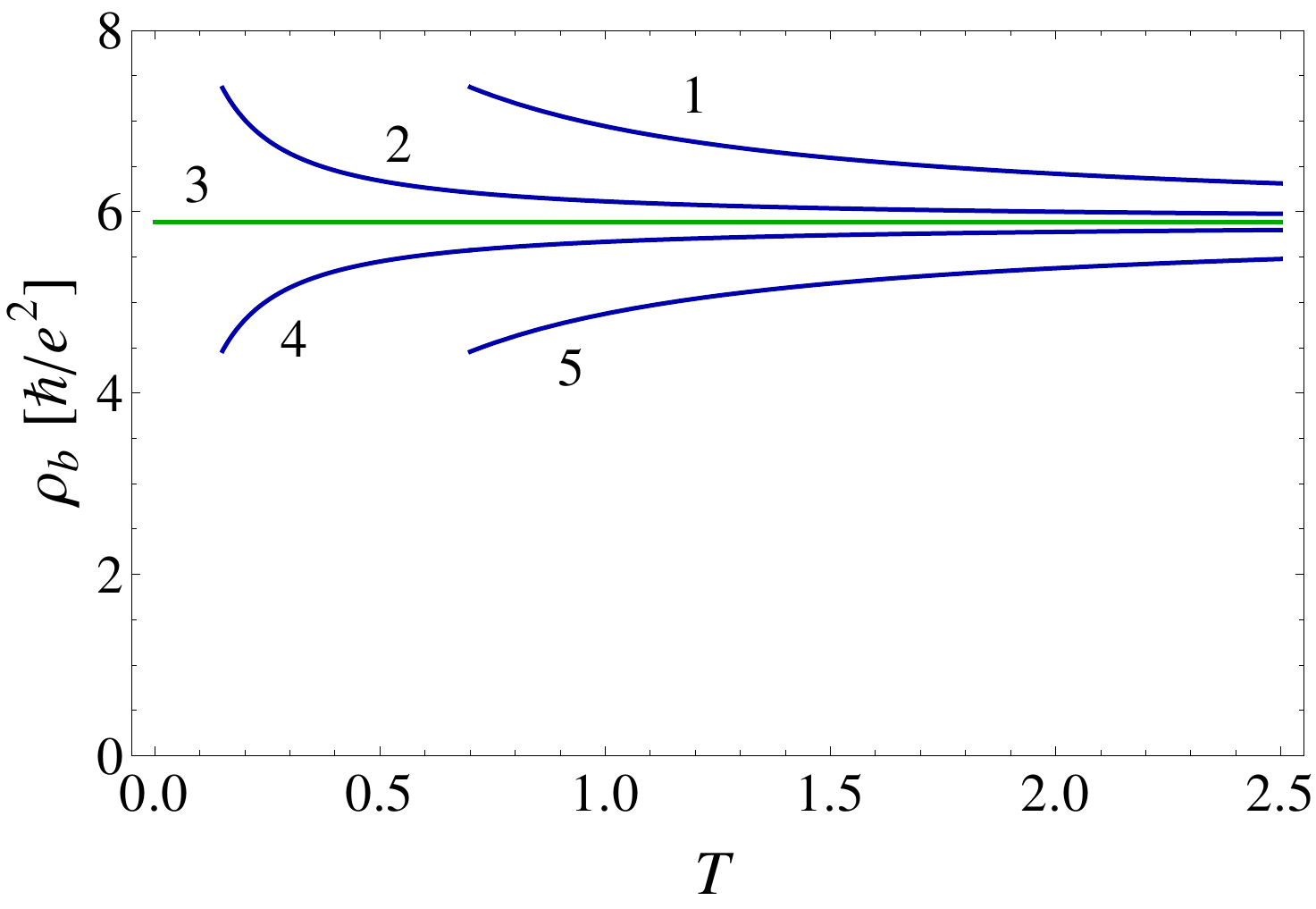}}
\subfigure[]{\label{fig:R-vs-P_no-gf} \includegraphics[scale=.46]{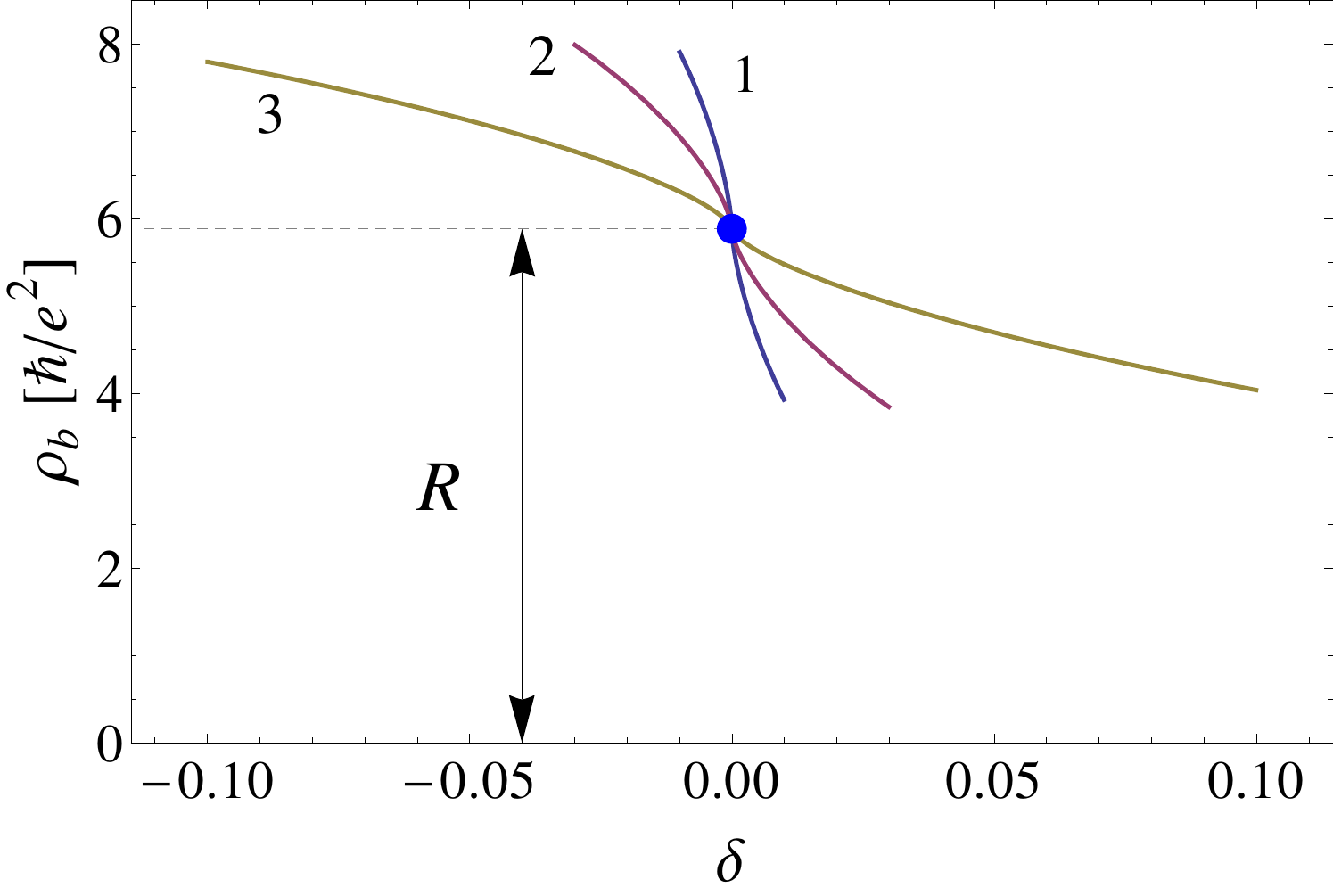}}
\caption{\label{fig:R_no-gf}
Behaviour of the low temperature DC resistivity near the quantum critical (QC)
superfluid--Mott-insulator phase transition. In a) and b), the solid line delimiting the SF indicates
a Kosterlitz-Thouless transition.
Panel c) shows the resistivity vs $T$ for different ratios of the non-thermal parameter $\de$,
which would tune the onsite repulsion, $U$, over the
tunneling amplitude, $t$, in a Bose-Hubbard model, for example.
The corresponding cuts are shown in the phase diagram in a)
and correspond to $\de=-0.01, -0.001, 0, 0.001, 0.01$ going from curve 1 to 5.
Panel d) shows the resistivity vs $\de$ at different temperatures,
with the corresponding cuts shown in the phase diagram in b). Curves 1,2,3 correspond to
$T=0.5, 1.0, 2.5$, respectively.
The universal resistivity at criticality is $\rho_b=R\hbar/e^2$, with $R=5.88$.
($\de$ and $T$ are given in a common and arbitrary unit of energy.)
}
\end{figure}

Focusing on $N=2$, \rfig{R_no-gf} shows the behaviour of the resistivity in the
QC region. We have indicated the presence of a Kosterlitz-Thouless transition
by a solid line at the interface between the QC regime and the SF. For $N>2$,
this finite-$T$ phase transition is converted to a crossover.
This figure should be compared with \rfig{R_num}, where there the bosons are coupled to
a Landau damped gauge field. We see that not only does the gauge field make the resistivity
larger, but its variations are also more pronounced as $\de$ and $T$ are changed.

\section{Thermal conductivity}
We discuss some aspects of the thermal conductivity of the rotors. In particular,
via the solution of a QBE, we explicitly show that in the absence of gauge fluctuations,
the rotors have infinite thermal conductivity, as expected based on conformal invariance.
We then go on to argue that with the addition of the damped gauge field, the conductivity
becomes finite, specifically $\ka_b=(k_B^2/\hbar)KT$, where $K$ is a dimensionless constant. In the
\largen formulation one needs to include effects of the thermal gradient on the constraint
field $\la$, complicating the analysis. We thus consider an alternate but
equivalent formulation of the critical theory where the hard constraint of the XY rotor model, $|b|=1$, is
traded for a soft one via the addition of quartic term $|b|^4$.

\subsection{Thermal response of pure $O(2)$ model: CFT and zero modes}
The critical theory for
the ``soft'' rotor model, without gauge field, is described by the action
\begin{align}
  S=\frac{1}{2}\int d^{d+1}x\, |\pd_\nu b|^2+m^2|b|^2 +\frac{u}{12}|b|^4\, ,
\end{align}
where $d=3-\ve$ is the space-time dimension, and $m^2=\ve(4\pi^2/15)T^2$ and
$u=\ve (24\pi^2/5)$. This is the standard Wilson-Fisher fixed point resulting from
a perturbative RG treatment (see for e.g. \onlinecite{damle97}, which studies the
electric transport properties of that model). We have thus
traded the \largen expansion for a dimensional expansion in $\ve$, $N=2$ being
fixed.

We are interested in the linear response of the bosons to
a thermal gradient. As we did above in the case of the electric transport,
we consider the QBE for the holon and doublon distribution functions, $f_\pm$.
The static thermal gradient is imposed
by including a position dependent temperature, $T(\b x)$, which in turn generates position
dependence for the distribution functions $f_\pm$. The corresponding QBE, which
can be found in Refs.\onlinecite{green-prl,green-prb}, reads:
\begin{align}\label{eq:thermal-qbe}
  \b v_k\cdot \frac{\pd f_\pm}{\pd \b x}=I_\pm[f_+,f_-]\,,
\end{align}
where $\b v_k=\pd \e_k/\pd \b k$ and $\e_k=\sqrt{k^2+m^2}$. The collision term is different from the
\largen expansion as it arises from the quartic interaction:
\begin{align}
  \label{eq:I-small-ep}
  I_\pm=-\frac{2u^2}{9}\int \prod_{i=1}^3\frac{d^2\b k_i}{(2\pi)^d 2\e_{k_i}}
  (\mc F_\pm^{\rm out}-\mc F_\pm^{\rm in})\frac{(2\pi)^{d+1}}{2\e_k}\de(\b k+\b k_1-\b k_2-\b k_3)
  \de(\e+\e_{k_1}-\e_{k_2}-\e_{k_3})
\end{align}
where scattering processes out of state $\{\b k,\pm\}$ have
\begin{align}
  \mc F_\pm^{\rm out} = 2f_\pm(\b k)f_\mp(\b k_1)[1+f_\pm(\b k_2)][1+f_\mp(\b k_3)] +
  f_\pm(\b k)f_\pm(\b k_1)[1+f_\pm(\b k_2)][1+f_\pm(\b k_3)]\, ,
\end{align}
$\mc F_\pm^{\rm in}$ is obtained by interchanging $f_\pm$ and $1+f_\pm$.
We now linearize the QBE to linear order in the temperature gradient.
From the LHS,
\begin{align}
  \frac{\pd f_\pm}{\pd \b x}=-\e_k\pd_{\e_k} n(\e_k) \frac{\b\nabla T}{T}\, ,
\end{align}
while the expanded distribution function reads
\begin{align}
  f_\pm(\b k)=n(\e_k)+\b k\cdot \frac{\b\nabla T}{T}\phi(k)\, ,
\end{align}
with $\phi$ characterizing the departure from equilibrium due to the applied thermal
gradient; it is analogous to $\varphi$ used above in the context of electric transport.
An important difference is that the thermal gradient leads to the same
non-equilibrium distribution functions for both the positive and negative charge
excitations: $f_+=f_-$, contrary to the case with an electric field.
In the linear-response regime, the heat current is simply the
energy current:
\begin{align}
  \b J_h=\int_{\b k} \b v_k \e_k[f_+(\b k)+f_-(\b k)] = 2\frac{\b\nabla T}{T} \int_{\b k} k_x^2\phi(k)\, ;
\end{align}
%The first equality shows that the heat current is equal to the momentum.
while we naturally get a
vanishing electric current because the latter involves the integral of
$f_+(\b k)-f_-(\b k)$, see \req{e-current}.
This is as it should be since the thermal conductivity is measured in an
open circuit setup in which there is no electric charge flow in the steady state.

The linearized QBE reads
\begin{align}\label{eq:cft-qbe}
  -\pd_{k}n_k = I[\phi(k)]\,,
\end{align}
where the functional for the linearized collision integral is
\begin{align}
   I[\phi(k)] = -\frac{\pi \ve^2}{75k^4}\Big\{ \frac{18k^2\phi_k}{n_k}
   \int dk_1dk_2\; I_1(k,k_1,k_2)n_{k_1}n_{k_2}(1+n_{k_2+k_1-k})\\
   -6(1+n_k)\int dk_1dk_2\frac{\phi_{k_1}}{n_{k_1}}I_2(k,k_1,k_2)n_{k_2}n_{k+k_1-k_2}\\
   -12n_k\int dk_1dk_2\frac{\phi_{k_1}}{n_{k_1}}I_3(k,k_1,k_2)n_{k_2}(1+n_{k_2+k-k_1}) \Big\}\,.
\end{align}
The functions $I_i$ result from the angle and $k_3$ angle integrations
of the $\de$-functions and are given in Ref.~\onlinecite{damle97}. We are using the shorthand
$n_k=n(k)$, having dropped the mass, $\mc O(\sqrt{\ve})$, in the dispersion relation,
$\e_k=\sqrt{k^2+m^2}\approx k$, to leading
order in $\ve$.

By conformal invariance we expect the thermal conductivity, which is given
by an integral over $\phi$, to be infinite as explained in Section~\ref{sec:thermal}. 
If we could invert the linear functional $I[\phi]$,
we would obtain $\phi(k)=I\inv [-\pd_k n_k]$, yielding a finite
function $\phi(k)$, as is the case for the electric conductivity. This would in turn imply a finite
thermal conductivity, running against general symmetry arguments. In fact,
$I[\phi]$ cannot be inverted. In other words, the linearized scattering integral has a zero mode, i.e.
a zero eigenvalue eigenfunction $\phi_0(k)$ such that $I[\phi_0(k)]=0$. This zero mode turns out to be the LHS of the
linearized QBE: 
\begin{align}
  \phi_0(k) &=-\pd_k n_k=n_k(1+n_k)\,, \label{eq:phi0} 
%    &={\rm conformal\; zero\; mode}\,, 
\end{align}
as we have verified by direct substitution. Hence, the
equation has no solution, or formally $\phi=\infty$, which implies that the CFT describing the Wilson-Fisher
fixed point has infinite thermal conductivity. It is an excellent check on the formalism,
as it is a priori not obvious how this divergence would come about within the QBE framework.

\subsection{Thermal response of gauged $O(2)$ model}
Let us now return to the case of relevance for the QC Mott transition. In that case the rotors
are coupled to a damped gauge field. As was explained in the main body, only the static gauge
fluctuations are effective at scattering the rotor excitations. These provide a simple elastic
scattering term to the QBE:
\begin{align}\label{eq:thermal-lin-qbe}
  -\pd_k n_k = \t I[\phi(k)], \qquad \t I[\phi]= I[\phi]-\frac{\phi}{\tau_a}
\end{align}
Although $I$ itself is not invertible, $\t I$ is for any finite value of the
scattering time $\tau_a$. In particular, the zero mode of $I$ no
longer is one for $\t I$: $\t I[-\pd_k n_k]=\pd_k n_k / \tau_a$, i.e. it now has
eigenvalue $-1/\tau_a$.

We have solved the QBE, \req{thermal-lin-qbe}, numerically and have verified
the presence of a singularity when the gauge scattering rate is absent. (For simplicity,
we have used a momentum-independent scattering rate but the same conclusions
will hold generically.) More precisely, the numerics yield a very large
value of $\kappa/T\sim 7\times 10^4$ in the CFT limit, i.e. when $1/\tau_a=0$.
This value grows as we increase the number of Chebyshev polynomials used to expand $\phi$
suggesting a divergence in the limit where an infinite number of basis polynomials is used.
Moreover, it becomes $\mc O(1)$ even for very small values of $1/\tau_a$, consistent with the 
fact that the gauge scattering rate moves the system away from the conformal point thus rendering the
thermal conductivity finite.

%\subsubsection{Numerical results}
\subsection{Similarity with fermionic CFT of Dirac fermions}
A similar analysis of conformal zero modes in the context of thermal
transport in a fermionic CFT of Dirac fermions was performed in Ref. \onlinecite{fritz09}.
A zero mode was identified in the corresponding linearized QBE, the analogue of \req{cft-qbe}.
It was was found that the zero mode is essentially given by $\phi_0(k)=-\pd_k n_F(k)=n_F(k)\left(1-n_F(k)\right)$,
where $n_F$ is the Fermi-Dirac distribution. This is the fermionic analogue of the bosonic
mode obtained above, \req{phi0}. It was further found that the introduction of anisotropy 
for the Dirac fermions breaks conformal invariance and makes the zero mode massive. As such, the anisotropy
can be seen as the analogue of the gauge scattering rate in our case.

\section{Rotor current polarization function}
\label{sec:Pi_b_j}
\begin{figure}
\centering
\includegraphics[scale=.5]{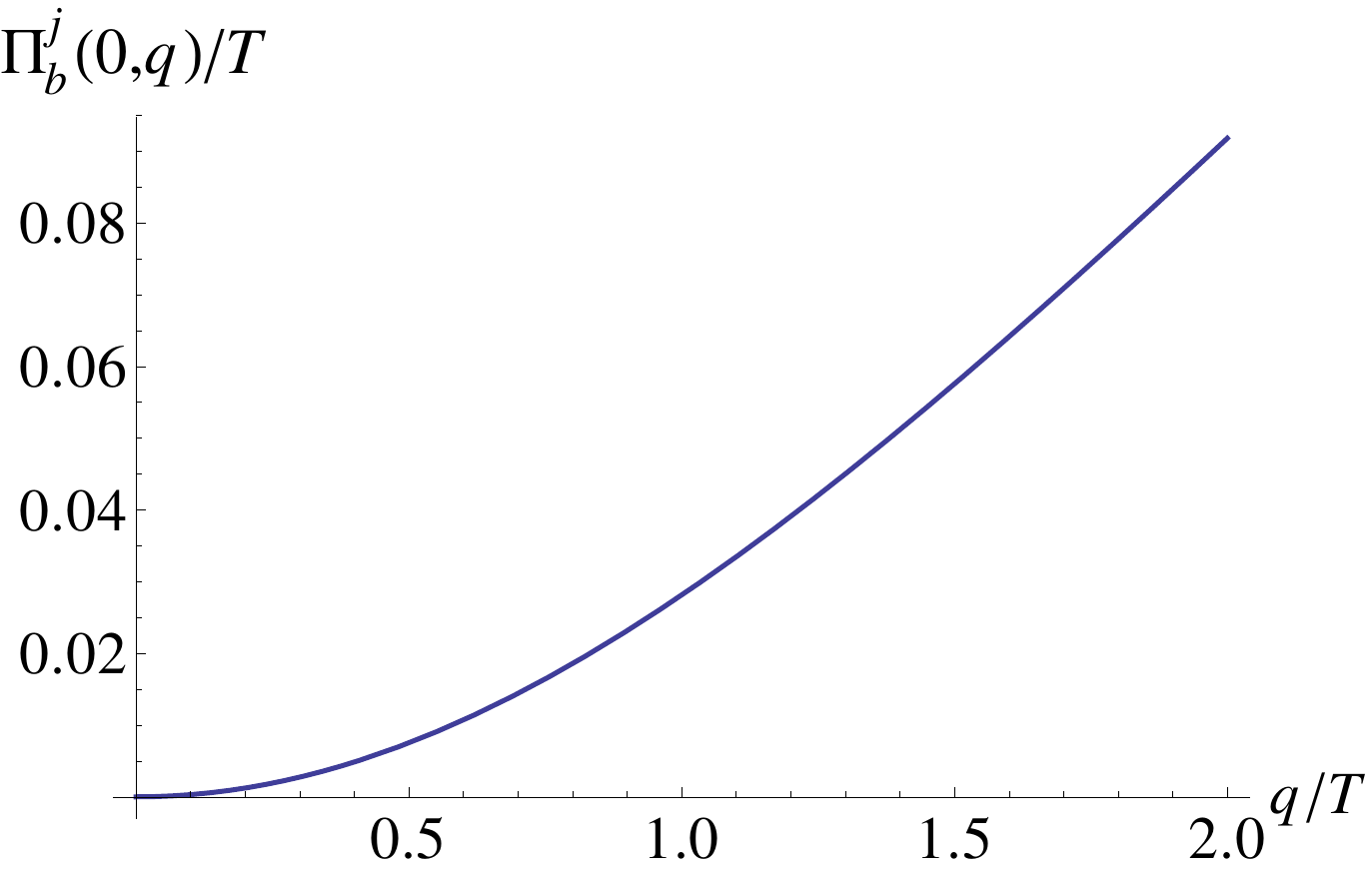}
\caption{\label{fig:Pi_b_j} Rotor current polarization function at $\de=0$ so that the rotor mass is $m=\Theta T$.
At small $q$, $\Pi_b^{\rm j}\sim q^2$, while for $q>T$ it is linear.}
\end{figure}

The static, non-regularized rotor current polarization function is obtained
from:
\begin{align}
  \Pi_b^{\rm j}(0,q)=-T\sum_{n}\int \frac{d^2\b p}{(2\pi)^2}
  \frac{(2\hat q\times\b p)^2}{\w_n^2+\e_p^2}\frac{1}{\w_n^2+\e_{p+q}^2}
\end{align}
where $\e_p=\sqrt{p^2+m^2}$ is the rotor energy. Note the presence of
a factor of 4 due to the (squared) vertex between
the transverse component of the gauge field and the rotor current.
Performing the Matsubara sum we obtain
\begin{align}
  \Pi_b^{\rm j}(0,q)=\int \frac{d^2\b p}{(2\pi)^2}\frac{1}{2}
  \frac{(2\hat q\times\b p)^2}{\e_{p+q}^2-\e_p^2}
  \left[\frac{1+2n(\e_{p+q})}{\e_{p+q}}-\frac{1+2n(\e_{p})}{\e_{p}}\right]
\end{align}
This integral is UV-divergent; putting a cutoff would lead to $\Pi_b^{\rm j}(0,0)\neq 0$,
which would violate the U(1) gauge invariance. We regulate by subtracting the integrand
evaluated at $q=0$. Slight care must be used in doing so because of an undetermined limit $0/0$.
The regulated expression reads
\begin{align}
  \Pi_b^{\rm j}(0,q)=2\int \frac{d^2\b p}{(2\pi)^2}
  (\hat q\times\b p)^2\left\{ \frac{1}{\e_{p+q}^2-\e_p^2}
    \left[ \frac{1+2n(\e_{p+q})}{\e_{p+q}}-\frac{1+2n(\e_{p})}{\e_{p}} \right]
    +\frac{1}{2\e_p^3} [1+2n(\e_{p})-2\e_p n'(\e_p)] \right\}
\end{align}
where $n'(\e)=\pd_\e n(\e)$. This integral can be evaluated numerically and the result is plotted
in Fig.~\ref{fig:Pi_b_j}. As expected, for $q\gg T$ the polarization function scales like $q$,
as in that regime the mass ($\sim T$) is negligible and we recover the zero temperature behaviour.
At small $q$, we obtain a $q^2$ scaling, consistent with the diamagnetic response of massive bosons.

We can numerically extract the slope of the polarization function at large $q/T$ to obtain the
result quoted in \req{Pib-j}, $\s_b^0\approx 0.063$. This corresponds to the rotor conductivity
in the $\w/T\gg 1$ limit, and agrees with the analytic expression $1/16=0.0625$ given in 
Section~\ref{sec:large-w}.
 
\bibliography{mit-ref}{}

%merlin.mbs apsrev4-1.bst 2010-07-25 4.21a (PWD, AO, DPC) hacked
%Control: key (0)
%Control: author (8) initials jnrlst
%Control: editor formatted (1) identically to author
%Control: production of article title (-1) disabled
%Control: page (0) single
%Control: year (1) truncated
%Control: production of eprint (0) enabled
\begin{thebibliography}{40}%
\makeatletter
\providecommand \@ifxundefined [1]{%
 \@ifx{#1\undefined}
}%
\providecommand \@ifnum [1]{%
 \ifnum #1\expandafter \@firstoftwo
 \else \expandafter \@secondoftwo
 \fi
}%
\providecommand \@ifx [1]{%
 \ifx #1\expandafter \@firstoftwo
 \else \expandafter \@secondoftwo
 \fi
}%
\providecommand \natexlab [1]{#1}%
\providecommand \enquote  [1]{``#1''}%
\providecommand \bibnamefont  [1]{#1}%
\providecommand \bibfnamefont [1]{#1}%
\providecommand \citenamefont [1]{#1}%
\providecommand \href@noop [0]{\@secondoftwo}%
\providecommand \href [0]{\begingroup \@sanitize@url \@href}%
\providecommand \@href[1]{\@@startlink{#1}\@@href}%
\providecommand \@@href[1]{\endgroup#1\@@endlink}%
\providecommand \@sanitize@url [0]{\catcode `\\12\catcode `\$12\catcode
  `\&12\catcode `\#12\catcode `\^12\catcode `\_12\catcode `\%12\relax}%
\providecommand \@@startlink[1]{}%
\providecommand \@@endlink[0]{}%
\providecommand \url  [0]{\begingroup\@sanitize@url \@url }%
\providecommand \@url [1]{\endgroup\@href {#1}{\urlprefix }}%
\providecommand \urlprefix  [0]{URL }%
\providecommand \Eprint [0]{\href }%
\providecommand \doibase [0]{http://dx.doi.org/}%
\providecommand \selectlanguage [0]{\@gobble}%
\providecommand \bibinfo  [0]{\@secondoftwo}%
\providecommand \bibfield  [0]{\@secondoftwo}%
\providecommand \translation [1]{[#1]}%
\providecommand \BibitemOpen [0]{}%
\providecommand \bibitemStop [0]{}%
\providecommand \bibitemNoStop [0]{.\EOS\space}%
\providecommand \EOS [0]{\spacefactor3000\relax}%
\providecommand \BibitemShut  [1]{\csname bibitem#1\endcsname}%
\let\auto@bib@innerbib\@empty
%</preamble>
\bibitem [{\citenamefont {Imada}\ \emph {et~al.}(1998)\citenamefont {Imada},
  \citenamefont {Fujimori},\ and\ \citenamefont {Tokura}}]{imada98-rev}%
  \BibitemOpen
  \bibfield  {author} {\bibinfo {author} {\bibfnamefont {M.}~\bibnamefont
  {Imada}}, \bibinfo {author} {\bibfnamefont {A.}~\bibnamefont {Fujimori}}, \
  and\ \bibinfo {author} {\bibfnamefont {Y.}~\bibnamefont {Tokura}},\ }\href
  {\doibase 10.1103/RevModPhys.70.1039} {\bibfield  {journal} {\bibinfo
  {journal} {Rev. Mod. Phys.}\ }\textbf {\bibinfo {volume} {70}},\ \bibinfo
  {pages} {1039} (\bibinfo {year} {1998})}\BibitemShut {NoStop}%
\bibitem [{\citenamefont {Florens}\ and\ \citenamefont
  {Georges}(2004)}]{florens04}%
  \BibitemOpen
  \bibfield  {author} {\bibinfo {author} {\bibfnamefont {S.}~\bibnamefont
  {Florens}}\ and\ \bibinfo {author} {\bibfnamefont {A.}~\bibnamefont
  {Georges}},\ }\href {\doibase 10.1103/PhysRevB.70.035114} {\bibfield
  {journal} {\bibinfo  {journal} {Phys. Rev. B}\ }\textbf {\bibinfo {volume}
  {70}},\ \bibinfo {pages} {035114} (\bibinfo {year} {2004})}\BibitemShut
  {NoStop}%
\bibitem [{\citenamefont {Lee}\ and\ \citenamefont {Lee}(2005)}]{sslee05}%
  \BibitemOpen
  \bibfield  {author} {\bibinfo {author} {\bibfnamefont {S.-S.}\ \bibnamefont
  {Lee}}\ and\ \bibinfo {author} {\bibfnamefont {P.~A.}\ \bibnamefont {Lee}},\
  }\href {\doibase 10.1103/PhysRevLett.95.036403} {\bibfield  {journal}
  {\bibinfo  {journal} {Phys. Rev. Lett.}\ }\textbf {\bibinfo {volume} {95}},\
  \bibinfo {pages} {036403} (\bibinfo {year} {2005})}\BibitemShut {NoStop}%
\bibitem [{\citenamefont {Senthil}(2008{\natexlab{a}})}]{senthil08-1}%
  \BibitemOpen
  \bibfield  {author} {\bibinfo {author} {\bibfnamefont {T.}~\bibnamefont
  {Senthil}},\ }\href {\doibase 10.1103/PhysRevB.78.035103} {\bibfield
  {journal} {\bibinfo  {journal} {Phys. Rev. B}\ }\textbf {\bibinfo {volume}
  {78}},\ \bibinfo {pages} {035103} (\bibinfo {year}
  {2008}{\natexlab{a}})}\BibitemShut {NoStop}%
\bibitem [{\citenamefont {Senthil}(2008{\natexlab{b}})}]{senthil08-2}%
  \BibitemOpen
  \bibfield  {author} {\bibinfo {author} {\bibfnamefont {T.}~\bibnamefont
  {Senthil}},\ }\href {\doibase 10.1103/PhysRevB.78.045109} {\bibfield
  {journal} {\bibinfo  {journal} {Phys. Rev. B}\ }\textbf {\bibinfo {volume}
  {78}},\ \bibinfo {pages} {045109} (\bibinfo {year}
  {2008}{\natexlab{b}})}\BibitemShut {NoStop}%
\bibitem [{\citenamefont {Podolsky}\ \emph {et~al.}(2009)\citenamefont
  {Podolsky}, \citenamefont {Paramekanti}, \citenamefont {Kim},\ and\
  \citenamefont {Senthil}}]{podolsky09}%
  \BibitemOpen
  \bibfield  {author} {\bibinfo {author} {\bibfnamefont {D.}~\bibnamefont
  {Podolsky}}, \bibinfo {author} {\bibfnamefont {A.}~\bibnamefont
  {Paramekanti}}, \bibinfo {author} {\bibfnamefont {Y.~B.}\ \bibnamefont
  {Kim}}, \ and\ \bibinfo {author} {\bibfnamefont {T.}~\bibnamefont
  {Senthil}},\ }\href {\doibase 10.1103/PhysRevLett.102.186401} {\bibfield
  {journal} {\bibinfo  {journal} {Phys. Rev. Lett.}\ }\textbf {\bibinfo
  {volume} {102}},\ \bibinfo {pages} {186401} (\bibinfo {year}
  {2009})}\BibitemShut {NoStop}%
\bibitem [{\citenamefont {{Potter}}\ \emph {et~al.}(2012)\citenamefont
  {{Potter}}, \citenamefont {{Barkeshli}}, \citenamefont {{McGreevy}},\ and\
  \citenamefont {{Senthil}}}]{potter12}%
  \BibitemOpen
  \bibfield  {author} {\bibinfo {author} {\bibfnamefont {A.~C.}\ \bibnamefont
  {{Potter}}}, \bibinfo {author} {\bibfnamefont {M.}~\bibnamefont
  {{Barkeshli}}}, \bibinfo {author} {\bibfnamefont {J.}~\bibnamefont
  {{McGreevy}}}, \ and\ \bibinfo {author} {\bibfnamefont {T.}~\bibnamefont
  {{Senthil}}},\ }\href@noop {} {\bibfield  {journal} {\bibinfo  {journal}
  {ArXiv e-prints}\ } (\bibinfo {year} {2012})},\ \Eprint
  {http://arxiv.org/abs/1204.1342} {arXiv:1204.1342 [cond-mat.str-el]}
  \BibitemShut {NoStop}%
\bibitem [{\citenamefont {{Nandkishore}}\ \emph {et~al.}(2012)\citenamefont
  {{Nandkishore}}, \citenamefont {{Metlitski}},\ and\ \citenamefont
  {{Senthil}}}]{rahul}%
  \BibitemOpen
  \bibfield  {author} {\bibinfo {author} {\bibfnamefont {R.}~\bibnamefont
  {{Nandkishore}}}, \bibinfo {author} {\bibfnamefont {M.~A.}\ \bibnamefont
  {{Metlitski}}}, \ and\ \bibinfo {author} {\bibfnamefont {T.}~\bibnamefont
  {{Senthil}}},\ }\href@noop {} {\bibfield  {journal} {\bibinfo  {journal}
  {ArXiv e-prints}\ } (\bibinfo {year} {2012})},\ \Eprint
  {http://arxiv.org/abs/1201.5998} {arXiv:1201.5998 [cond-mat.str-el]}
  \BibitemShut {NoStop}%
\bibitem [{\citenamefont {Anderson}(1973)}]{anderson73}%
  \BibitemOpen
  \bibfield  {author} {\bibinfo {author} {\bibfnamefont {P.}~\bibnamefont
  {Anderson}},\ }\href {\doibase 10.1016/0025-5408(73)90167-0} {\bibfield
  {journal} {\bibinfo  {journal} {Materials Research Bulletin}\ }\textbf
  {\bibinfo {volume} {8}},\ \bibinfo {pages} {153 } (\bibinfo {year}
  {1973})}\BibitemShut {NoStop}%
\bibitem [{\citenamefont {Shimizu}\ \emph {et~al.}(2003)\citenamefont
  {Shimizu}, \citenamefont {Miyagawa}, \citenamefont {Kanoda}, \citenamefont
  {Maesato},\ and\ \citenamefont {Saito}}]{kapEt_NMR_chi}%
  \BibitemOpen
  \bibfield  {author} {\bibinfo {author} {\bibfnamefont {Y.}~\bibnamefont
  {Shimizu}}, \bibinfo {author} {\bibfnamefont {K.}~\bibnamefont {Miyagawa}},
  \bibinfo {author} {\bibfnamefont {K.}~\bibnamefont {Kanoda}}, \bibinfo
  {author} {\bibfnamefont {M.}~\bibnamefont {Maesato}}, \ and\ \bibinfo
  {author} {\bibfnamefont {G.}~\bibnamefont {Saito}},\ }\href {\doibase
  10.1103/PhysRevLett.91.107001} {\bibfield  {journal} {\bibinfo  {journal}
  {Phys. Rev. Lett.}\ }\textbf {\bibinfo {volume} {91}},\ \bibinfo {pages}
  {107001} (\bibinfo {year} {2003})}\BibitemShut {NoStop}%
\bibitem [{\citenamefont {Yamashita}\ \emph {et~al.}(2011)\citenamefont
  {Yamashita}, \citenamefont {Yamamoto}, \citenamefont {Nakazawa},
  \citenamefont {Tamura},\ and\ \citenamefont {Kato}}]{sdmit_heat-cap}%
  \BibitemOpen
  \bibfield  {author} {\bibinfo {author} {\bibfnamefont {S.}~\bibnamefont
  {Yamashita}}, \bibinfo {author} {\bibfnamefont {T.}~\bibnamefont {Yamamoto}},
  \bibinfo {author} {\bibfnamefont {Y.}~\bibnamefont {Nakazawa}}, \bibinfo
  {author} {\bibfnamefont {M.}~\bibnamefont {Tamura}}, \ and\ \bibinfo {author}
  {\bibfnamefont {R.}~\bibnamefont {Kato}},\ }\href {\doibase
  10.1038/ncomms1274} {\bibfield  {journal} {\bibinfo  {journal} {Nature
  communications}\ }\textbf {\bibinfo {volume} {2}},\ \bibinfo {pages} {275+}
  (\bibinfo {year} {2011})}\BibitemShut {NoStop}%
\bibitem [{\citenamefont {Itou}\ \emph {et~al.}(2008)\citenamefont {Itou},
  \citenamefont {Oyamada}, \citenamefont {Maegawa}, \citenamefont {Tamura},\
  and\ \citenamefont {Kato}}]{kato-prb08}%
  \BibitemOpen
  \bibfield  {author} {\bibinfo {author} {\bibfnamefont {T.}~\bibnamefont
  {Itou}}, \bibinfo {author} {\bibfnamefont {A.}~\bibnamefont {Oyamada}},
  \bibinfo {author} {\bibfnamefont {S.}~\bibnamefont {Maegawa}}, \bibinfo
  {author} {\bibfnamefont {M.}~\bibnamefont {Tamura}}, \ and\ \bibinfo {author}
  {\bibfnamefont {R.}~\bibnamefont {Kato}},\ }\href {\doibase
  10.1103/PhysRevB.77.104413} {\bibfield  {journal} {\bibinfo  {journal} {Phys.
  Rev. B}\ }\textbf {\bibinfo {volume} {77}},\ \bibinfo {pages} {104413}
  (\bibinfo {year} {2008})}\BibitemShut {NoStop}%
\bibitem [{\citenamefont {Yamashita}\ \emph
  {et~al.}(2008{\natexlab{a}})\citenamefont {Yamashita}, \citenamefont
  {Nakazawa}, \citenamefont {Oguni}, \citenamefont {Oshima}, \citenamefont
  {Nojiri}, \citenamefont {Shimizu}, \citenamefont {Miyagawa},\ and\
  \citenamefont {Kanoda}}]{kapEt_heat-cap}%
  \BibitemOpen
  \bibfield  {author} {\bibinfo {author} {\bibfnamefont {S.}~\bibnamefont
  {Yamashita}}, \bibinfo {author} {\bibfnamefont {Y.}~\bibnamefont {Nakazawa}},
  \bibinfo {author} {\bibfnamefont {M.}~\bibnamefont {Oguni}}, \bibinfo
  {author} {\bibfnamefont {Y.}~\bibnamefont {Oshima}}, \bibinfo {author}
  {\bibfnamefont {H.}~\bibnamefont {Nojiri}}, \bibinfo {author} {\bibfnamefont
  {Y.}~\bibnamefont {Shimizu}}, \bibinfo {author} {\bibfnamefont
  {K.}~\bibnamefont {Miyagawa}}, \ and\ \bibinfo {author} {\bibfnamefont
  {K.}~\bibnamefont {Kanoda}},\ }\href {\doibase 10.1038/nphys942} {\bibfield
  {journal} {\bibinfo  {journal} {Nature Physics}\ }\textbf {\bibinfo {volume}
  {4}},\ \bibinfo {pages} {459} (\bibinfo {year}
  {2008}{\natexlab{a}})}\BibitemShut {NoStop}%
\bibitem [{\citenamefont {Yamashita}\ \emph
  {et~al.}(2008{\natexlab{b}})\citenamefont {Yamashita}, \citenamefont
  {Nakata}, \citenamefont {Kasahara}, \citenamefont {Sasaki}, \citenamefont
  {Yoneyama}, \citenamefont {Kobayashi}, \citenamefont {Fujimoto},
  \citenamefont {Shibauchi},\ and\ \citenamefont
  {Matsuda}}]{kapEt_thermal-transp}%
  \BibitemOpen
  \bibfield  {author} {\bibinfo {author} {\bibfnamefont {M.}~\bibnamefont
  {Yamashita}}, \bibinfo {author} {\bibfnamefont {N.}~\bibnamefont {Nakata}},
  \bibinfo {author} {\bibfnamefont {Y.}~\bibnamefont {Kasahara}}, \bibinfo
  {author} {\bibfnamefont {T.}~\bibnamefont {Sasaki}}, \bibinfo {author}
  {\bibfnamefont {N.}~\bibnamefont {Yoneyama}}, \bibinfo {author}
  {\bibfnamefont {N.}~\bibnamefont {Kobayashi}}, \bibinfo {author}
  {\bibfnamefont {S.}~\bibnamefont {Fujimoto}}, \bibinfo {author}
  {\bibfnamefont {T.}~\bibnamefont {Shibauchi}}, \ and\ \bibinfo {author}
  {\bibfnamefont {Y.}~\bibnamefont {Matsuda}},\ }\href {\doibase
  10.1038/nphys1134} {\bibfield  {journal} {\bibinfo  {journal} {Nature
  Physics}\ }\textbf {\bibinfo {volume} {5}},\ \bibinfo {pages} {44} (\bibinfo
  {year} {2008}{\natexlab{b}})}\BibitemShut {NoStop}%
\bibitem [{\citenamefont {Kanoda}\ and\ \citenamefont
  {Kato}(2011)}]{kato_kanoda11-rev}%
  \BibitemOpen
  \bibfield  {author} {\bibinfo {author} {\bibfnamefont {K.}~\bibnamefont
  {Kanoda}}\ and\ \bibinfo {author} {\bibfnamefont {R.}~\bibnamefont {Kato}},\
  }\href {\doibase 10.1146/annurev-conmatphys-062910-140521} {\bibfield
  {journal} {\bibinfo  {journal} {Annual Review of Condensed Matter Physics}\
  }\textbf {\bibinfo {volume} {2}},\ \bibinfo {pages} {167} (\bibinfo {year}
  {2011})}\BibitemShut {NoStop}%
\bibitem [{\citenamefont {Okamoto}\ \emph {et~al.}(2007)\citenamefont
  {Okamoto}, \citenamefont {Nohara}, \citenamefont {Aruga-Katori},\ and\
  \citenamefont {Takagi}}]{takagi}%
  \BibitemOpen
  \bibfield  {author} {\bibinfo {author} {\bibfnamefont {Y.}~\bibnamefont
  {Okamoto}}, \bibinfo {author} {\bibfnamefont {M.}~\bibnamefont {Nohara}},
  \bibinfo {author} {\bibfnamefont {H.}~\bibnamefont {Aruga-Katori}}, \ and\
  \bibinfo {author} {\bibfnamefont {H.}~\bibnamefont {Takagi}},\ }\href
  {\doibase 10.1103/PhysRevLett.99.137207} {\bibfield  {journal} {\bibinfo
  {journal} {Phys. Rev. Lett.}\ }\textbf {\bibinfo {volume} {99}},\ \bibinfo
  {pages} {137207} (\bibinfo {year} {2007})}\BibitemShut {NoStop}%
\bibitem [{\citenamefont {Kurosaki}\ \emph {et~al.}(2005)\citenamefont
  {Kurosaki}, \citenamefont {Shimizu}, \citenamefont {Miyagawa}, \citenamefont
  {Kanoda},\ and\ \citenamefont {Saito}}]{kapEt_p-nmr-res}%
  \BibitemOpen
  \bibfield  {author} {\bibinfo {author} {\bibfnamefont {Y.}~\bibnamefont
  {Kurosaki}}, \bibinfo {author} {\bibfnamefont {Y.}~\bibnamefont {Shimizu}},
  \bibinfo {author} {\bibfnamefont {K.}~\bibnamefont {Miyagawa}}, \bibinfo
  {author} {\bibfnamefont {K.}~\bibnamefont {Kanoda}}, \ and\ \bibinfo {author}
  {\bibfnamefont {G.}~\bibnamefont {Saito}},\ }\href {\doibase
  10.1103/PhysRevLett.95.177001} {\bibfield  {journal} {\bibinfo  {journal}
  {Phys. Rev. Lett.}\ }\textbf {\bibinfo {volume} {95}},\ \bibinfo {pages}
  {177001} (\bibinfo {year} {2005})}\BibitemShut {NoStop}%
\bibitem [{\citenamefont {Motrunich}(2006)}]{motrunich05}%
  \BibitemOpen
  \bibfield  {author} {\bibinfo {author} {\bibfnamefont {O.~I.}\ \bibnamefont
  {Motrunich}},\ }\href {\doibase 10.1103/PhysRevB.73.155115} {\bibfield
  {journal} {\bibinfo  {journal} {Phys. Rev. B}\ }\textbf {\bibinfo {volume}
  {73}},\ \bibinfo {pages} {155115} (\bibinfo {year} {2006})}\BibitemShut
  {NoStop}%
\bibitem [{\citenamefont {Senthil}\ \emph {et~al.}(2004)\citenamefont
  {Senthil}, \citenamefont {Vojta},\ and\ \citenamefont {Sachdev}}]{svslong}%
  \BibitemOpen
  \bibfield  {author} {\bibinfo {author} {\bibfnamefont {T.}~\bibnamefont
  {Senthil}}, \bibinfo {author} {\bibfnamefont {M.}~\bibnamefont {Vojta}}, \
  and\ \bibinfo {author} {\bibfnamefont {S.}~\bibnamefont {Sachdev}},\ }\href
  {\doibase 10.1103/PhysRevB.69.035111} {\bibfield  {journal} {\bibinfo
  {journal} {Phys. Rev. B}\ }\textbf {\bibinfo {volume} {69}},\ \bibinfo
  {pages} {035111} (\bibinfo {year} {2004})}\BibitemShut {NoStop}%
\bibitem [{\citenamefont {Sachdev}(2011)}]{sachdev-book}%
  \BibitemOpen
  \bibfield  {author} {\bibinfo {author} {\bibfnamefont {S.}~\bibnamefont
  {Sachdev}},\ }\href@noop {} {\emph {\bibinfo {title} {Quantum Phase
  Transitions}}},\ \bibinfo {edition} {2nd}\ ed.\ (\bibinfo  {publisher}
  {Cambridge University Press},\ \bibinfo {address} {England},\ \bibinfo {year}
  {2011})\BibitemShut {NoStop}%
\bibitem [{\citenamefont {Damle}\ and\ \citenamefont
  {Sachdev}(1997)}]{damle97}%
  \BibitemOpen
  \bibfield  {author} {\bibinfo {author} {\bibfnamefont {K.}~\bibnamefont
  {Damle}}\ and\ \bibinfo {author} {\bibfnamefont {S.}~\bibnamefont
  {Sachdev}},\ }\href {\doibase 10.1103/PhysRevB.56.8714} {\bibfield  {journal}
  {\bibinfo  {journal} {Phys. Rev. B}\ }\textbf {\bibinfo {volume} {56}},\
  \bibinfo {pages} {8714} (\bibinfo {year} {1997})}\BibitemShut {NoStop}%
\bibitem [{\citenamefont {Lee}(2009)}]{lee-largeN}%
  \BibitemOpen
  \bibfield  {author} {\bibinfo {author} {\bibfnamefont {S.-S.}\ \bibnamefont
  {Lee}},\ }\href {\doibase 10.1103/PhysRevB.80.165102} {\bibfield  {journal}
  {\bibinfo  {journal} {Phys. Rev. B}\ }\textbf {\bibinfo {volume} {80}},\
  \bibinfo {pages} {165102} (\bibinfo {year} {2009})}\BibitemShut {NoStop}%
\bibitem [{\citenamefont {Nayak}\ and\ \citenamefont
  {Wilczek}(1994)}]{Nayak94}%
  \BibitemOpen
  \bibfield  {author} {\bibinfo {author} {\bibfnamefont {C.}~\bibnamefont
  {Nayak}}\ and\ \bibinfo {author} {\bibfnamefont {F.}~\bibnamefont
  {Wilczek}},\ }\href {\doibase 10.1016/0550-3213(94)90477-4} {\bibfield
  {journal} {\bibinfo  {journal} {Nuclear Physics B}\ }\textbf {\bibinfo
  {volume} {417}},\ \bibinfo {pages} {359 } (\bibinfo {year}
  {1994})}\BibitemShut {NoStop}%
\bibitem [{\citenamefont {Mross}\ \emph {et~al.}(2010)\citenamefont {Mross},
  \citenamefont {McGreevy}, \citenamefont {Liu},\ and\ \citenamefont
  {Senthil}}]{Mross10}%
  \BibitemOpen
  \bibfield  {author} {\bibinfo {author} {\bibfnamefont {D.~F.}\ \bibnamefont
  {Mross}}, \bibinfo {author} {\bibfnamefont {J.}~\bibnamefont {McGreevy}},
  \bibinfo {author} {\bibfnamefont {H.}~\bibnamefont {Liu}}, \ and\ \bibinfo
  {author} {\bibfnamefont {T.}~\bibnamefont {Senthil}},\ }\href {\doibase
  10.1103/PhysRevB.82.045121} {\bibfield  {journal} {\bibinfo  {journal} {Phys.
  Rev. B}\ }\textbf {\bibinfo {volume} {82}},\ \bibinfo {pages} {045121}
  (\bibinfo {year} {2010})}\BibitemShut {NoStop}%
\bibitem [{\citenamefont {Kaul}\ \emph {et~al.}(2008)\citenamefont {Kaul},
  \citenamefont {Metlitski}, \citenamefont {Sachdev},\ and\ \citenamefont
  {Xu}}]{kaul08}%
  \BibitemOpen
  \bibfield  {author} {\bibinfo {author} {\bibfnamefont {R.~K.}\ \bibnamefont
  {Kaul}}, \bibinfo {author} {\bibfnamefont {M.~A.}\ \bibnamefont {Metlitski}},
  \bibinfo {author} {\bibfnamefont {S.}~\bibnamefont {Sachdev}}, \ and\
  \bibinfo {author} {\bibfnamefont {C.}~\bibnamefont {Xu}},\ }\href {\doibase
  10.1103/PhysRevB.78.045110} {\bibfield  {journal} {\bibinfo  {journal} {Phys.
  Rev. B}\ }\textbf {\bibinfo {volume} {78}},\ \bibinfo {pages} {045110}
  (\bibinfo {year} {2008})}\BibitemShut {NoStop}%
\bibitem [{\citenamefont {Ioffe}\ and\ \citenamefont {Larkin}(1989)}]{ioffe89}%
  \BibitemOpen
  \bibfield  {author} {\bibinfo {author} {\bibfnamefont {L.~B.}\ \bibnamefont
  {Ioffe}}\ and\ \bibinfo {author} {\bibfnamefont {A.~I.}\ \bibnamefont
  {Larkin}},\ }\href {\doibase 10.1103/PhysRevB.39.8988} {\bibfield  {journal}
  {\bibinfo  {journal} {Phys. Rev. B}\ }\textbf {\bibinfo {volume} {39}},\
  \bibinfo {pages} {8988} (\bibinfo {year} {1989})}\BibitemShut {NoStop}%
\bibitem [{\citenamefont {Cha}\ \emph {et~al.}(1991)\citenamefont {Cha},
  \citenamefont {Fisher}, \citenamefont {Girvin}, \citenamefont {Wallin},\ and\
  \citenamefont {Young}}]{cha91}%
  \BibitemOpen
  \bibfield  {author} {\bibinfo {author} {\bibfnamefont {M.-C.}\ \bibnamefont
  {Cha}}, \bibinfo {author} {\bibfnamefont {M.~P.~A.}\ \bibnamefont {Fisher}},
  \bibinfo {author} {\bibfnamefont {S.~M.}\ \bibnamefont {Girvin}}, \bibinfo
  {author} {\bibfnamefont {M.}~\bibnamefont {Wallin}}, \ and\ \bibinfo {author}
  {\bibfnamefont {A.~P.}\ \bibnamefont {Young}},\ }\href {\doibase
  10.1103/PhysRevB.44.6883} {\bibfield  {journal} {\bibinfo  {journal} {Phys.
  Rev. B}\ }\textbf {\bibinfo {volume} {44}},\ \bibinfo {pages} {6883}
  (\bibinfo {year} {1991})}\BibitemShut {NoStop}%
\bibitem [{\citenamefont {Sachdev}(1998)}]{sachdev98}%
  \BibitemOpen
  \bibfield  {author} {\bibinfo {author} {\bibfnamefont {S.}~\bibnamefont
  {Sachdev}},\ }\href {\doibase 10.1103/PhysRevB.57.7157} {\bibfield  {journal}
  {\bibinfo  {journal} {Phys. Rev. B}\ }\textbf {\bibinfo {volume} {57}},\
  \bibinfo {pages} {7157} (\bibinfo {year} {1998})}\BibitemShut {NoStop}%
\bibitem [{\citenamefont {Witczak-Krempa}\ and\ \citenamefont
  {Sachdev}(2012)}]{will-subir}%
  \BibitemOpen
  \bibfield  {author} {\bibinfo {author} {\bibfnamefont {W.}~\bibnamefont
  {Witczak-Krempa}}\ and\ \bibinfo {author} {\bibfnamefont {S.}~\bibnamefont
  {Sachdev}},\ }\href {\doibase 10.1103/PhysRevB.86.235115} {\bibfield
  {journal} {\bibinfo  {journal} {Phys. Rev. B}\ }\textbf {\bibinfo {volume}
  {86}},\ \bibinfo {pages} {235115} (\bibinfo {year} {2012})}\BibitemShut
  {NoStop}%
\bibitem [{\citenamefont {Chubukov}\ \emph {et~al.}(1994)\citenamefont
  {Chubukov}, \citenamefont {Sachdev},\ and\ \citenamefont {Ye}}]{chubukov94}%
  \BibitemOpen
  \bibfield  {author} {\bibinfo {author} {\bibfnamefont {A.~V.}\ \bibnamefont
  {Chubukov}}, \bibinfo {author} {\bibfnamefont {S.}~\bibnamefont {Sachdev}}, \
  and\ \bibinfo {author} {\bibfnamefont {J.}~\bibnamefont {Ye}},\ }\href
  {\doibase 10.1103/PhysRevB.49.11919} {\bibfield  {journal} {\bibinfo
  {journal} {Phys. Rev. B}\ }\textbf {\bibinfo {volume} {49}},\ \bibinfo
  {pages} {11919} (\bibinfo {year} {1994})}\BibitemShut {NoStop}%
\bibitem [{\citenamefont {Lee}\ and\ \citenamefont {Nagaosa}(1992)}]{lee92}%
  \BibitemOpen
  \bibfield  {author} {\bibinfo {author} {\bibfnamefont {P.~A.}\ \bibnamefont
  {Lee}}\ and\ \bibinfo {author} {\bibfnamefont {N.}~\bibnamefont {Nagaosa}},\
  }\href {\doibase 10.1103/PhysRevB.46.5621} {\bibfield  {journal} {\bibinfo
  {journal} {Phys. Rev. B}\ }\textbf {\bibinfo {volume} {46}},\ \bibinfo
  {pages} {5621} (\bibinfo {year} {1992})}\BibitemShut {NoStop}%
\bibitem [{\citenamefont {Kim}\ \emph {et~al.}(1995)\citenamefont {Kim},
  \citenamefont {Lee},\ and\ \citenamefont {Wen}}]{yb-qbe}%
  \BibitemOpen
  \bibfield  {author} {\bibinfo {author} {\bibfnamefont {Y.~B.}\ \bibnamefont
  {Kim}}, \bibinfo {author} {\bibfnamefont {P.~A.}\ \bibnamefont {Lee}}, \ and\
  \bibinfo {author} {\bibfnamefont {X.-G.}\ \bibnamefont {Wen}},\ }\href
  {\doibase 10.1103/PhysRevB.52.17275} {\bibfield  {journal} {\bibinfo
  {journal} {Phys. Rev. B}\ }\textbf {\bibinfo {volume} {52}},\ \bibinfo
  {pages} {17275} (\bibinfo {year} {1995})}\BibitemShut {NoStop}%
\bibitem [{\citenamefont {Nave}\ and\ \citenamefont {Lee}(2007)}]{nave}%
  \BibitemOpen
  \bibfield  {author} {\bibinfo {author} {\bibfnamefont {C.~P.}\ \bibnamefont
  {Nave}}\ and\ \bibinfo {author} {\bibfnamefont {P.~A.}\ \bibnamefont {Lee}},\
  }\href {\doibase 10.1103/PhysRevB.76.235124} {\bibfield  {journal} {\bibinfo
  {journal} {Phys. Rev. B}\ }\textbf {\bibinfo {volume} {76}},\ \bibinfo
  {pages} {235124} (\bibinfo {year} {2007})}\BibitemShut {NoStop}%
\bibitem [{\citenamefont {Senthil}()}]{senthil-cft}%
  \BibitemOpen
  \bibfield  {author} {\bibinfo {author} {\bibfnamefont {T.}~\bibnamefont
  {Senthil}},\ }\href@noop {} {}\bibinfo {howpublished}
  {unpublished}\BibitemShut {NoStop}%
\bibitem [{\citenamefont {Vojta}\ \emph {et~al.}(2000)\citenamefont {Vojta},
  \citenamefont {Zhang},\ and\ \citenamefont {Sachdev}}]{vojta00}%
  \BibitemOpen
  \bibfield  {author} {\bibinfo {author} {\bibfnamefont {M.}~\bibnamefont
  {Vojta}}, \bibinfo {author} {\bibfnamefont {Y.}~\bibnamefont {Zhang}}, \ and\
  \bibinfo {author} {\bibfnamefont {S.}~\bibnamefont {Sachdev}},\ }\href
  {\doibase 10.1142/S0217979200004271} {\bibfield  {journal} {\bibinfo
  {journal} {Int. J. Mod. Phys. B}\ }\textbf {\bibinfo {volume} {14}},\
  \bibinfo {pages} {3719} (\bibinfo {year} {2000})}\BibitemShut {NoStop}%
\bibitem [{\citenamefont {Kanoda}()}]{kanoda-up}%
  \BibitemOpen
  \bibfield  {author} {\bibinfo {author} {\bibfnamefont {K.}~\bibnamefont
  {Kanoda}},\ }\href@noop {} {}\bibinfo {howpublished} {\emph{et al.},
  unpublished}\BibitemShut {NoStop}%
\bibitem [{\citenamefont {Kato}()}]{kato-up}%
  \BibitemOpen
  \bibfield  {author} {\bibinfo {author} {\bibfnamefont {R.}~\bibnamefont
  {Kato}},\ }\href@noop {} {}\bibinfo {howpublished} {\emph{et al.},
  unpublished}\BibitemShut {NoStop}%
\bibitem [{\citenamefont {Bhaseen}\ \emph {et~al.}(2007)\citenamefont
  {Bhaseen}, \citenamefont {Green},\ and\ \citenamefont {Sondhi}}]{green-prl}%
  \BibitemOpen
  \bibfield  {author} {\bibinfo {author} {\bibfnamefont {M.~J.}\ \bibnamefont
  {Bhaseen}}, \bibinfo {author} {\bibfnamefont {A.~G.}\ \bibnamefont {Green}},
  \ and\ \bibinfo {author} {\bibfnamefont {S.~L.}\ \bibnamefont {Sondhi}},\
  }\href {\doibase 10.1103/PhysRevLett.98.166801} {\bibfield  {journal}
  {\bibinfo  {journal} {Phys. Rev. Lett.}\ }\textbf {\bibinfo {volume} {98}},\
  \bibinfo {pages} {166801} (\bibinfo {year} {2007})}\BibitemShut {NoStop}%
\bibitem [{\citenamefont {Bhaseen}\ \emph {et~al.}(2009)\citenamefont
  {Bhaseen}, \citenamefont {Green},\ and\ \citenamefont {Sondhi}}]{green-prb}%
  \BibitemOpen
  \bibfield  {author} {\bibinfo {author} {\bibfnamefont {M.~J.}\ \bibnamefont
  {Bhaseen}}, \bibinfo {author} {\bibfnamefont {A.~G.}\ \bibnamefont {Green}},
  \ and\ \bibinfo {author} {\bibfnamefont {S.~L.}\ \bibnamefont {Sondhi}},\
  }\href {\doibase 10.1103/PhysRevB.79.094502} {\bibfield  {journal} {\bibinfo
  {journal} {Phys. Rev. B}\ }\textbf {\bibinfo {volume} {79}},\ \bibinfo
  {pages} {094502} (\bibinfo {year} {2009})}\BibitemShut {NoStop}%
\bibitem [{\citenamefont {Fritz}\ and\ \citenamefont
  {Sachdev}(2009)}]{fritz09}%
  \BibitemOpen
  \bibfield  {author} {\bibinfo {author} {\bibfnamefont {L.}~\bibnamefont
  {Fritz}}\ and\ \bibinfo {author} {\bibfnamefont {S.}~\bibnamefont
  {Sachdev}},\ }\href {\doibase 10.1103/PhysRevB.80.144503} {\bibfield
  {journal} {\bibinfo  {journal} {Phys. Rev. B}\ }\textbf {\bibinfo {volume}
  {80}},\ \bibinfo {pages} {144503} (\bibinfo {year} {2009})}\BibitemShut
  {NoStop}%
\end{thebibliography}%
\end{document}